\pdfoutput=1
\documentclass[a4paper,fleqn,usenatbib]{mnras}
\usepackage{kpfonts}
\usepackage{bm}
\usepackage{amsmath}
\usepackage{natbib}
\usepackage{graphicx}
\usepackage[T1]{fontenc}
\usepackage{lmodern}
\usepackage{amssymb}
\usepackage{lipsum}
\usepackage{flushend}

\title[Extending VCA for studying turbulence anisotropies]{Extending velocity channel analysis for studying turbulence anisotropies}

\author[Kandel, Lazarian \& Pogosyan]{
D. Kandel,$^{1}$\thanks{E-mail: dkandel@ualberta.ca}
A. Lazarian,$^{2}$
D. Pogosyan$^{1, 3}$
\\
$^{1}$Physics Department, University of Alberta, Edmonton, T6G 2E1, Canada\\
$^{2}$Department of Astronomy, University of Wisconsin, 475 North Charter Street, Madison, WI 53706, USA\\
$^{3}$CNRS and UPMC, UMR 7095, Institut d'Astrophysique de Paris, F-75014, Paris, France\\
}

\date{Accepted 2016 May 26. Received 2016 May 20; in original form 2016 April 15}

\pubyear{2016}

\begin{document}
\label{firstpage}
\pagerange{\pageref{firstpage}--\pageref{lastpage}}
\maketitle

\begin{abstract}
We extend the velocity channel analysis (VCA), introduced by Lazarian \& Pogosyan, of the  intensity fluctuations in the velocity slices of position-position-velocity (PPV) spectroscopic data from Doppler broadened lines to study statistical anisotropy of the underlying velocity and density that arises in a turbulent medium from the presence of magnetic field. In particular, we study analytically how the anisotropy of the intensity correlation in the channel maps changes with the thickness of velocity channels. In agreement with the earlier VCA studies we find that the anisotropy in the thick channels reflects the anisotropy of the density field, while the relative contribution of density and velocity fluctuations to the thin velocity channels depends on the density spectral slope. We show that the anisotropies arising from Alfv\'en, slow and fast magnetohydrodynamical modes are different, in particular, the anisotropy in PPV created by fast modes is opposite to that created by Alfv\'en and slow modes, and this can be used to separate their contributions. We successfully compare our results with the recent numerical study of the PPV anisotropies measured with synthetic observations. We also extend our study to the medium with self-absorption as well as to the case of absorption lines. In addition, we demonstrate how the studies of anisotropy can be performed using interferometers. 
\end{abstract}


\section{Introduction}

The interstellar medium (ISM) is turbulent on scales ranging from au to kpc. The Big Power Law in the sky obtained with electron scattering and scintillations \citep{armstrong1995electron} and extended with Wisconsin H$\alpha$ Mapper data in  \cite{chepurnov2010extending}
presents a notable example of the turbulent interstellar cascade. Numerous examples include the studies of non-thermal Doppler broadening of spectral lines, fluctuations of density and synchrotron emission (see reviews by \citealt{cho2003mhd}; \citealt{elmegreen2004interstellar}, \citealt{mac2004control}; \citealt{ballesteros2007molecular}; \citealt{2007ARA&A..45..565M}; \citealt{2009SSRv..143..357L}). 

Magnetohydrodynamical (MHD) turbulence is accepted to be
of key importance for fundamental astrophysical processes, e.g. star formation (see, e.g., \citealt{2007ARA&A..45..565M}; \citealt{federrath2012star}; \citealt{federrath2013origin};  \citealt{salim2015universal}), propagation and acceleration of cosmic rays (see \citealt{brandenburg2013astrophysical} and references therein). Therefore, understanding turbulence is important for both galactic and extragalactic research. 

How to study astrophysical turbulence? A number of recent papers demonstrated the crucial importance of
observational studies and obtaining quantitative measure from observations (see \citealt{chepurnov2009turbulence};   \citealt{brunt2010method}; \citealt{chepurnov2010velocity,chepurnov2015turbulence}; \citealt{gaensler2011low}; \citealt{burkhart2012properties};  \citealt{brunt2013principal}; \citealt{federrath2013star};   
\citealt{kainulainen2014unfolding}; \citealt{burkhart2015phase}).  We feel that this balances the field where the significant progress of numerical modelling of astrophysical turbulence shifted somewhat the attention of the astrophysical
community from observational studies. Therefore, we believe that stressing of the synergy of the observational and numerical studies is due. Indeed, present codes can
produce simulations that resemble observations (see e.g.  \citealt{federrath2013universality}) in terms of structures and scaling laws, but because of their limited numerical resolution, they cannot reach the observed Reynolds\footnote{The Reynolds number is $\text{Re}\equiv L_fV/\nu=(V/L_f)/(\nu/L^2_f)$ which is the ratio of an eddy 
turnover rate $\tau^{-1}_{\text{eddy}}=V/L_f$ to the viscous dissipation rate $\tau_{\text{dis}}^{-1}=\eta/L^2_f$. Therefore, large $\text{Re}$ correspond to 
negligible viscous dissipation of large eddies over the cascading time $\tau_{\text{casc}}$ which is equal to $\tau_{\text{eddy}}$ in Kolmogorov turbulence.} numbers of the ISM.

Statistical studies represent the best hope to bridge the gap between simulations and observations. Thus, many techniques beyond the traditional turbulence power spectrum have been developed to study and parametrize observational magnetic
turbulence. These include higher order spectra, such as
the bispectrum (\citealt{burkhart2009density}), higher order statistical moments (\citealt{kowal2007density};  \citealt{burkhart2009density}), density/column-density PDF analyses (\citealt{federrath2008density}; \citealt{burkhart2012column}), topological techniques (such as genus, see \citealt{2008ApJ...688.1021C}), clump and hierarchical structure algorithms (such as dendrograms, see \citealt{rosolowsky2008structural}; 
\citealt{burkhart2013hierarchical}), Delta variance analysis (\citealt{stutzki1998fractal}; \citealt{ossenkopf2008structure}), principal component analysis (PCA; \citealt{heyer1997application}; \citealt{heyer2008magnetically}; \citealt{roman2011turbulence}; \citealt{0004-637X-818-2-118}), Tsallis function studies for ISM turbulence (\citealt{esquivel2010tsallis}; \citealt{tofflemire2011interstellar}), 
velocity channel analysis and velocity coordinate spectrum (\citealt{lazarian2004velocity,lazarian2006studying,lazarian2008studying}),
and structure/correlation functions as tests of intermittency and anisotropy (\citealt{cho2003compressible}; \citealt{esquivel2005velocity}; \citealt{kowal2010velocity}; see also \citealt{federrath2009fractal}; \citealt{federrath2010comparing}; \citealt{konstandin2012statistical}), analysis of turbulence phase information (\citealt{burkhart2015phase}), and also recent work on filament detection (see \citealt{smith2014nature}; \citealt{federrath2016universality}) that links the structure and formation of filaments in the ISM to the statistics of turbulence.  

The turbulence spectrum, which is a statistical measure of turbulence, can be used to compare observations with both numerical simulations and theoretical predictions. Note that statistical descriptions are nearly indispensable strategy when dealing with turbulence. The big advantage of statistical techniques is that they extract underlying regularities of the flow and reject incidental details. The energy spectrum 
$E(k)\mathrm{d}k$ of turbulence characterizes how much
energy resides at the interval of scales $k, k+\mathrm{d}k$. On one hand, at large scales $l$ which correspond to small wavenumbers $k$ (i.e. $l\sim 1/k$), one expects
to observe features reflecting energy injection, while at small scales one should see the scales corresponding to dissipation of kinetic energy. On the other hand, the spectrum at intermediate scales, often called inertial range, is determined by a complex process of energy transfer, which often leads to power-law spectra. For example, in the Kolmogorov description of unmagnetized incompressible turbulence, difference in velocities at different points in turbulent fluid increases on average with the separation between points as a cubic root of the separation,
i.e. $|\delta v| \sim l^{1/3}$, which corresponds to the energy spectrum of $E(k)\sim k^{-5/3}$ in the inertial range. 
Thus, observational studies of the turbulence spectrum can determine sinks, sources and energy transfer mechanisms of astrophysical turbulence. 

There have been lot of attempts to obtain the turbulence spectra (see \citealt{munch1958space}; \citealt{kleiner1985large}; \citealt{o1987evidence}; \citealt{miesch1999velocity}).
Velocity statistics is an extremely important turbulence measure. Although it is clear that Doppler broadened lines
are affected by turbulence, recovery of velocity statistics turned out to be extremely challenging without an adequate theoretical insight. Indeed, both line-of-sight (LOS) component of velocity and density contribute to fluctuations of the energy density $\rho_s (\bm{X}, v_z)$ in the  position-position-velocity (PPV) space. This motivated the study in \cite{lazarian2000velocity, lazarian2004velocity} (henceforth LP00 and LP04, respectively) which resulted in the analytical description of the statistical properties of the PPV energy density $\rho_s$. In those papers, the observed statistics of $\rho_s$ was related to the underlying 3D spectra of velocity and density in the astrophysical turbulent volume. Initially, the volume was considered transparent (LP00), but later the treatment was generalized for the volume with self-absorption (LP04). 

The technique  developed in LP00, LP04 was termed Velocity Channel Analysis (VCA) and this technique was proposed to analyse the spectra of velocity slices of PPV data cubes by gradually changing their thickness in order to find the underlying spectra of velocity and density of astrophysical turbulent motions. This technique has been successfully tested and elaborated in a number of subsequent papers (\citealt{lazarian2001emissivity}; \citealt{chepurnov2009turbulence}; \citealt{burkhart2013turbulence}) and the VCA analysis was successfully applied to a number of
observations (see an incomplete list in \citealt{2009SSRv..143..357L}). 

The statistical description of PPV data in LP00 and LP04 provided a way to develop a completely new technique to study turbulence via analysing the fluctuations of PPV intensity along the $v$-axis (\citealt{lazarian2006studying}, henceforth LP06). The corresponding technique was termed velocity coordinate spectrum (VCS) and was successfully applied to HI and CO data in e.g. \cite{padoan2009power}, \cite{chepurnov2010extending} and \cite{chepurnov2015turbulence} to obtain velocity spectra. However, this does not exhaust the potential of the analytical description of fluctuations in PPV space. Indeed, the MHD turbulence is known to be anisotropic  with magnetic field
defining the direction of anisotropy (\citealt{montgomery1981anisotropic}; 
\citealt{shebalin1983anisotropy}; \citealt{higdon1984density}). This opens prospects of studying
the direction of magnetic field using the observed velocity fluctuations.

For the first time, the possibility of studying magnetic field with observational data was discussed in \cite{lazarian2002seeing}. In particular,  the anisotropy was shown to exist for velocity channel maps obtained with MHD numerical simulations. The research that followed (see \citealt{esquivel2005velocity};  \citealt{heyer2008magnetically}; \citealt{burkhart2015alfvenic}) proved the utility of the suggested new technique to study magnetic fields in turbulence and to obtain the information about the Alfv\'en Mach number of turbulence $M_\text{A}\equiv V_L/V_\text{A}$,
where $V_L$ and $V_\text{A}$ are the injection and Alfv\'en velocities, respectively. Importantly, 
$M_\text{A}$ determines magnetization of turbulence, and this determines crucial properties of turbulent fluid including diffusion of cosmic rays (see \citealt{yan2002scattering, yan2004cosmic, yan2008cosmic}), 
heat (\citealt{narayan2001thermal}; \citealt{lazarian2006enhancement}), as well as reconnection diffusion (\citealt{2005AIPC..784...42L}; \citealt{santos2010diffusion, santos2014magnetic}; \citealt{lazarian2012magnetization}; \citealt{leao2013collapse}; \citealt{gonzalez2016magnetic}; see \citealt{lazarian2014reconnection} for a review),
which has been identified as a crucial process for star formation (see \citealt{li2015magnetized}).

In a recent study by \mbox{\cite{esquivel2015studying}}, the dependence of fluctuations anisotropy in velocity slices of PPV data cubes has been quantified using synthetic observations obtained with 3D MHD simulations. It confirmed the original finding in \cite{lazarian2001emissivity} that the anisotropy of the correlations of intensity in the velocity slice reflects the magnetic field direction and provided the empirical dependence of the observed anisotropy on the Alfv\'en Mach number $M_\text{A}$. This work motivates our present analytical study aimed at the analytical description of the anisotropies in the velocity slices of PPV data cubes.

The present study capitalizes on the recent analytical studies of anisotropy of synchrotron fluctuations and its polarization in \cite{lazarian2012statistical, 0004-637X-818-2-178} (henceforth, LP12 and LP16 respectively). In those papers, the representation of MHD turbulence as the combination of three cascades, i.e. the Alfv\'en, fast and slow modes (see \citealt{goldreich1995toward}; \citealt{lithwick2001compressible}; \citealt{cho2002compressible, cho2003compressible};  \citealt{kowal2010velocity}), was used. For the purpose of  observational studies, magnetic fluctuations were described using tensors in the frame of the mean field, which is different from the local magnetic field of reference used in the theory of turbulence (cf. \citealt{cho2003compressible}). 

In what follows, we will use the correspondence between magnetic and velocity fluctuations in MHD turbulence to provide the description of fluctuations of intensity in the velocity slices. Similar to LP12 we will also provide the decomposition of the observed correlation function anisotropies into multipoles and, similar to LP12, we will focus on the quadrupole anisotropy.
We will also discuss in what sense the fluctuations of magnetic field and velocity field are different and what this difference entails for the analysis of astrophysical turbulence. We stress the synergetic nature of different ways of statistical studies of turbulence using various observational data sets, including magnetic anisotropy studies in this paper and in LP12 and LP16. 

Anisotropy allows one to study magnetic field direction as well as magnetization of the media (see \citealt{lazarian2001emissivity}; \citealt{esquivel2005velocity, esquivel2011velocity}; \citealt{esquivel2015studying}). However, in analogy with LP12, we should expect the anisotropies produced by different MHD modes to be different which opens a way to separate the contributions from these different modes. Note that this possibility is different from what one expects by studying turbulence based on the dispersion of probability distribution functions (see \citealt{federrath2009fractal, federrath2010comparing}; \citealt{burkhart2012column}). 

VCA provides a way of studying astrophysical turbulence by making use
of extensive spectroscopic surveys, in particular HI and CO data. The present study
significantly enhances its value and abilities. Below in Section 2 we present the qualitative
 discussion of VCA study, introduce the properties of MHD turbulence that we require
 for our study. In Section 3, we review the turbulence statistics in PPV space. In Section 4, we derive the tensor structure of different MHD modes, and in Section 5 we describe anisotropy in the intensity statistics due to anisotropy in tensor structure of density and velocity field. In Section 6, we show our results by considering pure velocity effects, as well as density effects and also carry out absorption line study and the study on effects of spatial and spectroscopic resolution. In Section 7, we discuss the effects of self-absorption on the observed anisotropy. In Section 8, we present practical guide to the results of our study, and in Section 9, we present an example to handle data from an anisotropic PPV space. In Section 10, we present some of the discussion of past works that relate to our study. The detailed derivations of velocity correlation tensors in real space for individual modes, and some of the important derivation for intensity anisotropy, are provided in  Appendix (\ref{general}-\ref{za}).

\section{Nature of PPV Space and Velocity Channel Analysis}

The nature of the PPV space has been a source of numerous confusions, with many researchers trying to identify the density enhancements in PPV with the actual density fluctuations in real space. The study in LP00 clearly showed that this is erroneous and velocity fluctuations can be responsible for a significant part of the PPV structures (see also \citealt{2009SSRv..143..357L}; 
\citealt{burkhart2013turbulence}) 

\begin{figure*}
  \centering
  \includegraphics[width=.4\textwidth]{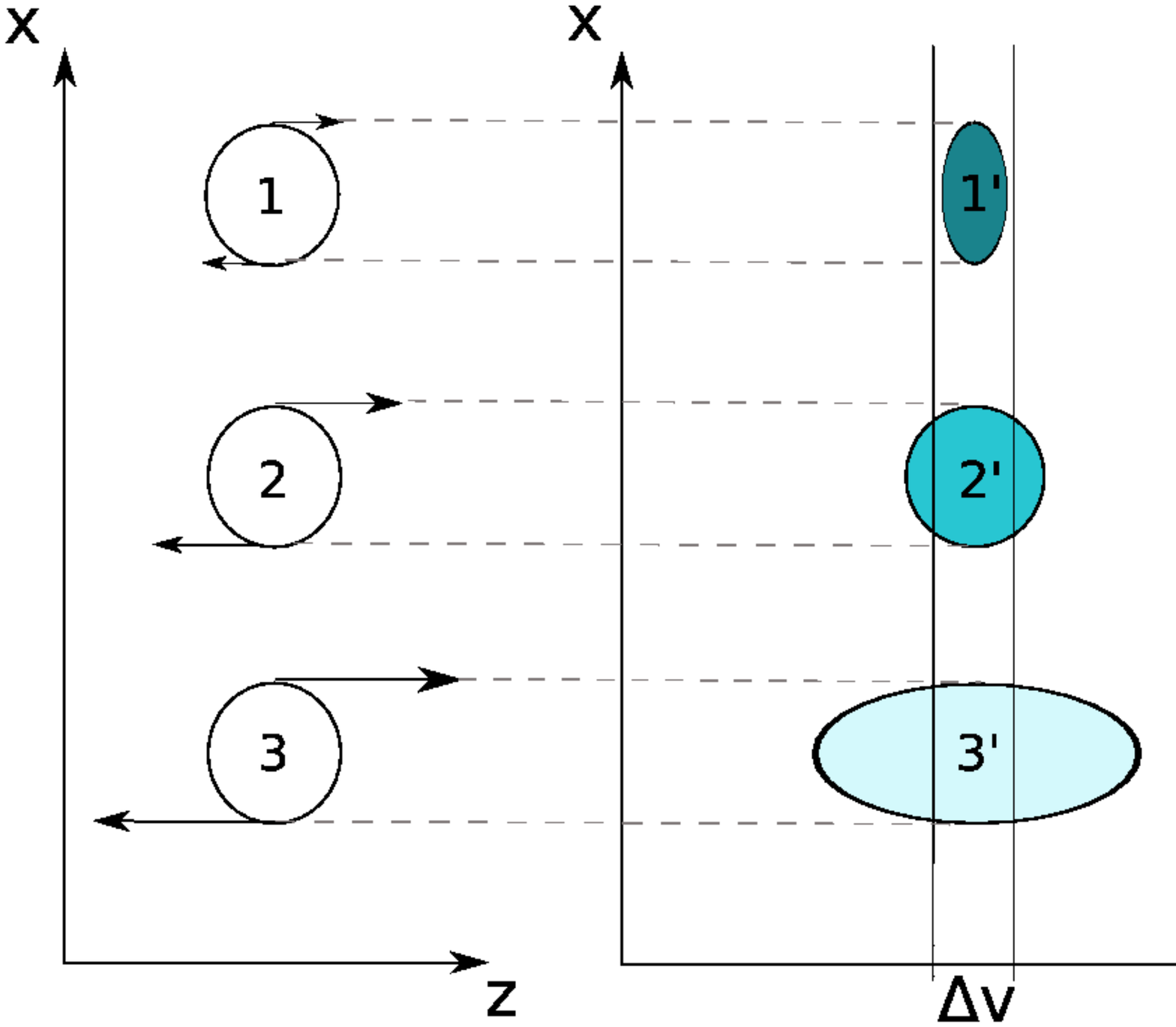}\hspace*{2cm}
   \includegraphics[width=.3\textheight]{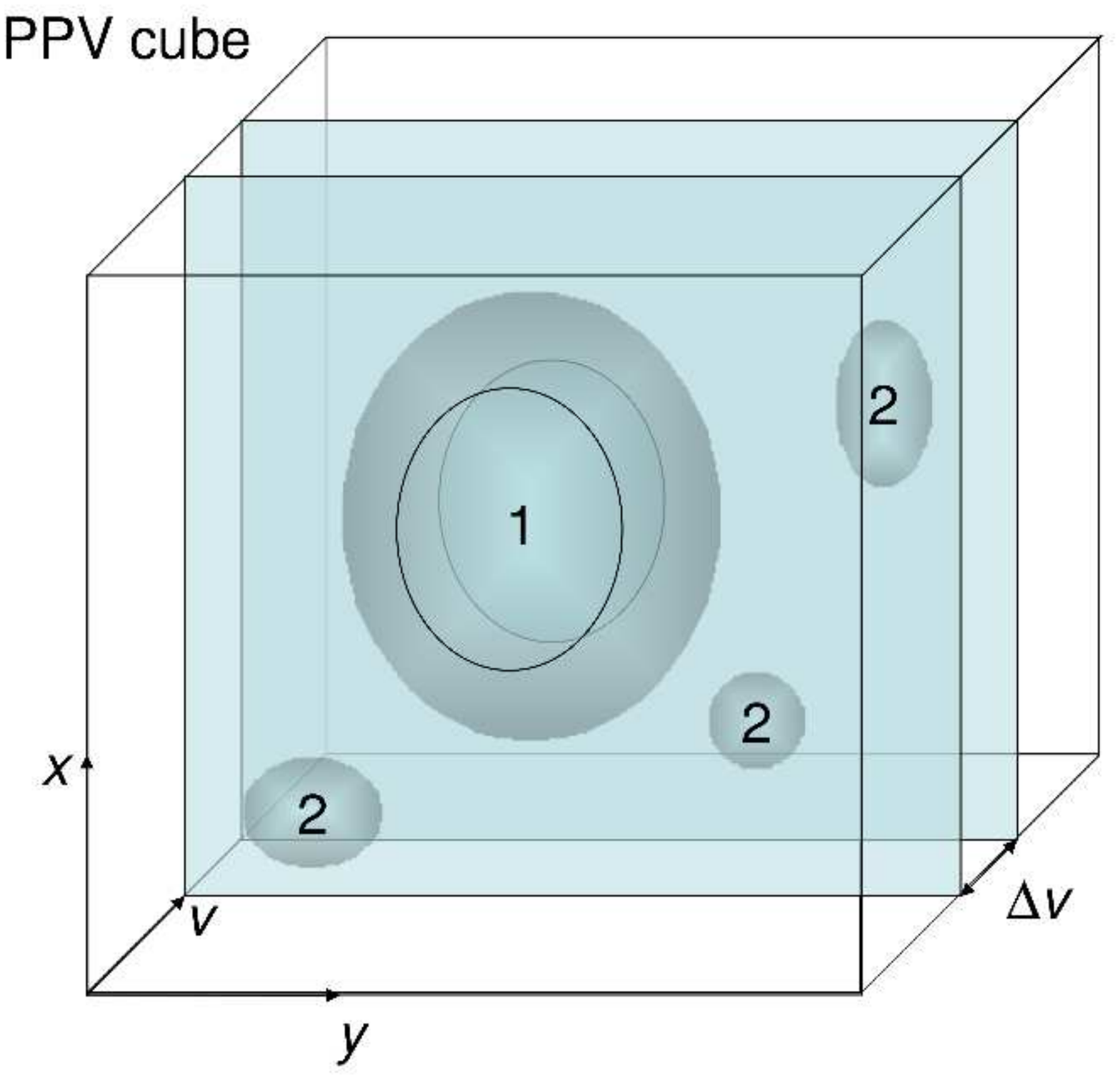}
   \caption{Left: an illustration of the mapping from the real space to the PPV space. In the real `PPP' space, the three eddies have the same
  size, the {\it same density} of emitting material, but different velocities. They are being mapped to the PPV space and there they have the same PP dimensions, but a different
  $v$-size. The larger  the velocity of eddies, the larger the $v$-extent of the eddies, which in turn implies less density of emitting atoms {\it over the image of the eddy}. 
  Therefore, in terms of the intensity of fluctuations in the velocity channel $\Delta v$, the largest contribution comes from the eddy with the least velocity dispersion, i.e.
  eddy 1,  while the eddy with the largest velocity dispersion, i.e. eddy 3, produces the faintest PPV image.  Right: PPV data cube. Illustration of the concepts of the thick and thin velocity
slices. The slices are thin for the PPV images of the large eddies, and thick for the images of small
eddies. From \protect\cite{2009SSRv..143..357L}.}.
\label{fig1}
\end{figure*}

The non-trivial nature of the statistics of the eddies in the PPV space is illustrated in Fig. \ref{fig1}. The figure illustrates the fact that from three equal-sized and equal-density eddies, the one with the smallest
velocity provides the largest contribution to the PPV intensity. Jumping forward in our presentation, we can mention that this explains the scalings of power spectra obtained in LP00, which indicates that
  a spectrum of eddies that corresponds to most of turbulent energy at {\it large} scales corresponds to the spectrum of thin channel map intensity fluctuations having most of the energy at {\it small} scales. It is also clear that if the channel map or velocity slice of PPV data gets thicker than the velocity extent of the eddy 3, all the eddies contribute to the intensity fluctuations in the same way, i.e. in proportion to the total number of atoms within the eddies. Similarly, in terms of the spectrum of fluctuations along the $v$-axis, the weak velocity eddy 1 provides the most singular small-scale contribution.
  
The PPV statistics can be sampled by exploring the fluctuations of intensity within velocity slices or channel maps of a given thickness $\Delta v$ (see the right-hand panel of Fig. \ref{fig1}) which is the essence of VCA technique. This was the way observers traditionally attempted to quantify the properties of PPV fluctuations. The alternative way of studying PPV fluctuations is by analysing the fluctuations along the $v$-axis. This new way of study was introduced in LP00 and elaborated in LP06 (see also \citealt{chepurnov2009turbulence}); it was termed VCS. Our current study is devoted to elaborating the VCA technique. 

The right-hand panel of Fig. \ref{fig1} illustrates the studies of turbulence using the VCA technique. The PPV space is presented by XYV cube where a velocity slice $\Delta v$ is shown. In turbulence, large eddies have larger velocities, for e.g. in Kolmogorov turbulence the velocity of eddies $\delta v_l$ increases with eddy size as $l^{1/3}$. Therefore,  larger eddies like eddy 1 have velocities associated with it larger than $\Delta v$,
while smaller eddies like eddy 2 have their velocities less than $\Delta v$. As a result, the slice
$\Delta v$ fully samples eddy 2, but samples only a part of the eddy 1. Therefore, we say that the slice is {\it thin} for eddy 1  and {\it thick} for eddy 2. The notion of thin and thick velocity slices was introduced in LP00,  with slices  being `thick' for eddies with velocity ranges less than $\Delta v$ and `thin' otherwise. The right-hand panel of Fig. \ref{fig1} illustrates the relativity of this notion for eddies of different sizes. 

The VCA was formulated in LP00 for the purposes of obtaining spectra of velocity and density turbulence, and therefore the anisotropy of turbulence was disregarded. Our present work is focused on studying turbulence anisotropies.

\section{MHD Turbulence Statistics Employed}

In this section, we present the description of the velocity mapping PPV space based on our earlier studies (LP00; LP04) but explicitly account for the turbulence anisotropy following the description of magnetized turbulence we presented in LP12. 

\subsection{Turbulence statistics in PPV space}

As we mentioned earlier, the point-wise measurements in XYZ space and therefore the direct measurements of the statistics of magnetized turbulence are not available with spectroscopic measurements. Instead, the measurements of intensity of emission are defined in PPV space (see the right-hand panel of Fig. \ref{fig1}) or XYV volume, where the turbulence information along LOS, which we assume to be aligned along the $z$-axis, is subject to a non-linear transformation due to the mapping to the LOS. Doppler shifts are affected only by the line of sight component of turbulence velocities, which to simplify our notations we denote as $v$. 

The theory of PPV space was pioneered in LP00 and was later extended for special cases in LP04, LP06, LP08. The main expressions of the theory that we are going to use within our study are summarized in Appendix \ref{ppvspace}. These expressions describe the non-linear velocity mapping of turbulence irrespective of the degree of turbulence anisotropy. 

In this paper, we are studying how intensity statistics reflects the anisotropic nature of the velocity and density fields in magnetized turbulence. The intensity from an emitting medium in PPV space is dependent on density of emitters and their velocity distribution in PPV space. Therefore, intensity correlation function for a turbulent field is dependent on both the correlation of density as well as velocity correlation, and for optically thin lines 
is given by (LP04):
\begin{align}\label{eq:intensitycorrelation}
\xi_\text{I}(R,\phi,\Delta v)\propto \frac{\epsilon^2\bar{\rho}^2}{2\upi}\int_{-S}^S\mathop{\mathrm{d}z}[1+\tilde{\xi}_\rho(R,\phi)]\left[D_z(\bm{r})+2\beta_{\text{T}}\right]^{-1/2}\nonumber\\
\int_{-\Delta v/2}^{\Delta v/2}\mathop{\mathrm{d}v}W(v)\exp\left[-\frac{v^2}{2(D_z(\bm{r})+2\beta_\text{T})}\right],
\end{align}
where $\bm{r}=\bm{x}_1-\bm{x}_2$ is the spatial separation of two turbulent points, $\bm{R}=\bm{X}_1-\bm{X}_2$ is their separation in two dimensional sky, $\phi$ is the angle that $R$ makes with sky-projected magnetic field, $D_z$ is the $z$-projection of velocity structure function, $\Delta v$ is the thickness of velocity slice, $\tilde{\xi}_\rho$ is the over-density correlation, $\beta_\text{T}\equiv k_BT/m$ is the thermal broadening, and $W(v)$ is a window function which describes how the integration over velocities is carried out. 

From Eq.\eqref{eq:intensitycorrelation}, one can observe that because of the presence of the factor $1+\tilde{\xi}_\rho(R,\phi)$, the integral can be separated into two parts such that one part contains only the contribution from velocity effects whereas other part contains the contribution from density effects as well. Formally, we can write
\begin{equation}\label{eq:separating}
\xi_\text{I}(R,\phi,\Delta v)=\xi_\text{I}^{v}(R,\phi,\Delta v)+\xi_\text{I}^{\rho}(R,\phi,\Delta v),
\end{equation}
where the superscripts $v$ and $\rho$ are to remind ourselves which effects these term comprise of. Naturally, in the absence of any density fluctuations, only the first term in the above equation survives. The usefulness of the above expression comes from the fact that at various regions of interest, one or the other term becomes unimportant, as we shall see this in more detail in this paper. 

To describe intensity statistics at small scales, it is more convenient to use the intensity structure function, 
\begin{equation}\label{eq:intesityst}
\mathcal{D}(R,\phi,\Delta v)=2\left[\xi_\text{I}(R=0,\Delta v=0)-\xi_\text{I}(R,\phi,\Delta v)\right].
\end{equation}
The above two equations are the main equations that we will use for our subsequent analysis.

\subsection{Velocity correlation tensor for MHD turbulence}

To describe turbulence in ISM, one should account for the magnetization of the media. 
 In MHD turbulence, there exists a preferred direction pointing towards the direction of mean magnetic field; therefore, the concept of isotropy applicable to hydrodynamic Kolmogorov turbulence breaks down, and the turbulent statistics are anisotropic. The problem of describing anisotropic turbulence was addressed in LP12 in the framework of studies of anisotropies of synchrotron intensities. Here we adopt the same representation of the anisotropic MHD turbulence using axisymmetric tensors (see more justification in LP12).

In what follows, we are describing the statistics of anisotropic velocity field, which has many similarities with the statistics of anisotropic turbulent magnetic field described in LP12. We would like to stress that the deeply entrenched in the literature the description of MHD turbulence based on a model having mean magnetic field plus {\it isotropic} fluctuations contradicts theoretical, numerical and observational studies of magnetized turbulence and therefore should be discarded\footnote{We note that the present day models of the cosmic microwave background foreground still use this erroneous model for representing magnetic fields.}. Indeed, MHD turbulence is neither isotropic nor can it be represented by mean field with isotropic fluctuations. The correct description of MHD turbulence involves the combination of three different cascades with different degree of fluctuation anisotropies, and this is the description we use in the present work. 
 
Following the notation of \citet{chandrasekhar1950theory},  the velocity correlation tensor of axisymmetric turbulence is 
\begin{equation}\label{correlationfunction}
\langle v_i(\bm{x}_1)v_j(\bm{x}_2)\rangle=
A\hat{r}_i\hat{r}_j + B\delta_{ij}+C\hat{\lambda}_i\hat{\lambda}_j
+D\left(\hat{\lambda}_i\hat{r}_j+\hat{\lambda}_i\hat{r}_j\right)~,
\end{equation}
where $\hat \lambda$ unit vector specifies the preferred direction
\footnote{All results are invariant under replacement of $\hat\lambda$
by $-\hat\lambda$ that specify the same axis.}.
For isotropic turbulence, the coefficients $C$ and $D$ of the velocity correlation tensor are zero.
At zero separation, $\bm{r}\rightarrow 0$, the correlation function gives the variance tensor
\begin{equation}
\langle v_i(\bm{x}_1)v_j(\bm{x}_1
)\rangle=B(0)\delta_{ij}+C(0)\hat{\lambda}_i\hat{\lambda}_j.
\end{equation}

Similarly, we can define the structure function tensor
for the velocity field
\begin{equation}\label{eq:velocitystr}
D_{ij}(\bm{r})\equiv
\left\langle\left(v_i(\bm{x}_1)-v_i(\bm{x}_2)\right)
\left(v_j(\bm{x}_1)-v_j(\bm{x}_2)\right)\right\rangle.
\end{equation}
The main quantity that will appear in our analysis
is the $z$- projection of the velocity structure function
\begin{align}\label{structure}
D_z(\bm{r})=D_{ij}\hat{z}_i\hat{z}_j=
2[(B(0)-B(r,\mu))+(C(0)-C(r,\mu))\cos^2\gamma\nonumber\\
-A(r,\mu)\cos^2\theta
-2 D(r,\mu)\cos\theta\cos\gamma]~.
\end{align}
The variables and parameters that appear in the definition of $D_z(\bm{r})$ are summarized in Table~\ref{tab:params}. Among them there are four angles that we
keep track of. First, we have $\theta$ and $\phi$ which are spherical coordinates of the separation vector in the frame where the $z$-axis is aligned with the LOS and the $x$-axis is aligned with projection of the symmetry axis on the plane of the sky. Dependence of the observed intensity correlation on $\phi$ is the main focus of the paper, while $\theta$ get integrated along the LOS. Angle $\gamma$ is a fixed parameter of the problem that describes the direction of the mean magnetic field with respect to the $z$-axis. Lastly, $\mu$ is angle between the separation vector and the symmetry axis. The local axisymmetric properties of the turbulence models depend explicitly on $\mu$ only.  Between these four angles there is a relation
\begin{equation}\label{eq:mu}
\mu(\gamma,\theta,\phi)=\sin\gamma\sin\theta\cos\phi+\cos\gamma\cos\theta.
\end{equation}

\begin{table*}
\centering
\caption{List of notations used in this paper.}
\begin{tabular}{l c c}  
\hline
\hline  
\emph{Parameter} & \emph{Meaning} & \emph{First appearance}\\
\hline
$\bm{r}$ & 3-D separation &  Eq. \eqref{eq:intensitycorrelation}\\ 
$\bm{R}$ & 2-D sky separation & Eq. \eqref{eq:intensitycorrelation}\\
$\hat{\lambda}$ & Mean direction of magnetic field &  Eq. \eqref{correlationfunction} \\ 
$\phi$ & 2-D angle between sky-projected $\bm{r}$ and sky-projected $\hat{\lambda}$  & Eq. \eqref{eq:mu} \\
$\theta$ &  Angle between LOS $\hat{z}$ and $\hat{r}$ & Eq. \eqref{eq:intensitycorrelation} \\
$\gamma$ & angle between line of sight $\hat z$ and symmetry axis $\hat\lambda$ & Eq. \eqref{structure}\\
$\mu$ & Cosine of the angle between $\bm{r}$ and $\hat\lambda$ & Eq. \eqref{correlationfunction}\\
$\mu_k=\hat{\bm{k}}\cdot\hat{\bm{\lambda}}$ & Angle between $\hat{\bm{k}}$ and $\hat\lambda$ & Eq. \eqref{correlationfunction}\\
$D_{ij}(\bm{r})$ & Velocity structure function & Eq. \eqref{eq:velocitystr}\\
$D_z(\bm{r})$ & $z$-projection of velocity structure function & Eq. \eqref{structure}\\
$a_{\bm{k}}$ & Random amplitude of a mode & Eq. \eqref{fouriercorr}\\
$\hat{\xi}_{\bm{k}}$ & Direction of allowed displacement in the mode & Eq. \eqref{fouriercorr}\\ 
$\mathcal{A}(k,\hat{\bm{k}}\cdot\hat{\lambda})$ & Power spectrum of a mode & Eq. \eqref{fouriercorr}\\
$\tilde{d}_m(R)$ & Multipole moment of intensity structure function & Eq. \eqref{eq:multipolemoment}\\
$r_c$ & Cut-off scale for density correlation & Eq. \eqref{shallow}\\
$c_\rho$ & Density anisotropy parameter & Eq. \eqref{shallow}\\
$M_\text{A}$ & Alfv\'en Mach number & Eq. \eqref{eq:alfvenpower}\\
$\beta_\text{T}$ & Thermal velocity & Eq. \eqref{eq:intensitycorrelation}\\
$W(v)$ & Window function & Eq. \eqref{eq:intensitycorrelation}
\\
$P_m(K)$ & Intensity moments in 2d Fourier space & Eq. \eqref{eq:interferometermoment}\\
$B(\bm{X})$ & Beam of an instrument & Eq. \eqref{eq:interferometerbeam}\\
$\theta_0$ & Diagram of an instrument & Eq. \eqref{eq:diagram}\\
$\beta$ & Plasma constant & -\\
 \hline
 \hline
\end{tabular}
\label{tab:params}
\end{table*}

\section{Representing MHD Turbulence Modes}
\label{mhdmodes}
Before we proceed with the formal mathematical description, a few statements about the properties of MHD turbulence are due (see a more detailed discussion in \citealt{brandenburg2013astrophysical}). It is natural to accept that the properties of MHD turbulence depend on the degree of magnetization. Those can be characterized by the Alfv\'en Mach number $M_\text{A}=V_L/V_\text{A}$, where $V_L$ is the injection velocity at the scale $L$ and $V_\text{A}$ is the Alfv\'en velocity.  It is intuitively clear that for $M_\text{A}\gg 1$ magnetic forces should not be important in the vicinity of injection scale. This is the limiting case of super-Alfv\'enic turbulence. The case of $M_\text{A}=1$ is termed trans-Alfv\'enic and the case of $M_\text{A}<1$ sub-Alfv\'enic turbulence. Naturally, $M_\text{A}\ll 1$ should correspond to magnetic field with only marginally perturbed field direction.

The modern theory of MHD turbulence started with the seminal paper by \cite{goldreich1995toward}, (henceforth GS95). They suggested the theory of turbulence of Alfv\'enic waves or Alfv\'enic modes, as in turbulence non-linear interactions modify wave properties significantly. For instance, in GS95 theory Alfv\'enic perturbations cascade to a smaller scale in just about one period ($\equiv l/v_l$, $l$ being the eddy size), which is definitely not a type of wave behaviour. The GS95 was formulated for trans-Alfv\'enic turbulence, e.g. for $M_\text{A}=1$. The generalization of GS95 for $M_\text{A}<1$ and $M_\text{A}>1$ can be found in \cite{lazarian1999reconnection} (henceforth LV99). 

The original GS95 theory was also augmented by the concept of local system of reference (LV99; \citealt{cho2000anisotropy}; \citealt{maron2001simulations}; \citealt{cho2002simulations}) that specifies that the turbulent motions should be viewed not in the system of reference of the mean magnetic field, but in the system of reference of magnetic field comparable with the size of the eddies. From the point of view of the observational study that we deal with in this paper, the local system of reference is not available. Therefore, we should view Alfv\'enic turbulence in the global system of reference which for sub-Alfv\'enic turbulence is related to the mean magnetic field (see the discussions in \citealt{cho2002compressible}; \citealt{esquivel2005velocity}; LP12). In this system of reference, the observed statistics of turbulence is somewhat different. While in GS95 there are two different energy spectra, namely, the parallel and perpendicular, in the global system of reference the perpendicular fluctuations dominate which allows us to use  a single spectral index for the two directions in our treatment. Similarly, if in the local system of reference the anisotropy is increasing with the decrease of size of the eddies, it stays constant in the global system of reference. It is this property that allowed us to use in LP12 the theoretical description for axisymmetric turbulence by \citet{chandrasekhar1950theory} in order to describe observed turbulent fluctuations\footnote{The `detection' of the {\it scale-dependent} GS95 anisotropy in the numerical study by \cite{vestuto2003spectral} and the subsequent observational studies influenced by the aforementioned work (e.g. \citealt{heyer2008magnetically}) is a result of misinterpretation of numerical data as it is discussed e.g. in \cite{xu2015damping}.}.

For super-Alfv\'enic turbulence, the turbulent motions are essentially hydrodynamic up to the scale of
$l_{A}=LM_\text{A}^{-3}$ and after that scale they follow along the GS95 cascade. If we observe Alfv\'enic turbulence at scales larger than $l_A$ we will not see anisotropy. However, if our tracers are clustered on scales less than $l_{\text{A}}$, we will see the anisotropy corresponding to the field of the large eddy. For instance, turbulence in a molecular cloud with the scales less that $l_{\text{A}}$ will show anisotropy. 

For sub-Alfv\'enic turbulence, the original cascade is {\it weak} with parallel scale of perturbations of magnetic field not changing, while the perpendicular scale getting smaller and smaller as the turbulence cascades (see LV99). However, at scale $l_{\text{trans}}\sim  LM_\text{A}^2$
the turbulence gets {\it strong} in terms of its non-linear interactions, with the modified GS95 scalings (see LV99) being applicable. 

To obtain the full description of MHD turbulence, one has to include the turbulence of compressible, i.e. slow and fast, modes (\citealt{lithwick2001compressible}; \citealt{cho2002compressible, cho2003compressible}). While the entrenched notion in literatures is that for compressible turbulence Alfv\'en, slow and fast modes are strongly coupled and therefore cannot be considered separately, the numerical study in \cite{cho2003compressible} provided a decomposition of the modes and proved that they form cascades of their own (see a bit more sophisticated method of decomposition employed in \citealt{kowal2010velocity}).\footnote{A similar decomposition has been recently performed for relativistic MHD in Takamoto \& Lazarian (in preparation).} This was used in LP12 to provide the representation of these modes for the observational studies of magnetic field. In what follows, we discuss the turbulent velocity field, which entails some modifications compared to LP12. 

As we have already mentioned, 
the motions in an isothermal turbulent plasma can be decomposed into three
types of MHD modes
\textemdash Alfv\'en, fast and slow modes. Fast and slow modes are compressible 
while Alfv\'en mode is incompressible. 
Each of these three modes of turbulence forms its own cascade (see GS95; \citealt{2015ASSL..407..163B}). 
The power laws of the modes are defined by the theory but the properties of modes can change. Therefore, following the tradition of VCA development (LP00)
and our synchrotron studies (LP00; LP16), for the purpose of our observational
study, we keep the indices of velocity and density as parameters that can be established by observations. This is intended to provide a test using the VCA of the modern MHD theory and induce its further development. Nevertheless, to compare the observations of anisotropy with predictions, we keep the structure of the tensors corresponding to the modes. In doing so, we follow the approach in LP12, but modify the treatment to account for the difference of the fast and slow modes in terms of magnetic field and in terms of velocity. Indeed,
the magnetic field that was dealt with in LP12 must satisfy an additional solenoidality constraint, while there is no such a constraint for the turbulent velocity.

In this paper, our focus is to understand how turbulence anisotropies transfer into the anisotropy of the statistics of intensity fluctuations within PPV slices and how the latter statistics changes with the thickness of the slices. As was shown in LP00, the statistics of intensity fluctuations within a PPV slice can be affected by both the velocity statistics and density statistics, and there are regimes when only velocity fluctuations determine the fluctuations of intensity within a thin slice.  

In this section we shall discuss correlation tensors of velocity fields generated by each of the MHD modes above. 
The details of the velocity correlation tensor of each mode depend on the
allowed displacement of plasma in the mode
and the distribution of power among different wavelengths.

In general, the Fourier component of velocity in a mode is given by
$\bm{v}(\bm{k})=a_{\bm{k}}\hat{\xi}(\hat{\bm{k}},\hat{\bm{\lambda}}),$ where $\bm{k}$ is the wavevector, 
$a_{\bm{k}}$ is the random complex amplitude
of a mode and $\hat{\xi}$ is the direction of allowed displacement.
Therefore, the velocity correlation is given in Fourier space by
\begin{align}\label{fouriercorr}
\langle v_i(\bm{k}) v_j^*(\bm{k}') \rangle
=\langle a_{\bm{k}}a^*_{\bm{k}'}\rangle&
\left(\hat{\xi}_{\bm{k}} \otimes \hat{\xi}^*_{\bm{k}'}\right)_{ij}\nonumber\\
&\equiv \mathcal{A}(k,\hat{\bm{k}}\cdot\hat{\bm{\lambda}})\left(\hat{\xi}_{\bm{k}} \otimes \hat{\xi}^*_{\bm{k}}\right)_{ij}\delta(\bm{k}-\bm{k}'),
\end{align}
where $\mathcal{A}(k,\hat{\bm{k}}\cdot\hat{\bm{\lambda}})=\langle \hat{a}_{\bm{k}}\hat{a}^*_{\bm{k}}\rangle$ is the power spectrum which in our
case depends on the angle 
$\mu_{k}\equiv \hat{\bm{k}}\cdot\hat{\bm{\lambda}}$.
Fourier transform of Eq. \eqref{fouriercorr} gives velocity correlation tensor in the real space
\begin{equation}\label{eq:realspacevcorr}
\left\langle v_i(\bm{x}_1) v_j(\bm{x}_1+\bm{r})\right\rangle =\int \mathop{\mathrm{d}k}k^2 \mathop{\text{d$\Omega $}_k}  \mathrm{e}^{i\bm{k}\cdot\bm{r}}\mathcal{A}(k,\hat{\bm{k}}\cdot\hat{\bm{\lambda}})\left(\hat{\xi}_{\bm{k}} \otimes \hat{\xi}^*_{\bm{k}}\right)_{ij}.
\end{equation}

The power spectrum can be decomposed into spherical harmonics as
\begin{equation}\label{randomcorr}
\mathcal{A}(k,\hat{\bm{k}}\cdot\hat{\bm{\lambda}})=
\sum_{\ell_1 m_1}\frac{4\upi}{2\ell_1+1}\mathcal{A}_{\ell_1}(k)
Y_{\ell_1m_1}(\hat{\bm{k}})Y_{\ell_1 m_1}^*(\hat{\bm{\lambda}}),
\end{equation}
and similarly
\begin{equation}\label{displacecorr}
\left(\hat{\xi}_{\bm{k}}\otimes\hat{\xi}^*_{\bm{k}}\right)_{ij}=
\sum_{\ell_2 m_2}c^{ij}_{\ell_2m_2}(\hat\lambda)Y_{\ell_2 m_2}(\hat{\bm{k}})~,
\end{equation}
where $c^{ij}_{\ell_2m_2}$ coefficients depend on the mode structure, and are 
tabulated further in this section for each mode. With these definitions, Eq. \eqref{eq:realspacevcorr} can be expressed as 
\begin{align}
\left\langle v_i v_j\right\rangle
=\sum_{\ell m}
4\upi \text{i}^{\ell}
Y_{\ell m}^{*}&(\hat{r})
\sum_{\ell_1m_1}
\frac{4\upi}{2\ell_1+1}
Y_{\ell_1 m_1}^*(\hat{\bm{\lambda}})\nonumber\\
&\sum_{\ell_2 m_2}
c^{ij}_{\ell_2 m_2}(\hat\lambda)\mathcal{T}_{\ell\ell_1}(r)\Psi_{\ell m,\ell_1 m_1,\ell_2 m_2}~,
\end{align}
where we have defined 
\begin{equation}
\mathcal{T}_{\ell\ell_1}(r) \equiv 
\int \mathop{\mathrm{d}k}k^2 j_{\ell}(kr)\mathcal{A}_{\ell_1}(k),
\end{equation}
and $\Psi$ is a shorthand notation for the combination of Wigner 3-j symbols 
given in detail in Appendix \ref{general}. On the other hand,
the velocity correlation tensor is given by Eq.\eqref{correlationfunction}, and
therefore, the above equations can be used to find the coefficients $A, B, C$ and $D$.
The procedure that we use to obtain them is also described in Appendix [\ref{general}].

The intensity statistics of a turbulent field is also affected by the density fluctuations. In a turbulent field, if density fluctuations are weak, it is easy to understand  density correlation for different modes. Assuming that the
density is given by $\rho\rightarrow\rho_o+\delta\rho$, where $\rho_0$ is the
mean density of the turbulent medium and $\delta\rho$ is the overdensity such
that $|\delta\rho| \ll \rho_0$, the continuity equation 
\begin{equation}
\frac{\partial \rho}{\partial t}+\nabla\cdot(\rho\bm{v})=0,
\end{equation}
gives (in Fourier space)
$\delta\rho_{\bm{k}}/\rho_0\sim \hat{a}(\hat{k}\cdot\hat{\xi}),$
using which we obtain the over-density power spectrum 
\begin{equation}
\langle\delta\rho_{\bm{k}}\delta\rho^*_{\bm{k}}\rangle
=\rho_0^2\mathcal{A}(k,\hat{\bm{k}}\cdot\hat{\bm{\lambda}})
|\hat{k}\cdot\hat{\xi}|^2.
\end{equation}
In real space, the overdensity correlation is given by
\begin{equation}\label{dencorr}
\left\langle \delta\rho(\bm{x}_1) \delta\rho(\bm{x}_1+\bm{r})\right\rangle =\rho_0^2\int \mathop{\mathrm{d}k}k^2 \mathop{\text{d$\Omega $}_k}  \mathrm{e}^{i\bm{k}\cdot\bm{r}}\mathcal{A}(k,\hat{\bm{k}}\cdot\hat{\bm{\lambda}})|\hat{k}\cdot\hat{\xi}|^2.
\end{equation}
These equations for density correlation are only valid when density
perturbations are weak. In the case when perturbations are not weak, we use the
ansatz discussed in Sec. [\ref{densitystatistics}].

Below we describe the properties of individual MHD modes. For compressible modes, these properties vary depending on the magnetization of the media,
which are determined by the parameter $\beta$, which is the ratio of thermal plasma energy density to the energy density of magnetic field. Thus this ratio, in addition to $M_\text{A}$ should be considered. It is important to note that the MHD modes are subject to strong non-linear damping. As a result, for instance, perturbations corresponding to Alfv\'en modes get damped over just one period. 

To describe correlation tensors of these modes we use their dispersion relations. Our treatment of MHD modes below is analogous to the one in LP12. Below we treat velocity fluctuations associated with MHD modes, while LP12 dealt with magnetic fluctuations. A brief summary of mode structures is also presented in Table \ref{tab:modes}. The properties of density perturbations in turbulent media are discussed in  \cite{cho2003compressible}, \cite{kowal2010velocity} and \cite{kowal2007density}.

\subsection{Alfv\'en mode}

Alfv\'en modes are essentially incompressible modes where displacement of plasma in an Alfv\'en wave is orthogonal to the plane spanned by the magnetic field and wavenumber, so that
\begin{equation}
\bm{v}_A\propto \hat{\xi}_{\bm{k}}=\frac{\hat{k}\times\hat{\lambda}}{\sqrt{1-(\hat{k}\cdot\hat{\lambda})^2}}.
\end{equation}
The corresponding tensor structure for Alfv\'en mode is then
\begin{align}\label{eq:alfvencorr}
(\hat{\xi}_{\bm{k}}\otimes\hat{\xi}^*_{\bm{k}}&)_{ij}=\nonumber\\
&(\delta_{ij}-\hat{k}_i\hat{k}_j)-\frac{(\hat{k}.\hat{\lambda})^2\hat{k}_i\hat{k}_j-\hat{k}.\hat{\lambda}(\hat{\lambda}_i\hat{k}_j+\hat{\lambda}_j\hat{k}_i)+\hat{\lambda}_i\hat{\lambda}_j}{1-(\hat{k}.\hat{\lambda})^2},
\end{align}
In the above equation, the part in the first parentheses is referred to as $E$-type correlation, and the second part is referred to as $F$-type correlation. The $E$-type correlation has been studied in detail in LP12. 

In the case of Alfv\'en mode, the power spectrum in the global system of reference is given by
\begin{equation}\label{eq:alfvenpower}
\mathcal{A}(k,\mu_k)=E(k,\mu_k)\propto k^{-11/3}\exp\left[-M_\text{A}^{-4/3}\frac{|\mu_k|}{(1-\mu_k^2)^{2/3}}\right],
\end{equation}
where $\mu_k=\hat{\bm{k}}\cdot\hat{\bm{\lambda}}$. 

The correlation tensor of Alfv\'en mode in real space is calculated in Appendix [\ref{alfven}]. The coefficients $A, B, C$ and $D$ are given by Eqs. \eqref{alfvenA}, \eqref{alfvenB}, \eqref{alfvenC} and \eqref{alfvenD}, respectively.

As Alfv\'en modes are incompressible, to the first-order approximation, they do not create any density fluctuations.  Indeed, for Alfv\'en waves, $\hat{\xi}$ is orthogonal to wavevector, and therefore the overdensity correlation must be zero (cf. Eq. \eqref{dencorr}).

\subsection{Fast mode}
Fast modes are compressible type of modes. In high-$\beta$ ($\equiv P_{\text{gas}}/P_{\text{mag}}$) plasma, they behave like acoustic waves, while in low-$\beta$ plasma they propagate with Alfv\'en speed irrespective of the magnetic field strength \citep{cho2005generation}. The power spectrum of this mode is isotropic and is given by
\begin{equation}\label{eq:fastpower}
\mathcal{A}(k,\mu_k)\propto k^{-7/2}.
\end{equation}
In this subsection, we will present the velocity correlation tensor as well as over-density correlation for fast modes in two regimes: high and low $\beta$.

\subsubsection{High-$\beta$ regime}

In the high-$\beta$ regime, displacement in fast modes is parallel to wave vector $\hat{k}$, and  the velocity is $\bm{v}\propto \hat{k}$. These are essentially sound waves compressing magnetic field. This mode is purely compressional type, and its tensor structure
in Fourier space is given by
\begin{equation}\label{eq:fasthbcorr}
\left(\hat{\xi}_{\bm{k}}\otimes\hat{\xi}^*_{\bm{k}}\right)_{ij}=\hat{k}_i\hat{k}_j.
\end{equation}

The correlation tensor structure of fast modes in real space is presented in Appendix \ref{FMHBe}. It has been shown that $C$ and $D$ parameters of this mode vanish, while  $A$ and $B$ are given by Eqs. \eqref{FMHbA} and \eqref{FMHbB}, respectively.

In the case when density perturbations are weak, the over-density correlation in fast modes in high-$\beta$ regime is (cf. Eq.\eqref{dencorr}) 
\begin{align}
\langle\delta\rho(\bm{x}_1)\delta\rho(\bm{r}+\bm{x}_1)\rangle=\int \mathrm{d}^3\bm{k}\mathrm{e}^{i\bm{k}.\bm{r}}&k^{-7/2}\nonumber\\
&=4\upi\int \mathop{\mathrm{d}k}k^{-3/2}j_0(kr).
\end{align}
Note that the above correlation represents steep density spectra for which structure function should be used for appropriate analysis to avoid divergence issues.

\subsubsection{Low $\beta$ regime}
In the low-$\beta$ regime, velocity is orthogonal to the direction of symmetry $\hat{\lambda}$, and therefore, the velocity is 
\begin{equation}
\bm{v}\propto \hat{\xi}_{\bm{k}}=\frac{(\hat{\lambda}\times\hat{k})\times \hat{\lambda}}{\sqrt{1-(\hat{k}\cdot\hat{\lambda})^2}}.
\end{equation}
This mode can be associated with compression of magnetic field. 
Using the above equation, we have
\begin{equation}\label{eq:fastlbcorr}
\left(\hat{\xi}_{\bm{k}}\otimes\hat{\xi}^*_{\bm{k}}\right)_{ij}\propto \frac{\hat{k}_i\hat{k_j}-(\hat{k}.\hat{\lambda})(\hat{k}_i\hat{\lambda}_j+\hat{k}_j\hat{\lambda}_i)+(\hat{k}.\hat{\lambda})^2\hat{\lambda}_i\hat{\lambda}_j}{1-(\hat{k}.\hat{\lambda})^2}.
\end{equation}
The velocity correlation function in real space for the above tensor is presented in the Appendix [\ref{FMLB}]. Because the power spectrum for this mode is isotropic, the correlation tensor is heavily simplified. The parameters $A, B, C$ and $D$ for this mode are presented in Eqs. \eqref{FMLBA}, \eqref{FMLBB}, \eqref{FMLBC} and \eqref{FMLBD}.

In the case when density perturbations are weak, the over-density correlation in fast modes in low $\beta$ regime is (cf. Eq.\eqref{dencorr}) 
\begin{align}
\langle&\delta\rho(\bm{x}_1)\delta\rho(\bm{r}+\bm{x}_1)\rangle=\int \mathrm{d}^3\bm{k}\mathrm{e}^{i\bm{k}.\bm{r}}k^{-7/2}(1-(\hat{k}.\hat{\lambda})^2)\nonumber\\
&=\frac{4\upi}{3}\int \mathop{\mathrm{d}k}k^{-3/2}j_0(kr)+\frac{8\upi}{3}\int \mathop{\mathrm{d}k}k^{-3/2}j_2(kr)P_2(\mu).
\end{align}

\begin{table*}
\centering
\caption{Summary of mode structure}
\begin{tabular}{l c c c c}  
\hline
\hline  
\emph{Mode} & \emph{Velocity tensor structure} & \emph{Power spectrum} & \emph{Type} & \emph{Equation}\\
\hline
Alfv\'en & $E-F$ & Anisotropic & Solenoidal & \ref{eq:alfvencorr}, \ref{eq:alfvenpower}\\
Fast (high $\beta$) & $P$ & Isotropic & Potential & \ref{eq:fasthbcorr}, \ref{eq:fastpower}\\
Fast (low $\beta$) & Mixed & Isotropic & Compressible & \ref{eq:fastlbcorr}, \ref{eq:fastpower}\\
Slow (high $\beta$) & $F$ & Anisotropic & Solenoidal &  \ref{eq:slowhbcorr}, \ref{eq:alfvenpower}\\
Slow (low $\beta$) & Mixed & Anisotropic & Compressible &  \ref{eq:slowhbcorr}, \ref{eq:alfvenpower} \\
Strong & $E$ & Anisotropic & Solenoidal &  \ref{eq:strong}\\
\hline
\end{tabular} 
\label{tab:modes}
\end{table*}

\subsection{Slow mode}
Slow modes in high-$\beta$ plasma are similar to pseudo-Alfv\'en modes in incompressible regime, while at low-$\beta$ they are density perturbations propagating with sonic speed parallel to magnetic field (see \citealt{cho2003compressible}). The power spectrum of this mode is the same as that of Alfv\'en mode (cf equation \ref{eq:alfvenpower}).

In this section, we will present the velocity correlation and over-density correlation of this mode in low- and high-$\beta$ regime.

\subsubsection{High-$\beta$}
In the high-$\beta$ regime, displacement is perpendicular to the wavevector $\hat{k}$, and therefore, $$\bm{v}\propto (\hat{k}\times\hat{\lambda})\times\hat{k}.$$
Therefore, this gives us a full tensor structure is
\begin{equation}\label{eq:slowhbcorr}
\left(\hat{\xi}_{\bm{k}}\otimes\hat{\xi}^*_{\bm{k}}\right)_{ij}= \frac{(\hat{k}.\hat{\lambda})^2\hat{k}_i\hat{k}_j-(\hat{k}.\hat{\lambda})^2(\hat{\lambda}_i\hat{k}_j+\hat{\lambda}_j\hat{k}_i)+\hat{\lambda}_i\hat{\lambda}_j}{1-(\hat{k}.\hat{\lambda})^2}.
\end{equation}
Slow modes are essentially incompressible types of mode in this regime. The above tensor structure is pure $F$-type, and the $F$-type  correlation tensor in real space is derived in Appendix \ref{vcorfsm}. The correlation parameters $A, B, C$ and $D$ are presented in Eqs. \eqref{SMHBA}, \eqref{SMHBB}, \eqref{SMHBC} and \eqref{SMHBD}.

Slow modes in high-$\beta$ regime have zero density fluctuations in a turbulent field where density perturbations are sufficiently weak (cf. Eq. \eqref{dencorr}). 

\subsubsection{Low $\beta$}
In this case, the displacement is parallel to the symmetry axis $\hat{\lambda}$, and therefore, the correlation tensor is $\langle v_iv_j\rangle \propto \hat{\lambda}_i\hat{\lambda_j}$. The real space correlation function of these modes is derived in Appendix \ref{SMLB}, and the result (Eq. \eqref{SMLBC})
\begin{equation}\label{eq:slowlbcorr}
\left\langle v_i(\bm{x}_1) v_j(\bm{x}_1+\bm{r})\right\rangle=\sum_\ell 4\upi i^\ell \mathcal{T}_{\ell\ell}(r)P_\ell(\mu)\hat\lambda_i\hat\lambda_j,
\end{equation}
where $\mathcal{T}_{\ell \ell}(r)$ is defined in Eq.\eqref{defineT}, and is related to the power spectrum of the mode.
Although the tensor structure of this mode is isotropic, the structure function is nevertheless anisotropic due to anisotropic power spectrum. 

In the case when density perturbations are weak, the over-density correlation in slow modes in low-$\beta$ regime is (cf. Eq.\eqref{dencorr}) 
\begin{align}
\langle&\delta\rho(\bm{x}_1)\delta\rho(\bm{r}+\bm{x}_1)\rangle=\int \mathrm{d}^3\bm{k}\mathrm{e}^{i\bm{k}.\bm{r}}k^{-11/3}(\hat{k}.\hat{\lambda})^2\nonumber\\
&=\frac{4\upi}{3}\int \mathop{\mathrm{d}k}k^{-5/3}j_0(kr)-\frac{8\upi}{3}\int \mathop{\mathrm{d}k}k^{-5/3}j_2(kr)P_2(\mu).
\end{align}

\subsection{Density fluctuations in MHD Turbulence}\label{densitystatistics}

In the previous subsections, we discussed a way of presenting density as a result of the compressions induced by compressible slow and fast modes. 
 However, for high sonic Mach number turbulence, this linear approximation is not good.  For example, linear model predicts steep density spectrum. However, in the case of supersonic turbulence, density perturbations are caused by shocks, and these perturbations are comparable to the density itself (\citealt{beresnyak2005density};  \citealt{kowal2007density}) . Therefore, the density statistics in this regime can be shallow\footnote{By shallow, we mean the power spectrum $P\propto k^n$, with $n>-3$. For instance, gravitational collapse can result to shallow power spectrum (see \citealt{federrath2013star}; \citealt{burkhart2015observational}).}. Therefore, a different representation of density modes is required.

To understand the effects of density fluctuations in the intensity statistics, we propose the following ansatz for the density correlation function $\xi(\bm{r},\mu)$. This ansatz is based on the results of \cite{jain28fluctuations} where density statistics is presented as a infinite series over spherical harmonics. We take only up to the second harmonics
and in the case of shallow spectrum:
\begin{equation}\label{shallow}
\xi(\bm{r},\mu)=\langle\rho\rangle^2\left[1+\left(\frac{r_c}{r}\right)^{\nu_\rho}\left(1+c_\rho P_2(\mu)\right)\right],\qquad \nu_\rho>0,
\end{equation}
whereas in the case of steep spectrum for $r\ll r_c$:
\begin{equation}\label{steep}
\xi_\rho(\bm{r},\mu)=\langle\rho\rangle^2\left[1-\left(\frac{r_c}{r}\right)^{\nu_\rho}\left(1+c_\rho P_2(\mu)\right)\right],\qquad \nu_\rho<0,
\end{equation}
where $r_c$ denotes a cut-off scale, and $c_\rho$ is a parameter, which depends on the details of the turbulent mode. An important criterion that the two ansatz presented above should satisfy in order to be called a `correlation function' is that their Fourier transform should be positive definite. It can be shown that this condition is true only when the following condition is satisfied:
\begin{equation}
c_\rho>\frac{2\nu_\rho}{3-\nu_\rho},
\end{equation}
for steep spectra whereas for shallow spectra the condition is
\begin{equation}
c_\rho<\frac{2\nu_\rho}{3-\nu_\rho}.
\end{equation}

Our representation above captures several essential features. First, the above correlation can be immediately broken into two parts: a constant and a part with spatial and angular dependence. With this it is natural to talk about pure velocity and pure density effects, and equations (\ref{eq:intensitycorrelation}) and (\ref{eq:velden}) become applicable. Second important feature of the above correlation is that it carries information about anisotropy. In an axisymmetric turbulence, only even harmonics survive due to symmetry and therefore, $P_2(\mu)$ is the dominant term which carries information on the anisotropy\footnote{For a highly anisotropic density fluctuation, higher order harmonics also contribute. We are, however, only concerned with the mild density anisotropy.}. 

\section{Anisotropic statistics of PPV velocity slices}
\label{anisotropicppvintensity}
In the previous sections, we have defined the tools that are required for our achieving our goal, i.e. describing the anisotropy of the PPV. 
In this section, we develop the analytical framework for the study of anisotropic turbulence through the intensity statistics of the PPV velocity slices. The scale $\Delta v$   in Eq. \eqref{eq:intensitycorrelation} is the slice thickness, and by comparing this slice thickness with the variance of velocity, we develop notion of thin and thick slice. If $\Delta v$ is  smaller than the velocity dispersion at the scale of study, it is a {\it thin} channel, whereas if $\Delta v$ is much larger than  the velocity dispersion, it is called a {\it thick} channel.

Anisotropy in intensity statistics is seen in the $\phi$ dependence of intensity structure function (see LP12). To study this angular dependence, similar to our study in LP12 and LP16, we will carry out a multipole expansion of the structure function in spherical harmonics. Such an expansion is useful as these multipoles can be studied observationally. In particular, for an isotropic turbulence, only monopole moment survives. In LP12 for synchrotron intensities, it was found that for studies of magnetic turbulence, the most important are monopole and quadrupole moments.  

\subsection{Intensity statistics in a thin slice regime}\label{sec: truncation}

 We first study intensity fluctuations in the thin slice limit, i.e. the case when the velocity-induced fluctuations are dominant. For that, we will only consider the velocity effects. Before we proceed to details we would like to make a remark on the usefulness of the results that we will obtain by just considering velocity effects. First, in the absence of any density fluctuations, our results  describe the intensity statistics with anisotropic effects. On the other hand, in the presence of steep density spectra, our results describe intensity statistics at small scales $R$. As will be shown later, though steep density spectra do not affect monopole term at larger $R$, they can significantly affect quadrupole term at large $R$; therefore by ignoring density effects one cannot account for observed anisotropy at these scales. In the case of shallow density spectra, however, density effects are important, and our results cannot describe full intensity statistics in this case. However, shallow spectra are not as common as steep, and we shall not worry about that in this section.
 
 In the case of thin velocity channel, the window function is defined by a narrow channel, $W(v)=\delta(v)$ and therefore, utilising equations (\ref{eq:intensitycorrelation}) and (\ref{eq:intesityst}), the intensity structure function can be expressed as
\begin{equation}\label{eq:narrowintensityst}
\mathcal{D}(R,\phi)\propto \frac{2\epsilon^2}{2\upi}\int_{-S}^S\mathop{\mathrm{d}z}\left[\frac{1+\tilde{\xi}_\rho(0,z)}{\sqrt{D_z(0,z)}}-\frac{1+\tilde{\xi}_\rho(\bm{R},z,\phi)}{\sqrt{D_z(\bm{R},z,\phi)}}\right],
\end{equation}
where we have ignored the thermal effects. This can be justified by taking thermal effect as a part of slice thickness (LP00). From Eq. \eqref{eq:separating}, the above equation can be broken into pure velocity and density terms. Here, we are only concerned about the pure velocity contribution which is 
\begin{equation}
\mathcal{D}_v(R,\phi)\propto \frac{2\epsilon^2}{2\upi}\int_{-S}^S\mathop{\mathrm{d}z}\left[\frac{1}{\sqrt{D_z(0,z)}}-\frac{1}{\sqrt{D_z(\bm{R},z,\phi)}}\right].
\end{equation}
To extract the non-trivial $\phi$ dependence from the above expression, we use multipole decomposition of the structure function in circular harmonics, and write the structure function as a series of sum of multipoles 
\begin{equation}\label{eq:multipolemoment}
\mathcal{D}_v(\bm{R},\phi,0)=\sum_m\tilde{d}_m(R)\text{e}^{\text{i}m\phi},
\end{equation}
where the multipole moments $\tilde{d}_m$, in the case of constant density, are given by
\begin{align}\label{multipolevelocity}
\tilde{d}_m(R)=\frac{\bar{\rho}^2}{2\upi}\Bigg[&2\upi\delta_{m0}\int_{-S}^S \mathrm{d}z\frac{1}{\sqrt{D_z(0,z)}}\nonumber\\
&-\int_0^{2\upi}\mathrm{d}\phi \mathrm{e}^{-im\phi}\int_{-S}^S \mathrm{d}z\frac{1}{\sqrt{D_z(\bm{R},z,\phi)}}\Bigg].
\end{align}
In writing the above equation, we have considered the fact that the integral of $1/\sqrt{D_z(0,z)}$ over $\phi$ for non-zero $m$ vanishes. 

We also introduce the a parameter called degree of Isotropy which is defined as
\begin{equation}\label{eq:isotropydeg}
\text{Isotropy Degree}=\frac{\mathcal{D}(R,\phi=0,\Delta v)}{\mathcal{D}(R,\phi=\upi/2,\Delta v)},
\end{equation}
where $\mathcal{D}$ is the intensity structure function. This parameter is particularly useful later to make comparisons with the numerical studies that have been carried out on anisotropic turbulence. It will be later shown that the isotropy degree has an interesting dependence on the thickness of velocity slice, which will be shown to be very useful in the study of intensity anisotropy.

We now proceed to find the multipole moments of intensity structure function in the thin slice limit at constant density. The most general form of  velocity structure function projected along LOS is given by equation \eqref{structure}. The coefficients $A, B, C$ and $D$ in this projected structure function are in general a function of $\mu$, and can be expressed through a multipole expansion  over Legendre polynomials $P_n(\mu)$ as discussed in Appendix \ref{sta}. Although, projected structure function in general contains  sum up to infinite order in multipole expansion, to obtain  analytical results, we  take the terms up to second order from the infinite sum for $A_n, B_n,\ddots$ and ignore the higher order terms. This approximation is
justified due to two reasons. First, these coefficients   all become exceedingly small for higher orders in the region of our interest, which is small $\textbf{r}$. Secondly, upon carrying out integral over the LOS, the effects of the higher order coefficients get diminished\footnote{This was tested numerically, and this statement is good as long as the power spectrum is not highly anisotropic.}. 
With this approximation, the $z$-projection of velocity structure function can be shown to be  (c.f Eq.\eqref{structure})
\begin{equation}
\label{phistructure1}
D_z(\bm{r})=f_1\left(1-f_2\cos\phi-f_3\cos^2\phi\right),
\end{equation}
where $f_1, f_2$ and $f_3$ are some other functions of $r,\gamma,\theta$ and
are independent of $\phi$. The details about $f_1, f_2$ and $f_3$ are provided in Appendix \ref{sta}. 

\begin{figure}
\begin{center}
\includegraphics[scale=0.5]{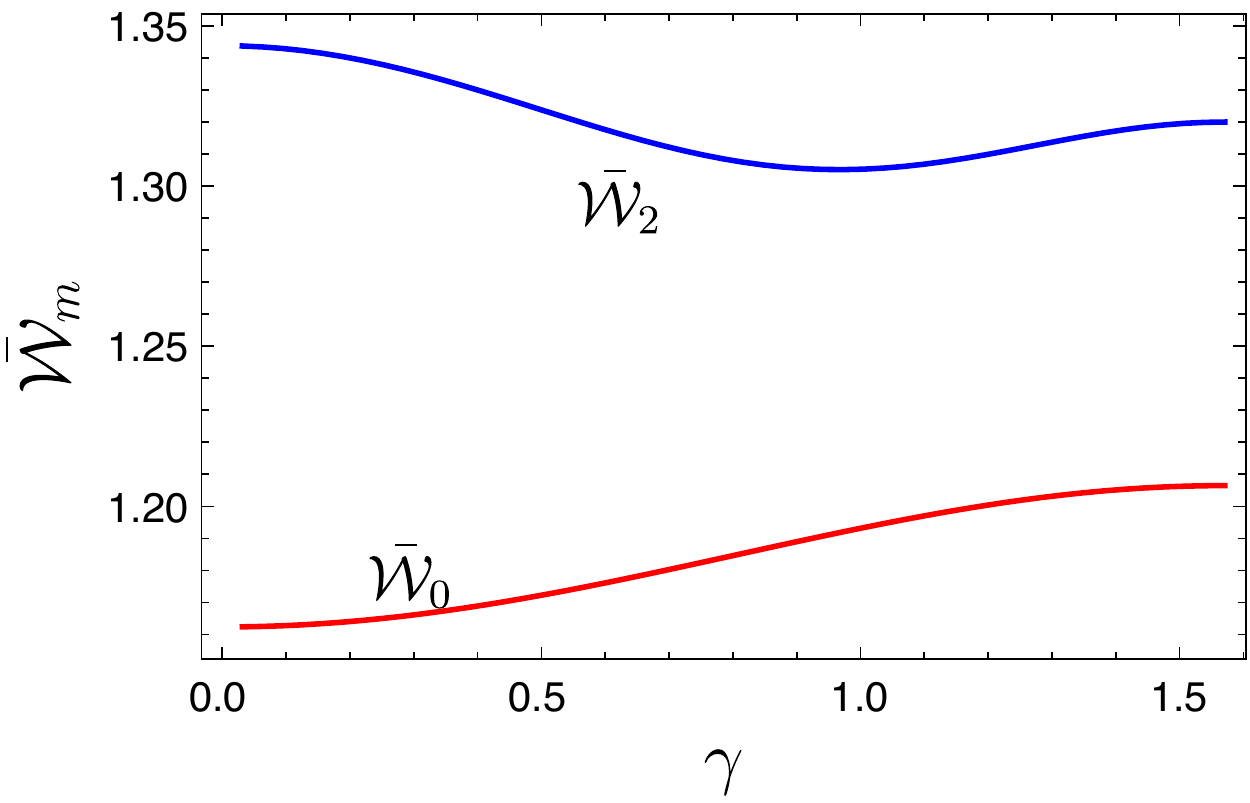}
\caption{The ratio of $\mathcal{W}_m$ obtained from analytical expressions (equations \ref{monopole} and \ref{quadrupole}) to the one obtained from numerical calculations for Alfv\'en mode at $M_\text{A}=0.7$. Note that the structure function is truncated to the same power for both numerical and analytical calculations.}
\label{fig:weightage}
\end{center}
\end{figure}

To evaluate Eq. \eqref{multipolevelocity}, it is usually convenient to carry out $\phi$-integral first and $z$-integral later. This has been done in Appendix \ref{phia} and \ref{za}. Utilizing equations (\ref{jainintegral}), (\ref{HI0result}) and (\ref{HIresult}) and Table \ref{tab:structurepara}, we arrive at the following form of the intensity structure function
\begin{equation}
\mathcal{D}_v(\textbf{R},\phi,0)\propto \sum_{m,2}^\infty\mathcal{W}_m(\sin\gamma)^m \mathrm{e}^{im\phi},
\end{equation}
where $\mathcal{W}_m$ is defined to be weightage function.  However, we are only interested in the monopole and quadrupole coefficients. Although Eq. \eqref{jainintegral} has sum that extends to infinity, for most of our purposes, it is enough to just take first few terms. Therefore, for monopole we take first two terms and for quadrupole term we only take the leading-order term in the sum. Note that the factors $f_1, f_2$ and $f_3$ in Eq. \eqref{phistructure1} are further written in terms of other factors $q_1, q_2,\ldots$ which are dependent on Alfv\'en Mach number $M_\text{A}$. The details of these factors are presented in  Appendix \ref{sta} and Table \ref{tab:structurepara}. Keeping this in mind, we have the monopole weightage function as
\begin{align}\label{monopole}
\mathcal{W}_0\approx-\Bigg\{\sqrt{\frac{\upi}{q_1+q_2}}\left(\frac{\Gamma \left(\frac{\nu}{4}-\frac{1}{2}\right)}{\Gamma \left(\frac{\nu}{4}\right)}-\frac{q_2}{2(q_1+q_2)}\frac{\Gamma \left(\frac{\nu}{4}+\frac{1}{2}\right)}{\Gamma \left(\frac{\nu}{4}+1\right)}\right)\nonumber\\
-\frac{\sqrt{\upi }}{4q_1^{3/2}} \left(\frac{\Gamma \left(\frac{\nu}{4}+\frac{1}{2}\right)}{\Gamma \left(\frac{\nu}{4}+1\right)}-\frac{3}{4}\frac{q_2}{q_1} \frac{\Gamma \left(\frac{\nu}{4}+\frac{3}{2}\right)}{\Gamma \left(\frac{\nu}{4}+2\right)}\right)u_1\sin^2\gamma\Bigg\}R^{1-\nu/2}.
\end{align}
Similarly, the quadrupole weightage function is given by
\begin{equation}\label{quadrupole}
\mathcal{W}_2\approx -R^{1-\nu/2}\frac{\sqrt{\upi }}{4q_1^{3/2}} \left(\frac{\Gamma \left(\frac{\nu}{4}+\frac{1}{2}\right)}{\Gamma \left(\frac{\nu}{4}+1\right)}-\frac{3}{4}\frac{q_2}{q_1} \frac{\Gamma \left(\frac{\nu}{4}+\frac{3}{2}\right)}{\Gamma \left(\frac{\nu}{4}+2\right)}\right)u_1.
\end{equation}
Eqs. \eqref{monopole} and \eqref{quadrupole} are only approximate and should be used with caution. In particular, equation \eqref{monopole} is good only when $(q_1+q_2)> q_2$, while equation \eqref{quadrupole} is good when $q_1>q_2$. However, even in the regime where these conditions do not hold, these equations are robust enough to predict approximate numerics that are not too far from the exact result. The ratio of weightage function $\mathcal{W}_m$, obtained from equations \eqref{monopole} and \eqref{quadrupole}, to that obtained numerically has been plotted in Fig. (\ref{fig:weightage}). As shown in the figure, the analytical results are close to the numerical results.  

It is interesting to note that the isocontours of intensity structure function can be elongated towards the direction parallel to the projection of magnetic field or perpendicular to it depending on the sign of $u_1$. For $u_1>0$, the isocontours should be aligned towards the parallel direction, while for $u_1<0$, they should be aligned towards the perpendicular direction. 

It is usually useful to obtain expressions for  quadrupole-to-monopole ratio, as this is the one which gives the measure of anisotropy. In our case, we have
\begin{equation}\label{eq:quadtomono}
\frac{\tilde{d}_2}{\tilde{d}_0}=\frac{\mathcal{W}_2\sin^2\gamma}{\mathcal{W}_0}.
\end{equation}
It is clear from the above equation that at $\gamma=0$, the anisotropy vanishes. 

\subsection{Intensity statistics in a thick slice regime}
\label{thick}
LP00 showed that density effects are dominant  if the velocity slice is `very thick'. In this limit, velocity effects get washed away in an optically thin medium. In this section, we derive expressions for intensity statistics in the case of very thick velocity slice. Using the results of LP04, we have the intensity correlation
\begin{align}
\xi_\text{I}(\bm{R},\phi)\propto \int_{-S}^S\mathop{\mathrm{d}z}&\frac{\xi_\rho(\bm{r},\phi)}{\sqrt{D_z(\bm{r,\phi}+2\beta)}}\nonumber\\
&\int_{-\infty}^\infty\mathop{\mathrm{d}v}\exp\left[-\frac{v^2}{2(D_z(\bm{r,\phi}+2\beta)}\right],
\end{align}
which upon carrying out the integration over $v$ gives
\begin{equation}\label{eq:intensitycorrel}
\xi_\text{I}(\bm{R},\phi)\propto \int_{-S}^S\mathop{\mathrm{d}z}\xi_\rho(\bm{r},\phi).
\end{equation}
This expression clearly shows that at thick slice, all the velocity information is erased, and density effects play a primary role in intensity statistics. Eq. \eqref{eq:intensitycorrel} allows us to obtain the intensity structure function as
\begin{equation}
\mathcal{D}(\bm{R},\phi)\propto \int_{-S}^S\mathop{\mathrm{d}z}\left[\xi_\rho(z)-\xi_\rho(\bm{r},\phi)\right].
\end{equation}

With some manipulations, it can be shown that
\begin{align}\label{eq:verythickintensity}
\mathcal{D}(\bm{R},\phi)\propto \pm(1+c_\rho P_2(\cos\gamma))\int_{-S}^S\mathop{\mathrm{d}z}\bigg((z^2+R^2)^{-\nu_\rho/2}\nonumber\\
-|z|^{-\nu_\rho/2}\bigg)
\pm\frac{3}{2}c_\rho(\sin^2\gamma\cos^2\phi-\cos^2\gamma) R^2\nonumber\\
\int_{-S}^S\mathop{\mathrm{d}z}(z^2+(R)^2)^{-\nu_\rho/2-1},
\end{align}
where $+$ sign is for steep density spectra whereas $-$ sign is for shallow density spectra.
The above expression can be evaluated analytically and yields

\begin{align}
\mathcal{D}(\bm{R},\phi)\propto \pm(1+c_\rho P_2(\cos\gamma))\Bigg(R^{-\nu_\rho} \, _2F_1\left(\frac{1}{2},\frac{\nu_\rho }{2};\frac{3}{2};-\frac{S^2}{R^2}\right)\nonumber\\
+\frac{S^{-\nu_\rho }}{\nu_\rho -1}\Bigg)
\pm \frac{3}{2}c_\rho(\sin^2\gamma\cos^2\phi-\cos^2\gamma)R^{-\nu_\rho}\nonumber\\
\, _2F_1\left(\frac{1}{2},\frac{\nu_\rho+2 }{2};\frac{3}{2};-\frac{S^2}{R^2}\right),
\end{align}
for $-1<\nu_\rho<1$. Note that for $\nu_\rho<-1$, intensity correlation function should be used. We are interested in small separation asymptote, i.e.  $R/S\ll 1$, given by
\begin{align}
\mathcal{D}(\bm{R},\phi)\propto \pm &\frac{\sqrt{\upi }  \Gamma \left(\frac{\nu_\rho-1}{2}\right)}{ \Gamma \left(\frac{\nu _\rho}{2}\right)}\Bigg[(1+c_\rho P_2(\cos\gamma))\nonumber\\
&+\frac{3}{2}c_\rho\frac{\nu_\rho-1}{\nu_\rho}(\sin^2\gamma\cos^2\phi-\cos^2\gamma)\Bigg]R^{1-\nu_\rho }.
\end{align}
The above equation gives some important qualitative features. First, the anisotropy vanishes at $\gamma=0$, which is again consistent with the fact that if the magnetic field is aligned along the LOS, then the statistics reduces to the isotropic statistics. Secondly, $c_\rho$ primarily determines the degree of anisotropy. Next, the iso-correlation contour is aligned towards the direction parallel to the sky-projected magnetic field if $c_\rho<0$ and towards direction orthogonal to the sky-projected magnetic field if $c_\rho>0$. It is expected that the fluctuations are elongated along the direction of magnetic field, and this would mean that for a {\it steep} spectrum, $c_\rho<0$, while for a {\it shallow} spectrum, it can be similarly shown that $c_\rho>0$. It has been shown that the density effects are isotropic at large sonic Mach number $M_s$ (\citealt{kowal2007density}). Therefore, we expect $c_\rho$ to approach 0 as $M_s$ goes large. Density anisotropy depend on Alfv\'en Mach number $M_\text{A}$ as well, although the dependence of anisotropy on sonic Mach number is more pronounced (\citealt{kowal2007density}). Therefore, $c_\rho$ should be a function of $M_s$ and $M_\text{A}$, and observational results might allow us in future to obtain good functional form of $c_\rho$. 

\section{Results}

\subsection{Effect of Velocity Fluctuations}

\subsubsection{Alfv\'en mode}
\label{sec:alvenresult}
For Alfv\'en modes, the component of velocity along the direction of the symmetry axis is zero and therefore $D=-A\mu$, and $C=A\mu^2-B$, or equivalently $\tilde{C}=-A\mu^2-\tilde{B}$, where $\tilde{C} \equiv C(0)-C(r)$. Therefore, the projection of structure function along the LOS is given by 
\begin{align}
D_z(\bm{r})=2[\tilde{B}+&\tilde{C}\cos^2\gamma -A\cos^2\theta-2D\cos\theta\cos\gamma]\nonumber\\
&=2\left[\tilde{B}(\bm{r})\sin^2\gamma-A(\bm{r})(\mu\cos\gamma-\cos\theta)^2\right].
\end{align}

It is clear that the above structure function vanishes at $\gamma=0$. In the limit when $\gamma\rightarrow 0$,  $(\mu\cos\gamma-\cos\theta)^2\rightarrow\sin^2\gamma\sin^2\theta\cos^2\phi$ and  $\mu\rightarrow \cos\theta $
and therefore
\begin{equation}\label{zerostf}
D_z(\bm{r})=2\sin^2\gamma\left[\tilde{B}-A_0\cos^2\theta\cos^2\phi\right]
~, \quad \gamma \to 0 ~.
\end{equation}
However, at $\gamma=0$, all the emitters have the same LOS velocity $v_z=0$. This implies that at this angle we are always in the thick slice regime
\footnote{In a thick slice regime, the intensity structure has a divergence of $S^2$, where $S$ is the  size of an emitting region. However, in a thin slice regime, the divergence is $S$. The fact that $1/\sin\gamma$ introduces an additional divergence is clear to explain that at $\gamma\sim 0$, thin slice approximation does not work.}.
With this observation, it is expected that the thin slice approximation
will not work whenever $\gamma$ is less than some critical angle $\gamma_c$.
The criterion for a slice to be thick is $\Delta v>\sqrt{D_z(R)}$, where $R$ is the separation between the two LOS. Therefore, 
we are in the thick slice regime whenever $\sin\gamma_c\lesssim\Delta v/(2\tilde{B})$. However, this only applies if the turbulent motions consist of only Alfv\'en modes. This situation is nevertheless quite rare because slow modes are also
of solenoidal type and usually come along with Alfv\'en modes.
Since slow modes have non-vanishing structure function at $\gamma=0$, 
thin slice approximation would 
still be valid if we consider the contribution of both slow and Alfv\'en modes, as at small $\gamma$ velocity structure function is dominated by
slow modes.
In a thin slice regime, calculating monopole and quadrupole terms primarily requires the knowledge of $q_1, q_2, q_3$ and $u_1$ (cf. equations \ref{monopole} and \ref{quadrupole}), which for the Alfv\'en modes are
\begin{align}\label{alfvencoeff}
q_1&=\left(2\tilde{B}_0+B_2\right)\sin^2\gamma, \qquad u_1=-2A_0\cos^2\gamma-3B_2\sin^2\gamma\nonumber\\
q_2&=-\left(3B_2\cos^2\gamma+2A_0\sin^2\gamma\right)\sin^2\gamma, \qquad q_3=0.
\end{align}

\begin{figure*}
\begin{center}
\includegraphics[scale=0.4]{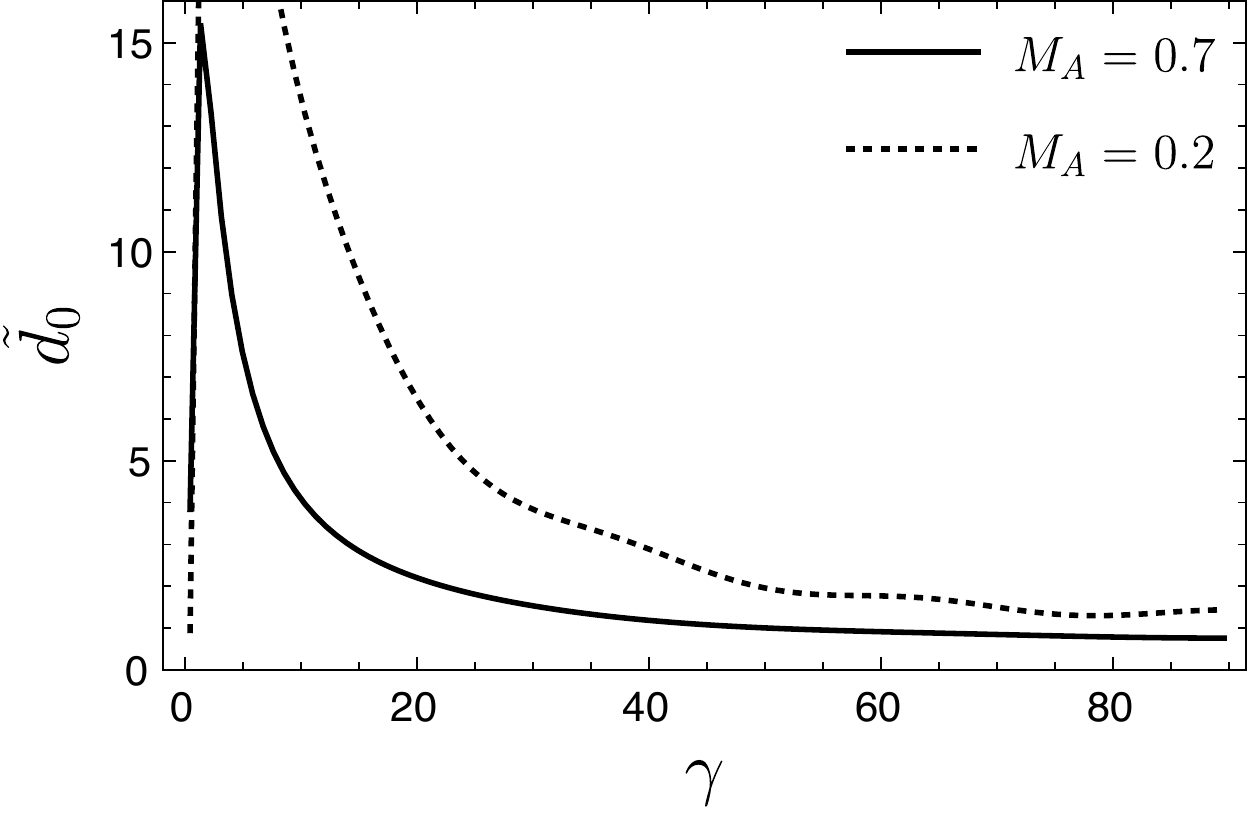}\hspace*{0.2cm}
\includegraphics[scale=0.4]{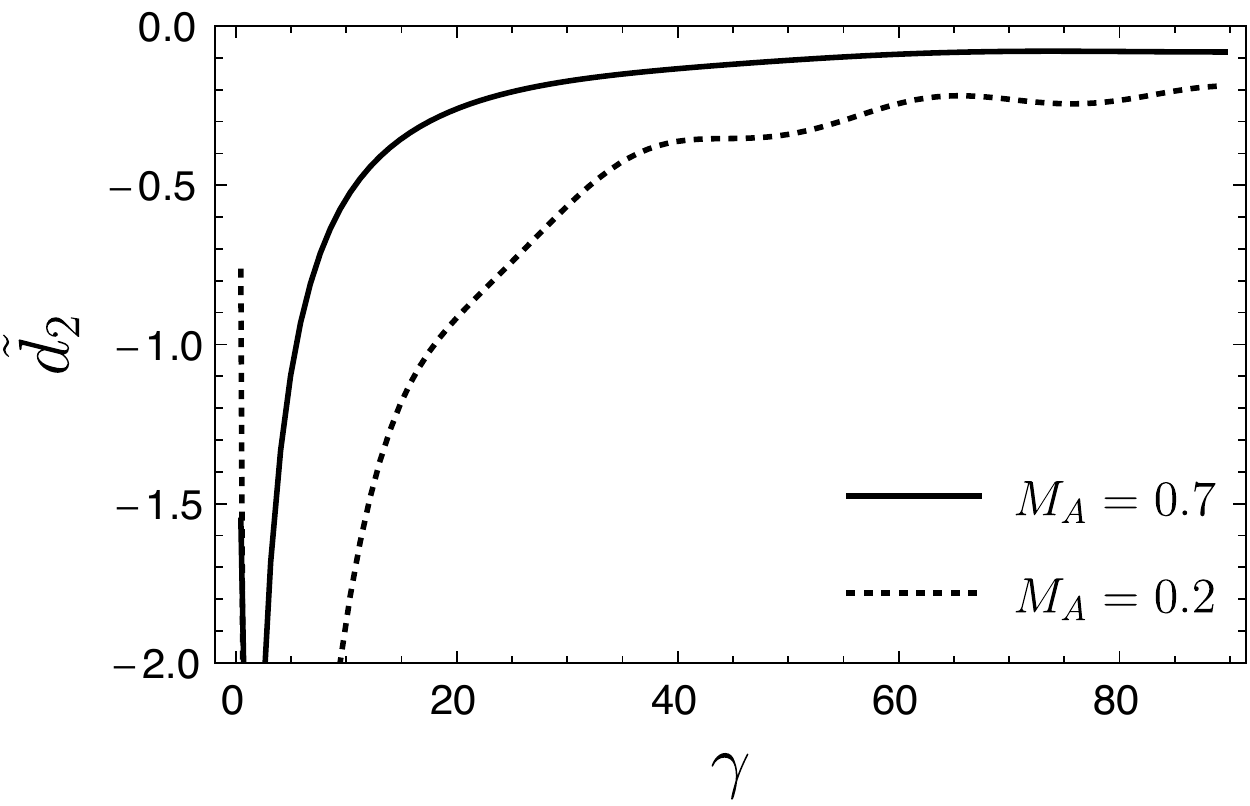}\hspace*{0.2cm}
\includegraphics[scale=0.4]{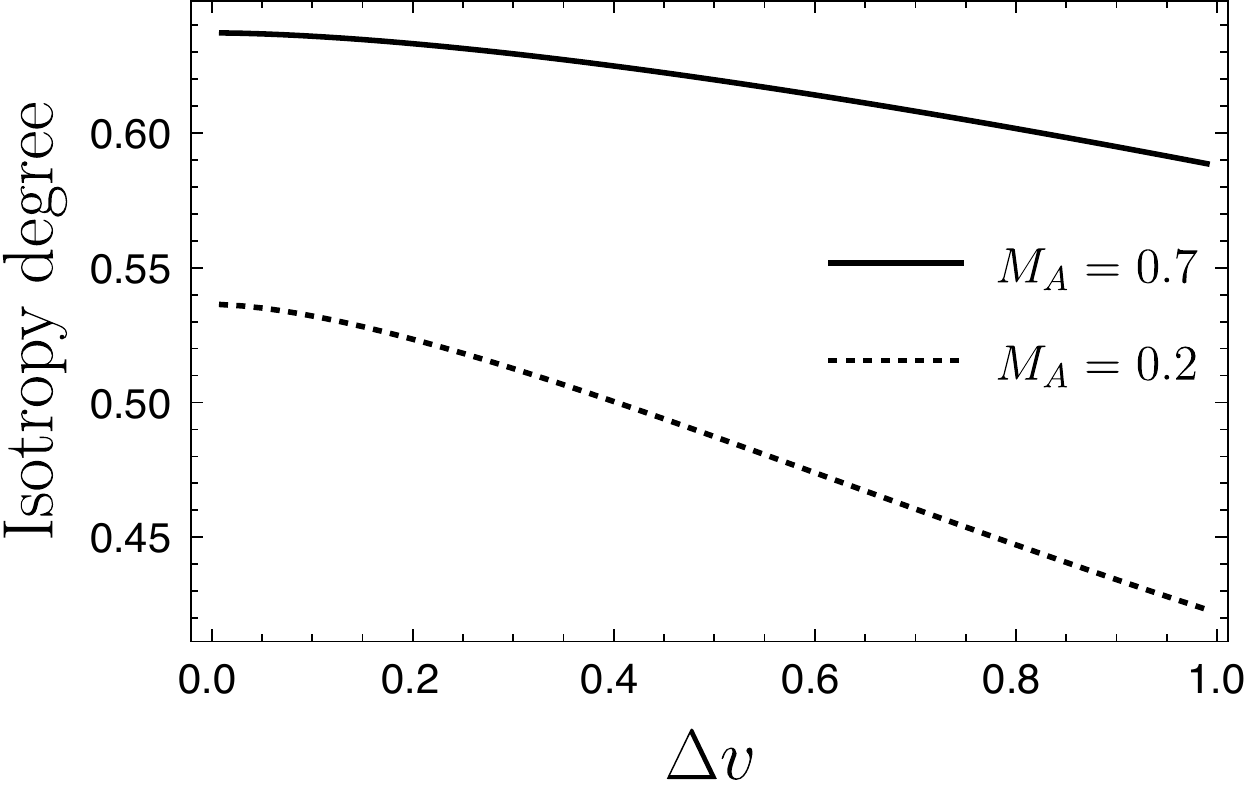}

\caption{Alfv\'en mode. From left to right: monopole, quadrupole and isotropy degree. Monopole and quadrupole are as a function of angle between LOS and magnetic field, $\gamma$. Isotropy degree as a function of $\Delta v$ is at $\gamma=\upi/2$. From top to bottom: top is at $M_\text{A}=0.4$, bottom at $M_\text{A}=0.7$ }
\label{amachp}
\end{center}
\end{figure*}

\begin{figure*}
\begin{center}
\includegraphics[scale=0.4]{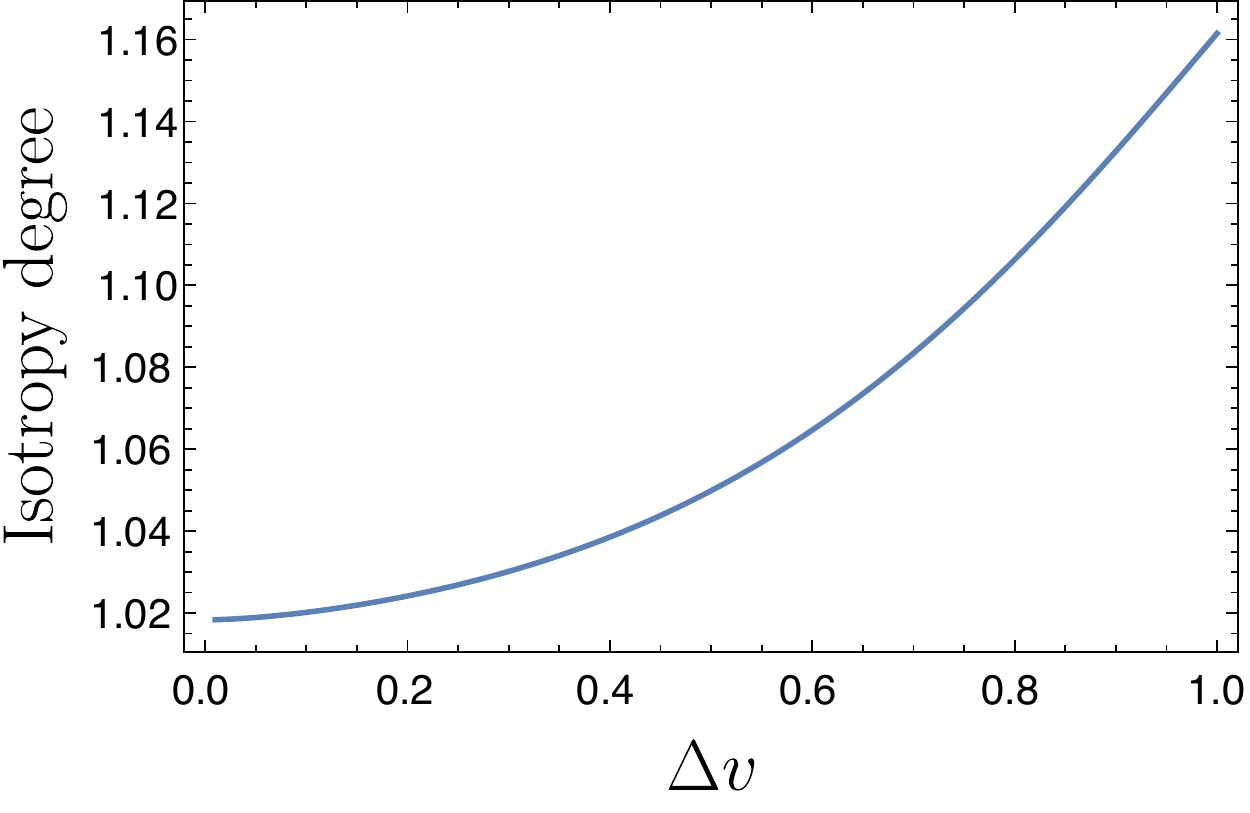}\hspace*{0.2cm}
\includegraphics[scale=0.4]{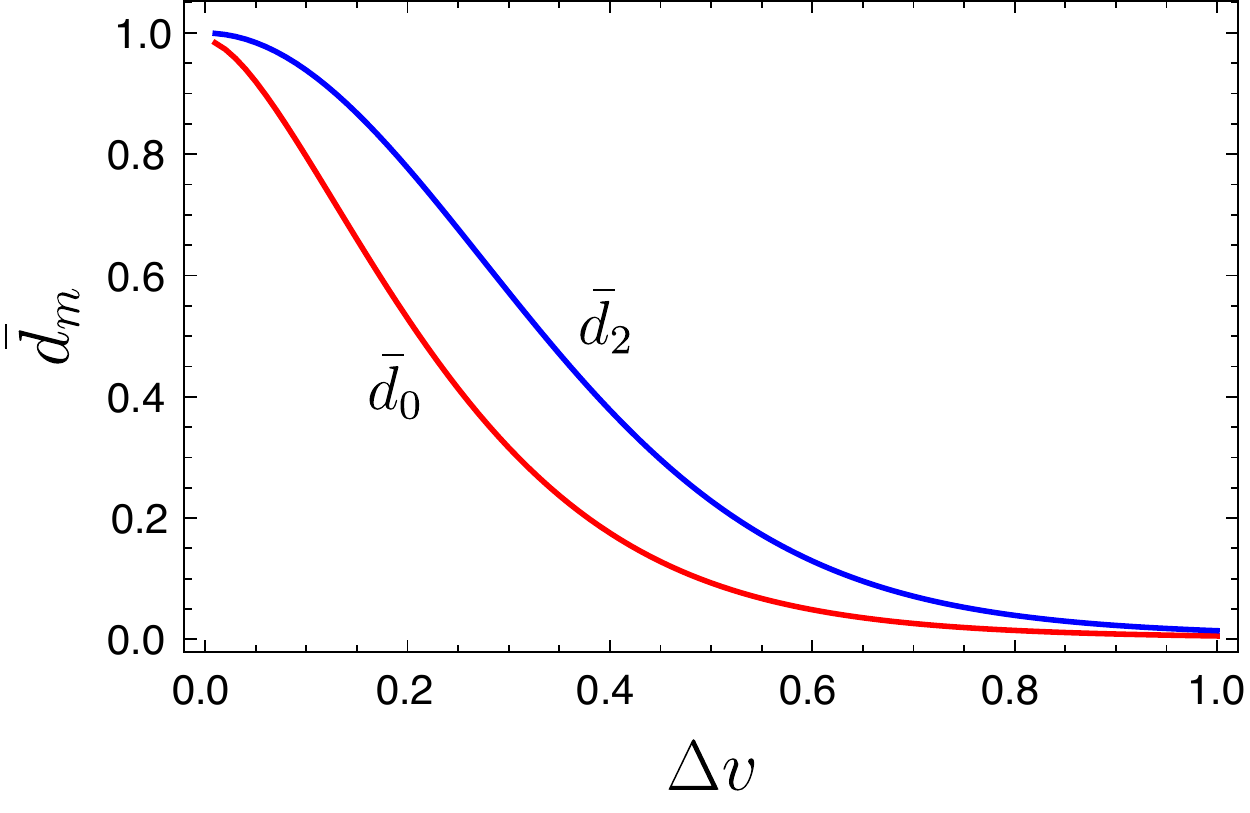}
\caption{Alfv\'en mode. Left: The degree of anisotropy assuming an isotropic power spectrum $\mathcal{A}(k,\mu_k)$ (cf. Eq. \ref{eq:alfvenpower}). The observed anisotropy in this case comes purely from the anisotropic tensor structure (cf. Eq \ref{eq:alfvencorr}). Right: quadrupole and monopole at $M_\text{A}=0.7$. Both are normalised by their highest value.}
\label{fig:isoalfven}
\end{center}
\end{figure*}

Fig. \ref{amachp} shows the monopole and quadrupole contributions 
as well as isotropy degree of intensity correlation from Alfv\'en modes.
We highlight several important properties.
First, both monopole and quadrupole components are decreasing with the increase in Alfv\'en Mach number, $M_\text{A}$. Secondly, the anisotropic feature decreases with the increase in Alfv\'en Mach number, which is expected as higher Alfv\'en Mach number corresponds to higher isotropy.  Moreover, both monopole and quadrupole are insensitive to $\gamma$ for $\gamma\gtrsim \upi/4$, which can be a useful feature to determine Alfv\'en Mach number $M_\text{A}$. In addition to that, it is clear from the figure that isotropy degree for Alfv\'en mode is less than 1. This implies that iso-correlation contours are elongated along the direction of sky projection of mean magnetic field. For Alfv\'en modes, this corresponds to the spectral suppression towards the direction parallel to the projected field. This effect is due to the structure of power spectrum of Alfv\'en modes. If this power spectrum was isotropic, the isocontours of this mode would be elongated along the direction orthogonal to the sky projection of mean-magnetic field (see the left-hand panel of Fig. \ref{fig:isoalfven}). Note that both the monopole and quadrupole are  increasing with the decrease in $\gamma$, which might look counter-intuitive. This increase is because of the fact that the structure function  $D_z\propto \sin^2\gamma$, and therefore, the intensity structure function $\mathcal{D}\propto \sin^{-1}\gamma$ which reflects that more and more emitters are occupying
the same velocity channel $v_z=0$.

By looking at the left-hand panel of Fig. [\ref{fig:isoalfven}], one can observe the decrease of isotropy degree for increasing slice thickness \footnote{Whenever we talk about slice thickness $\Delta v$, unless explicitly mentioned otherwise, we talk about slice thickness normalized by velocity dispersion.}, $\Delta v$. This decrease can be understood in the following sense: at small slice thickness, all emitters have similar LOS velocities and anisotropies are suppressed. But with the increase in slice thickness,  the correlations of velocity are better sampled, thus increasing the anisotropy. The change of anisotropy with slice thickness is an important result of this paper. This can be a useful tool in the study of MHD turbulence. It is however important to note that although the anisotropy increases with increasing $\Delta v$, the quadrupole and monopole individually approach to zero with increasing $\Delta v$. This is illustrated in the right-hand panel of Fig. \ref{fig:isoalfven}, which clearly shows that both the monopole and quadrupole are clearly approaching zero as $\Delta v$ approaches unity.

\subsubsection{Fast mode}
Fast modes in high-$\beta$ plasma correspond to sound waves, which are isotropic (see GS95; \citealt{cho2003compressible}) .

Fast modes in low-$\beta$ plasma have anisotropy in-built in the tensor structure, although their power spectrum is isotropic. For fast modes in low-$\beta$ regime, the component of velocity along the direction of symmetry axis is zero, and therefore, the projection of structure function along the LOS takes the same form as that for the Alfv\'en mode,
\begin{equation}
D_z(\bm{r})=2\left[\tilde{B}(\bm{r})\sin^2\gamma-A(\bm{r})(\mu\cos\gamma-\cos\theta)^2\right].
\end{equation}
The above structure function also vanishes at $\gamma=0$; therefore, the discussion about thin and thick slice applies to this mode as well. To find monopole and quadrupole terms, the coefficients $q_1, q_2, q_3$ and $u_1$ for this mode are given by equation \eqref{alfvencoeff}. 
\begin{figure*}
\begin{center}
\includegraphics[scale=0.4]{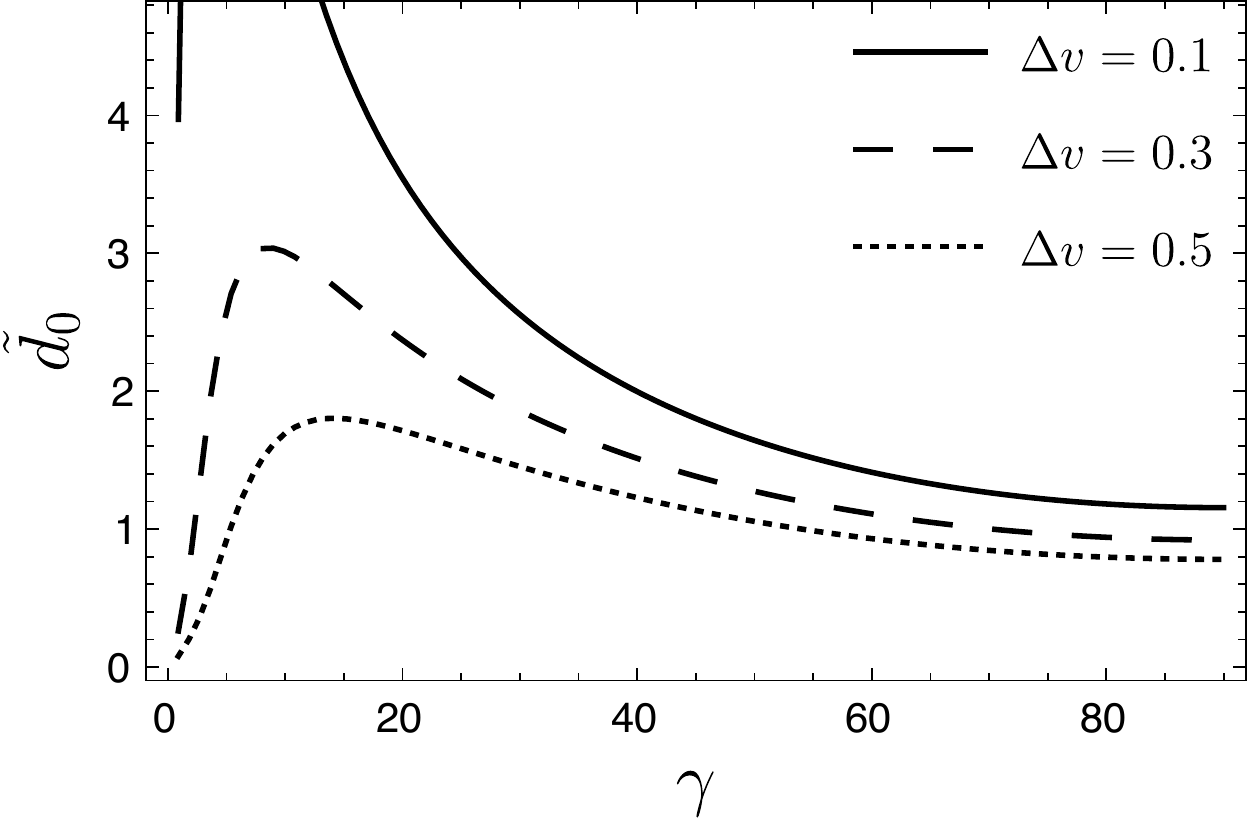}\hspace*{0.2cm}
\includegraphics[scale=0.4]{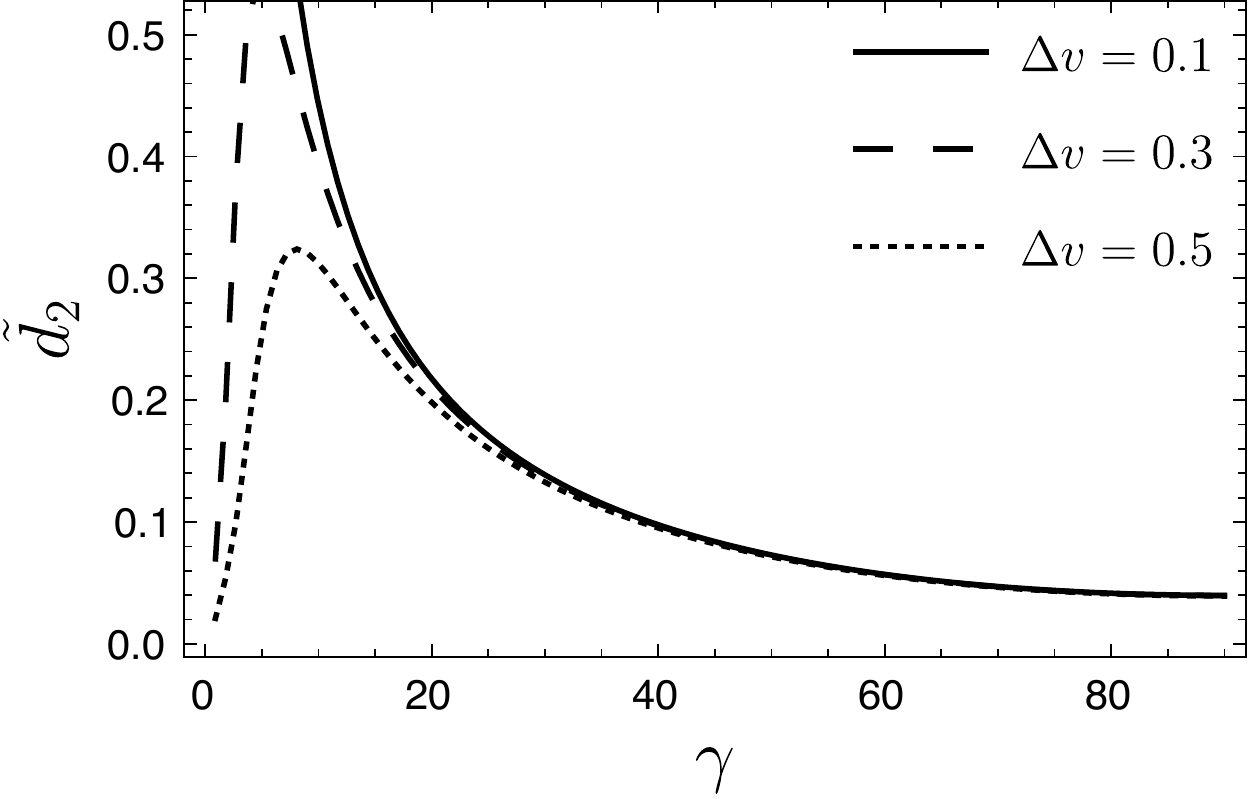}\hspace*{0.2cm}
\includegraphics[scale=0.4]{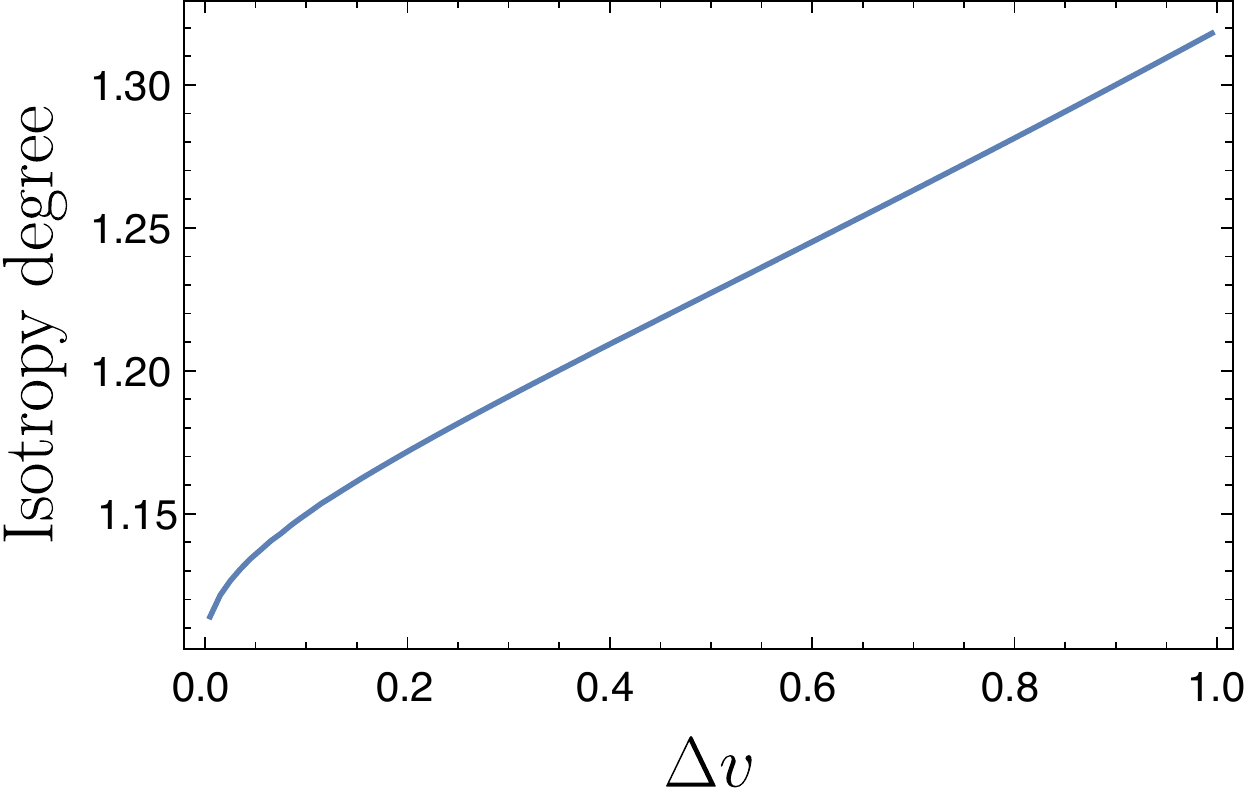}
\caption{Fast mode low-$\beta$. From left to right, monopole and quadrupole and degree of isotropy. The curves in left-hand and central panels represent from top to bottom: $\Delta v=0.1, \Delta v=0.3$ and $\Delta v=0.5$. The isotropy degree is at $\gamma=\upi/2$. All angles are in degrees.}
\label{fig:fast}
\end{center}
\end{figure*}

Fig. \ref{fig:fast} shows monopole, quadrupole and degree of anisotropy of low-$\beta$ fast modes. Of particularly interesting pattern is the degree of isotropy which is greater than 1, unlike Alfv\'en modes which had this isotropy degree less than 1. This clearly implies that intensity structure iso-contours of fast modes are elongated along the direction perpendicular to magnetic field projection in the 2-D plane. This in fact validates our previous assertion that for an isotropic power spectrum, the iso-contours should be elongated towards the direction perpendicular to sky-projected magnetic field. It is also interesting to note that even at $\gamma=\upi/2$ (which is the most anisotropic case), these modes are not so anisotropic. Therefore, observation of strong anisotropy signal could allow us to infer that fast modes are not  possibly unimportant (this cannot totally eliminate fast modes, because a mixture of fast and Alfv\'en modes can, for example, produce strong anisotropy as long as fast modes are subdominant ). Fig. [\ref{fig:fast}] shows while  monopole decreases rapidly with increasing slice thickness, the quadrupole is relatively less affected with the changing slice thickness; therefore, this increases the quadrupole-to-monopole ratio with increasing slice thickness.
\subsubsection{Slow mode}
Slow modes are anisotropic in both high and low-$\beta$. 
The detailed mode structure of this mode is studied in Appendix \ref{vcorfsm}. The structure function of low $\beta$ slow mode is given is 
\begin{equation}
D_z(\bm{r})=2(C(0)-C(\bm{r}))\cos^2\gamma.
\end{equation}
Analytical calculation of the monopole and quadrupole contribution to the intensity structure function requires knowledge of various parameters as shown in equations (\ref{monopole}) and (\ref{quadrupole}), and these parameters are summarized in Table \ref{tab:structurepara}.
\begin{figure*}
\begin{center}
\includegraphics[scale=0.4]{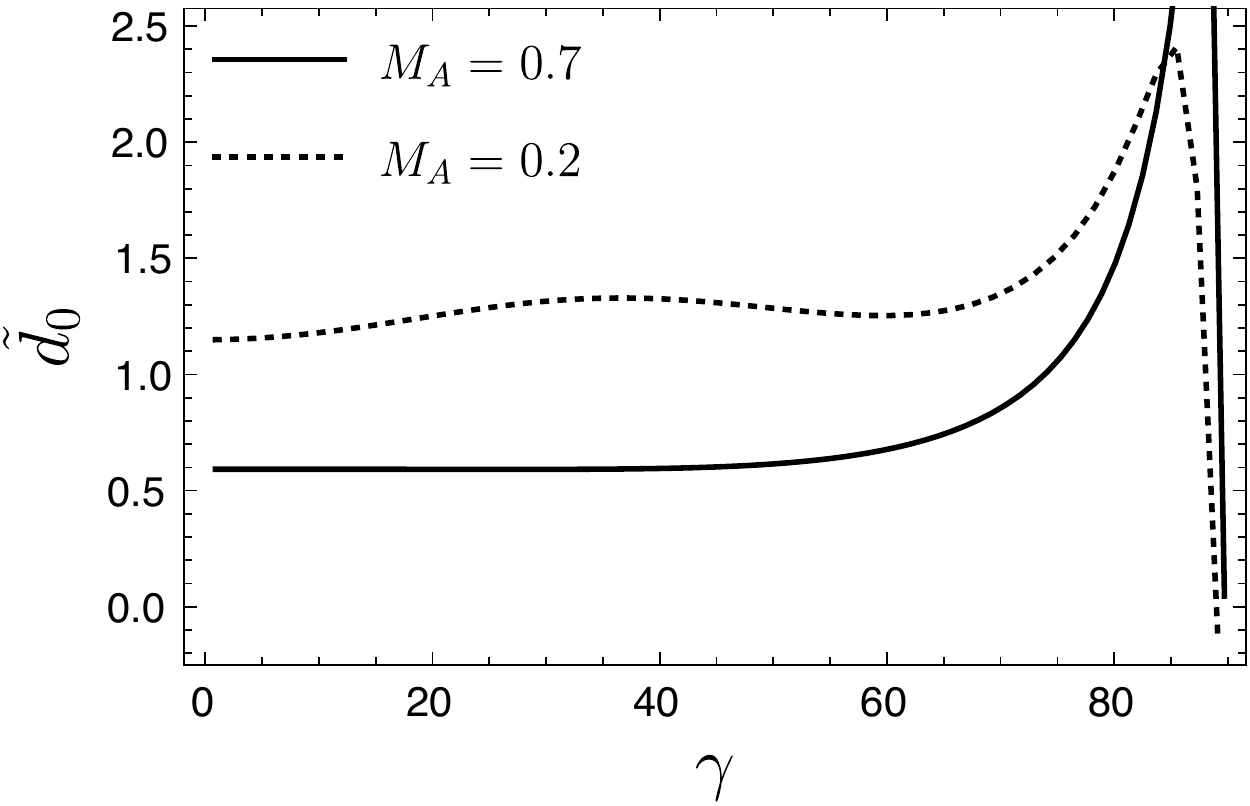}\hspace*{0.2cm}
\includegraphics[scale=0.4]{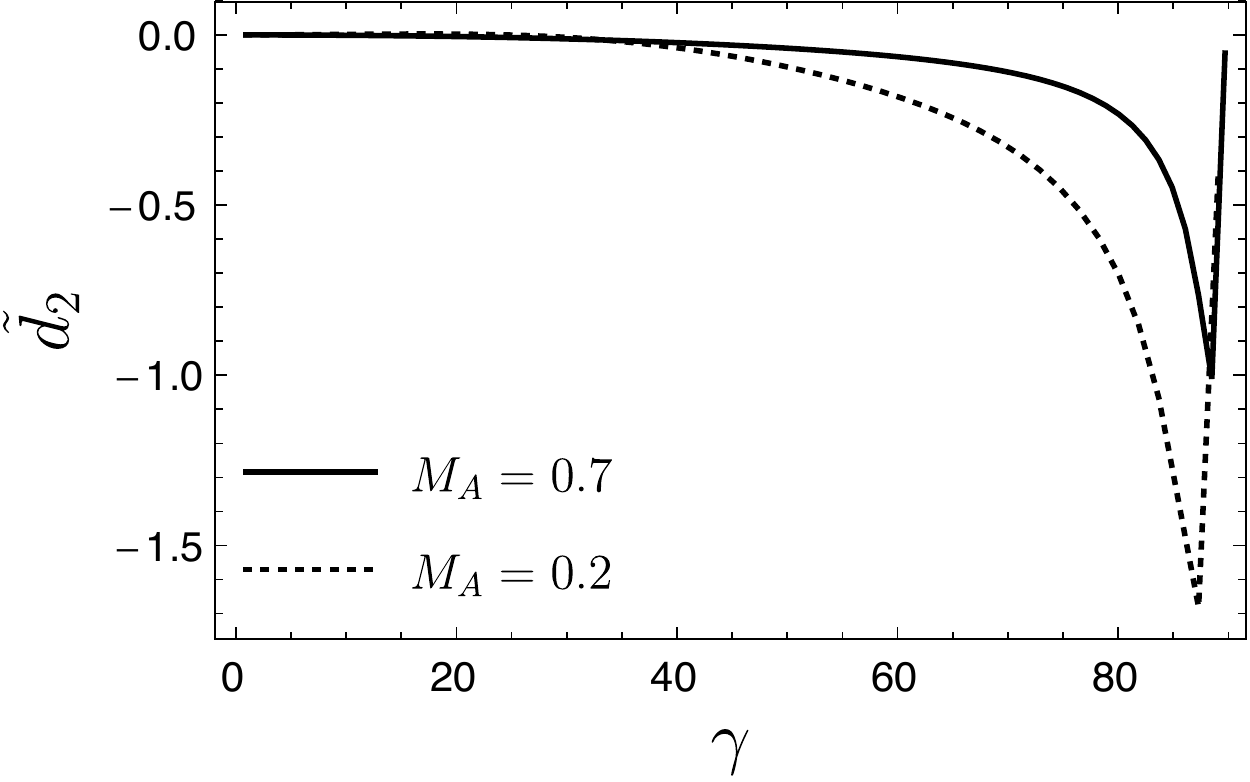}\hspace*{0.2cm}
\includegraphics[scale=0.4]{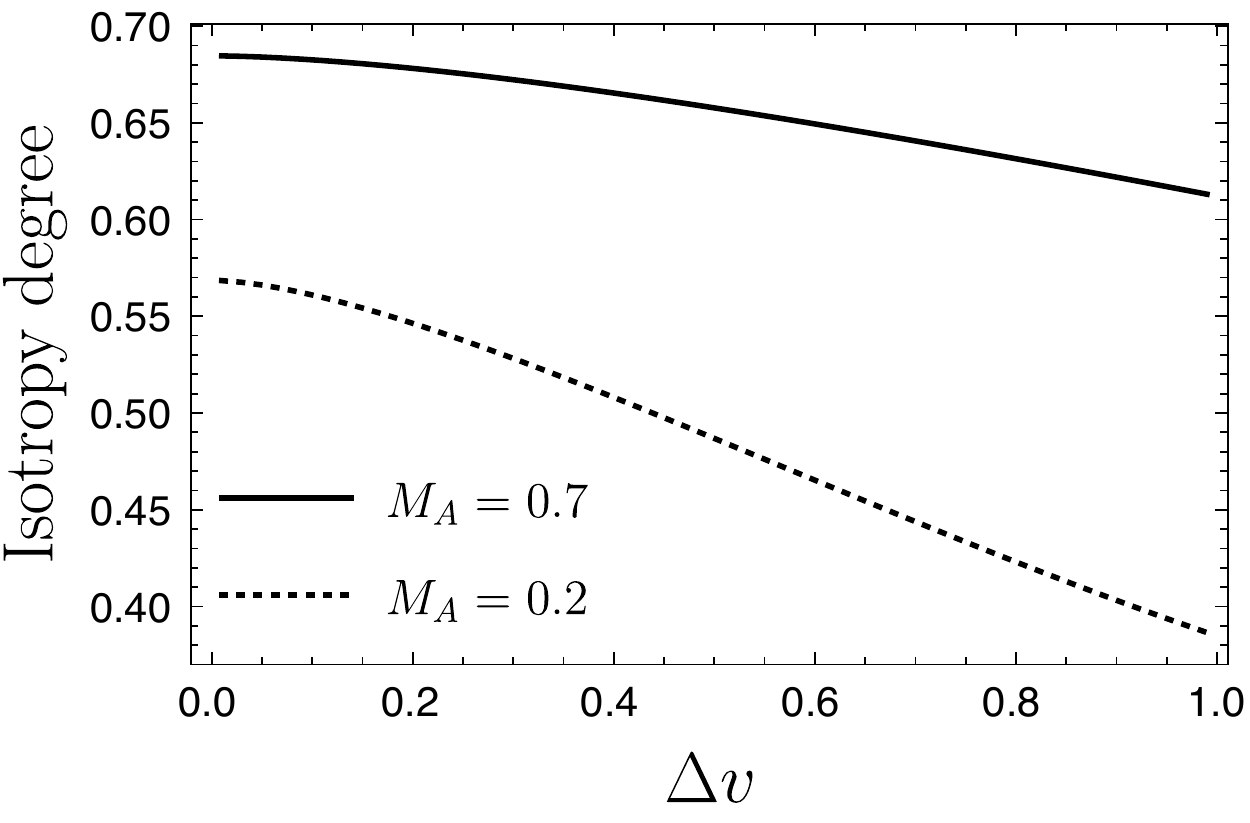}
\caption{Slow mode low-$\beta$. From left to right, monopole and quadrupole and degree of isotropy at $M_\text{A}=0.2$ and $0.7$, at $\Delta v=0.1$. The isotropy degree is calculated at $\gamma=\upi/3$.}
\label{fig:slowmodelowbeta}
\end{center}
\end{figure*}

Fig. \ref{fig:slowmodelowbeta} shows that slow modes in low $\beta$ are highly anisotropic at low Alfv\'en Mach number $M_\text{A}$. However, they become more isotropic at large $M_\text{A}$, which shows that the observed anisotropy of intensity fluctuations from these modes is primarily due to the anisotropy in power spectrum. The anisotropy is pronounced for $\gamma\gtrsim \upi/4$. Moreover, the iso-correlation contours in this limit are always elongated towards the direction of sky-projected magnetic field, which is similar to the Alfv\'en mode. Comparing Figs. \ref{amachp} and \ref{fig:slowmodelowbeta}], it is easy to see that in the regime $\gamma \gtrsim \upi/3$, slow modes in low $\beta$ are more anisotropic than Alfv\'en modes for same $M_\text{A}$. 

\begin{figure*}
\begin{center}
\includegraphics[scale=0.4]{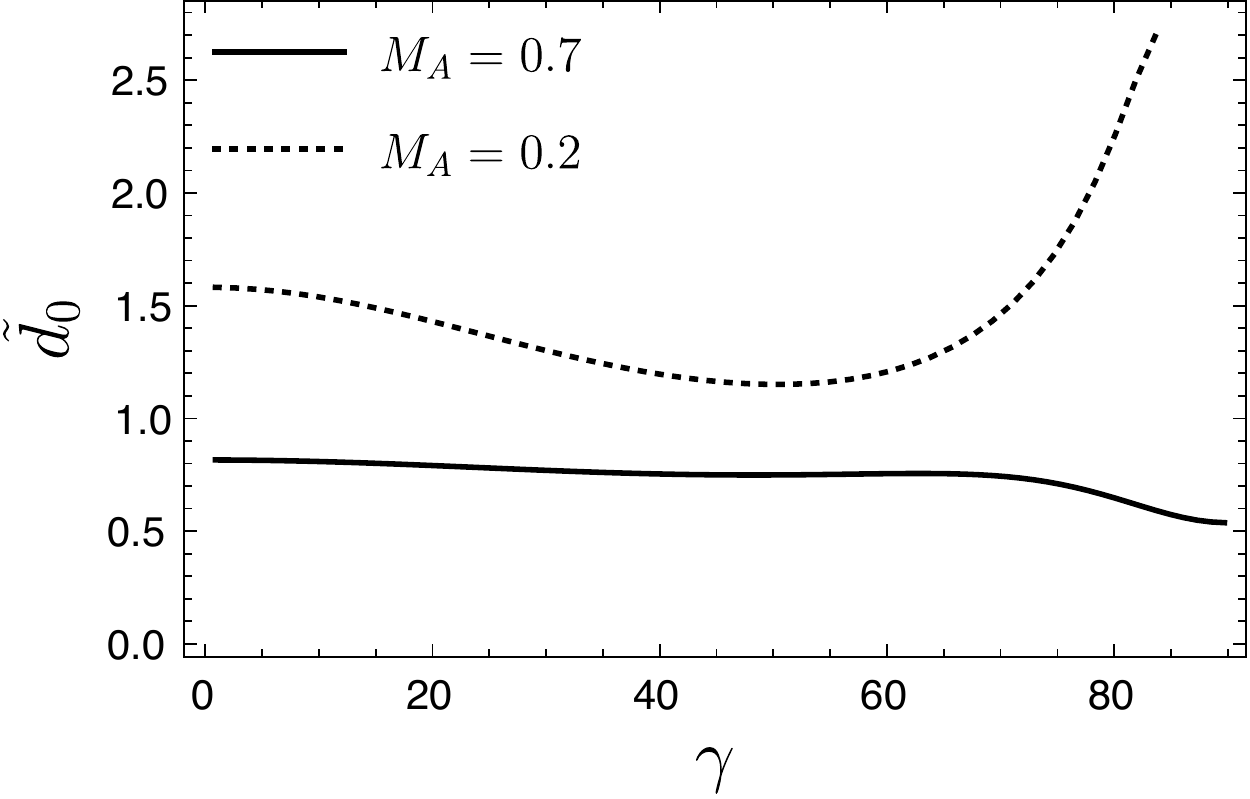}\hspace*{0.2cm}
\includegraphics[scale=0.4]{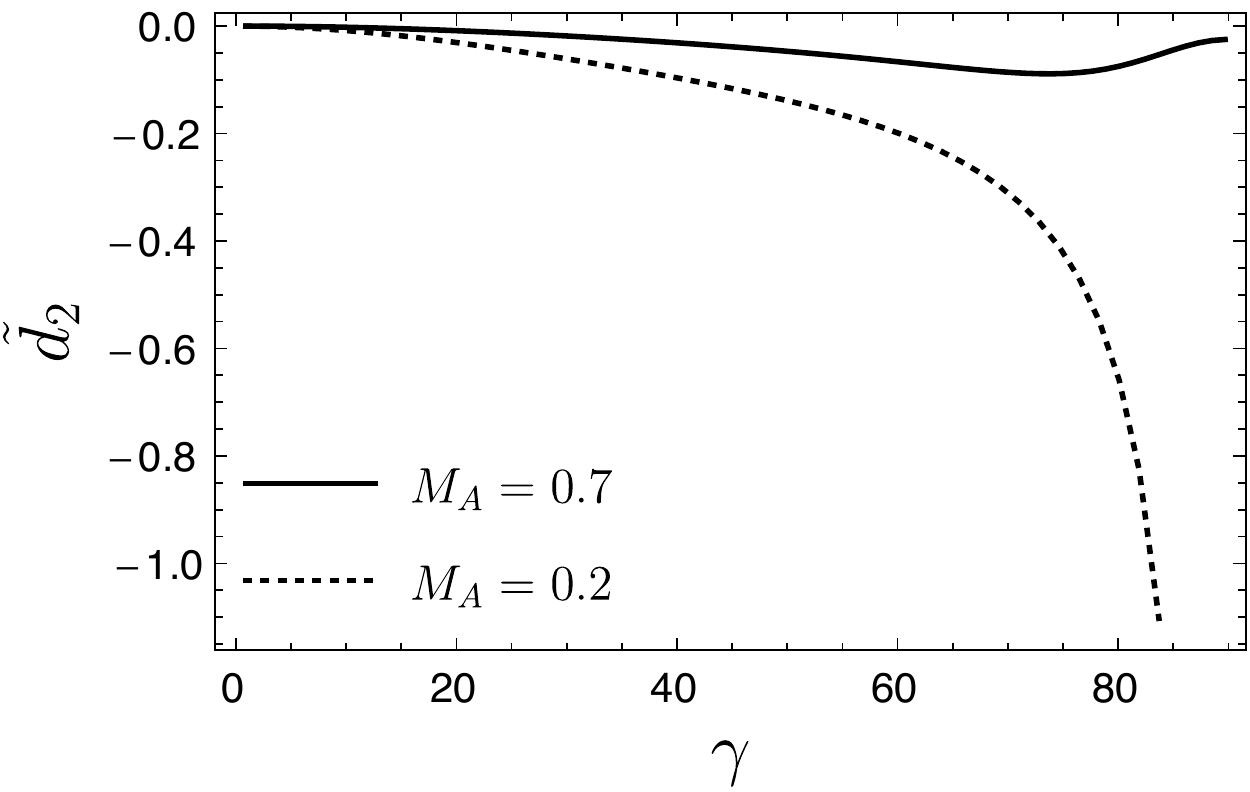}\hspace*{0.2cm}
\includegraphics[scale=0.4]{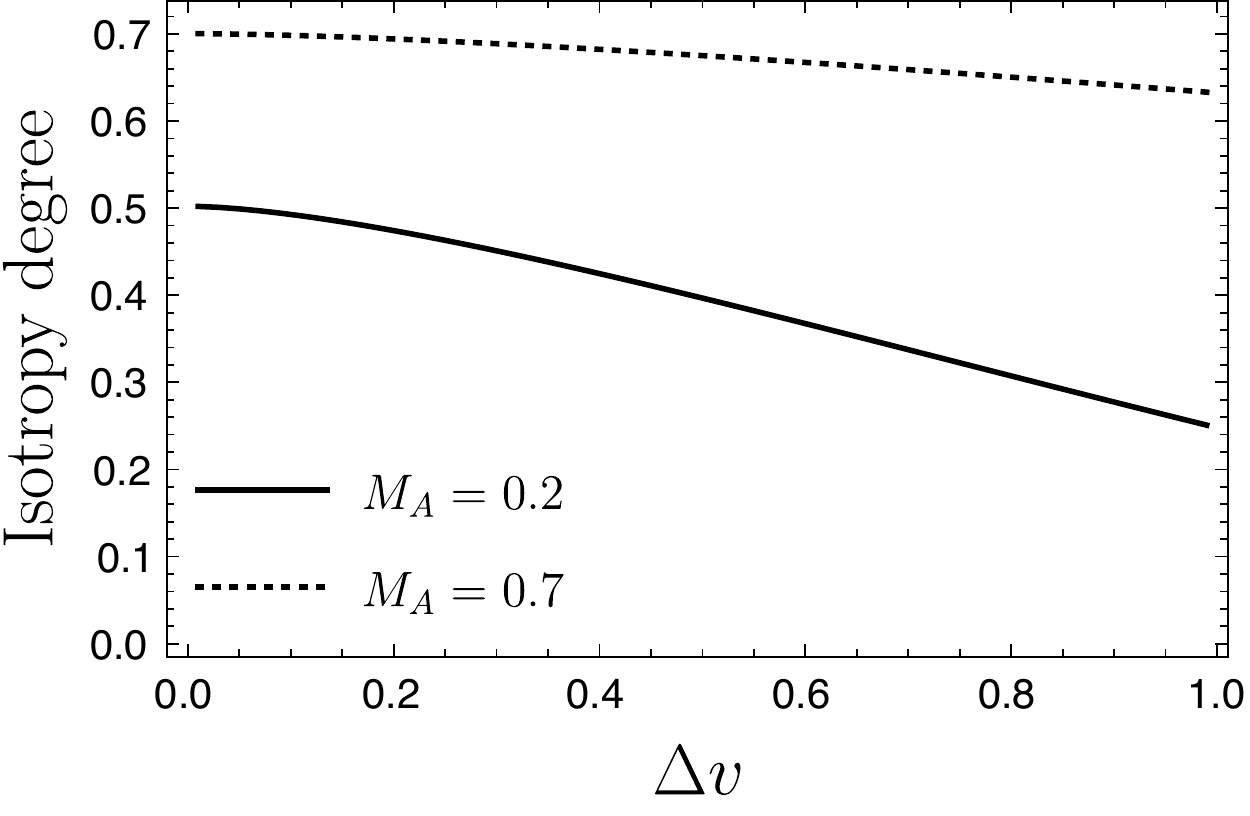}
\caption{Slow Mode high $\beta$. Monopole and quadrupole and degree of isotropy for slow mode at high $\beta$.  Left-hand and central panels are at $\Delta v=0.1$. The isotropy degree is calculated at $\gamma=\upi/3$.}
\label{fig:slowhighbeta}
\end{center}
\end{figure*}

Slow modes in high $\beta$ regime show more interesting properties as shown in Fig. \ref{fig:slowhighbeta}. The iso-correlation contours of this mode are aligned towards the direction parallel to the sky-projected magnetic field.  The anisotropy comes from the anisotropy in-built in the tensor structure of this mode as well as from the power spectrum (cf. equation \ref{eq:alfvenpower}) of this mode. Similar to Alfv\'en mode, the iso-correlation contours of this mode are aligned towards the direction perpendicular to the sky-projected magnetic field. 

However, our method of analysing the anisotropy by truncating the series of structure function (cf. Sec. \ref{sec: truncation}) does not work well for very small $M_\text{A}$. One reason is that the power spectrum in the regime of small $M_\text{A}$ becomes more or less like  $\delta(\hat{k}.\hat{\lambda})$, and therefore all $\mathcal{A}_n$ are important. Note that at small $M_\text{A}$, the intensity structure function, and hence the anisotropy, of high $\beta$ and low $\beta$ slow modes should behave in a similar way. This is because the power spectrum (cf. equation \ref{eq:alfvenpower}) of high $\beta$ slow mode behaves like $\delta(\hat{k}\cdot\hat{\lambda})$ at low $M_\text{A}$, and therefore, the tensor structure of slow modes at high $\beta$ (cf. equation \ref{eq:slowhbcorr}) should reduce to the same form as that of low $\beta$ slow modes, i.e. $D_z(\bm{r})\propto \hat{\lambda}_i\hat{\lambda}_j$ for both modes.

\subsubsection{Mixture of Different Modes}
In this section, we show effects of mixing of modes in the isotropy degree. Mixing effects are interesting as real world MHD turbulences have different modes and our observations are the result of the combined effects of these modes. 
\begin{figure*}
\begin{center}
\includegraphics[scale=0.4]{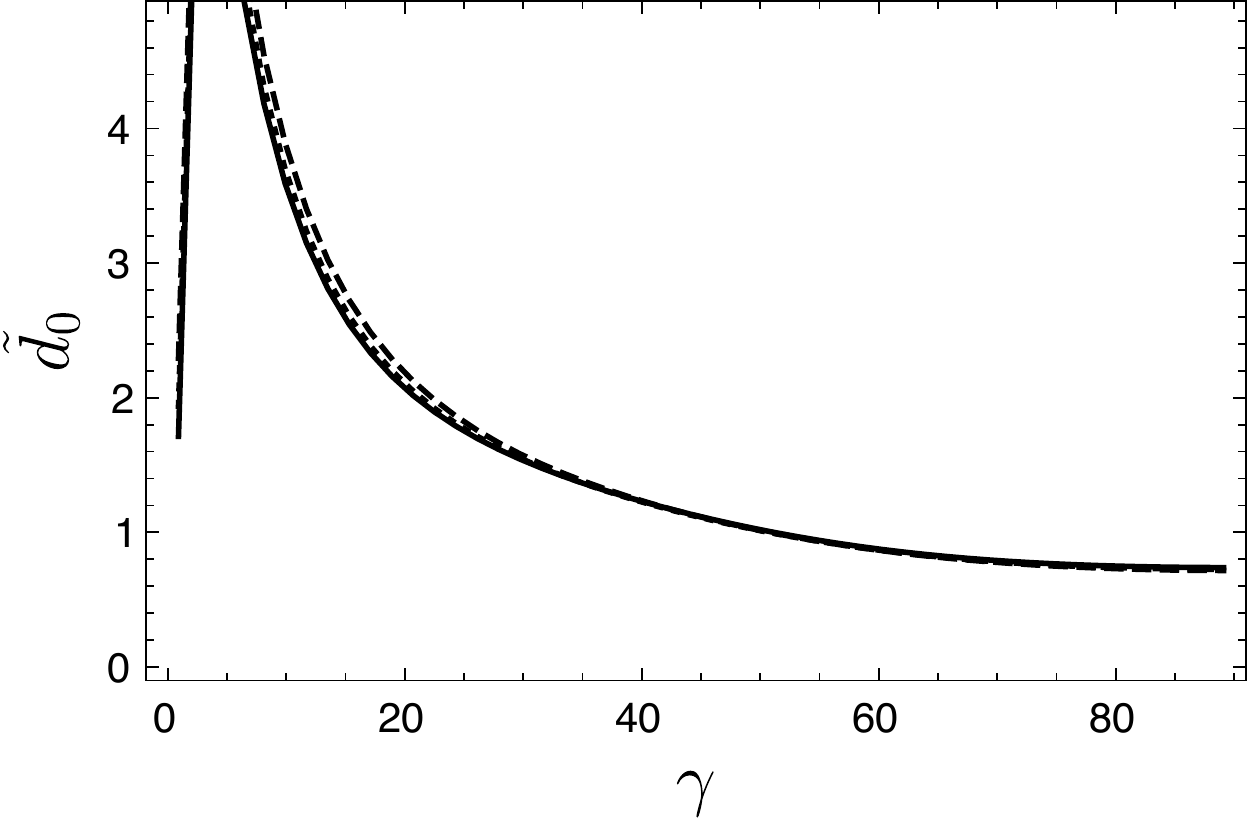}\hspace*{0.2cm}
\includegraphics[scale=0.4]{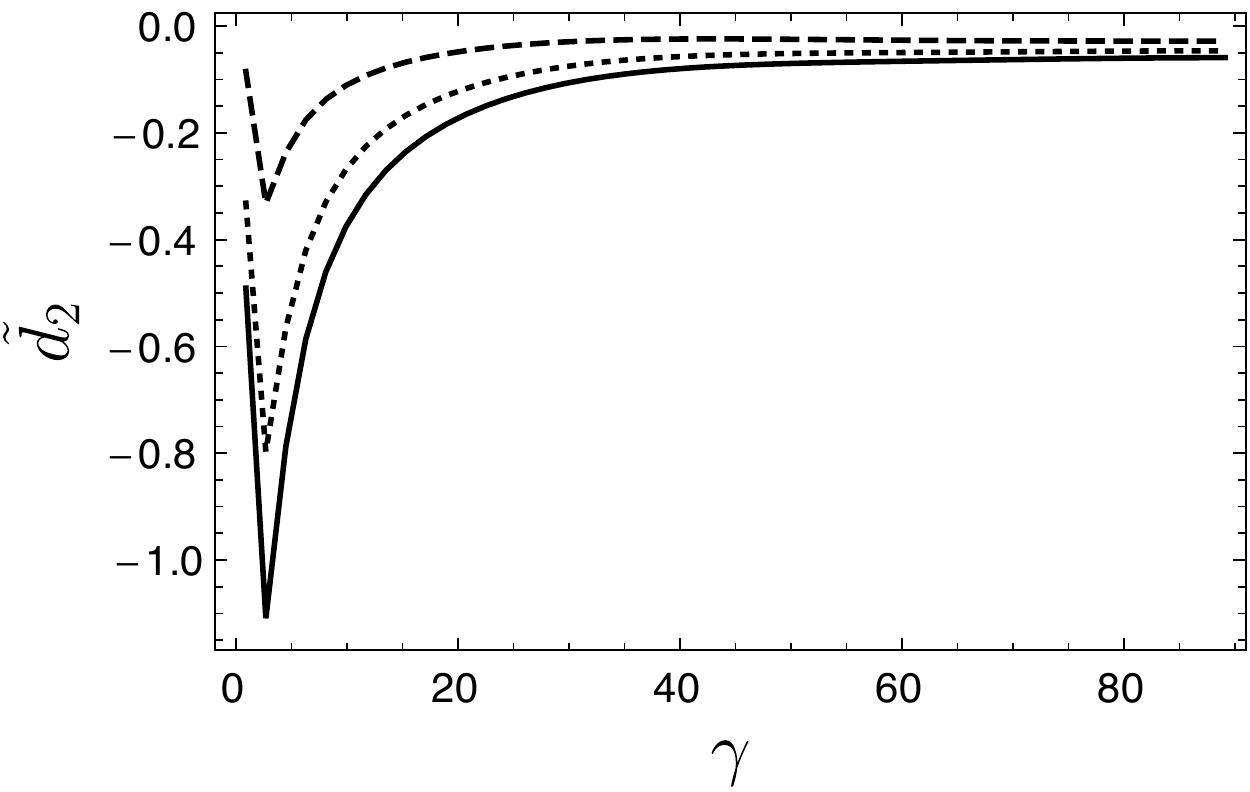}\hspace*{0.2cm}
\includegraphics[scale=0.4]{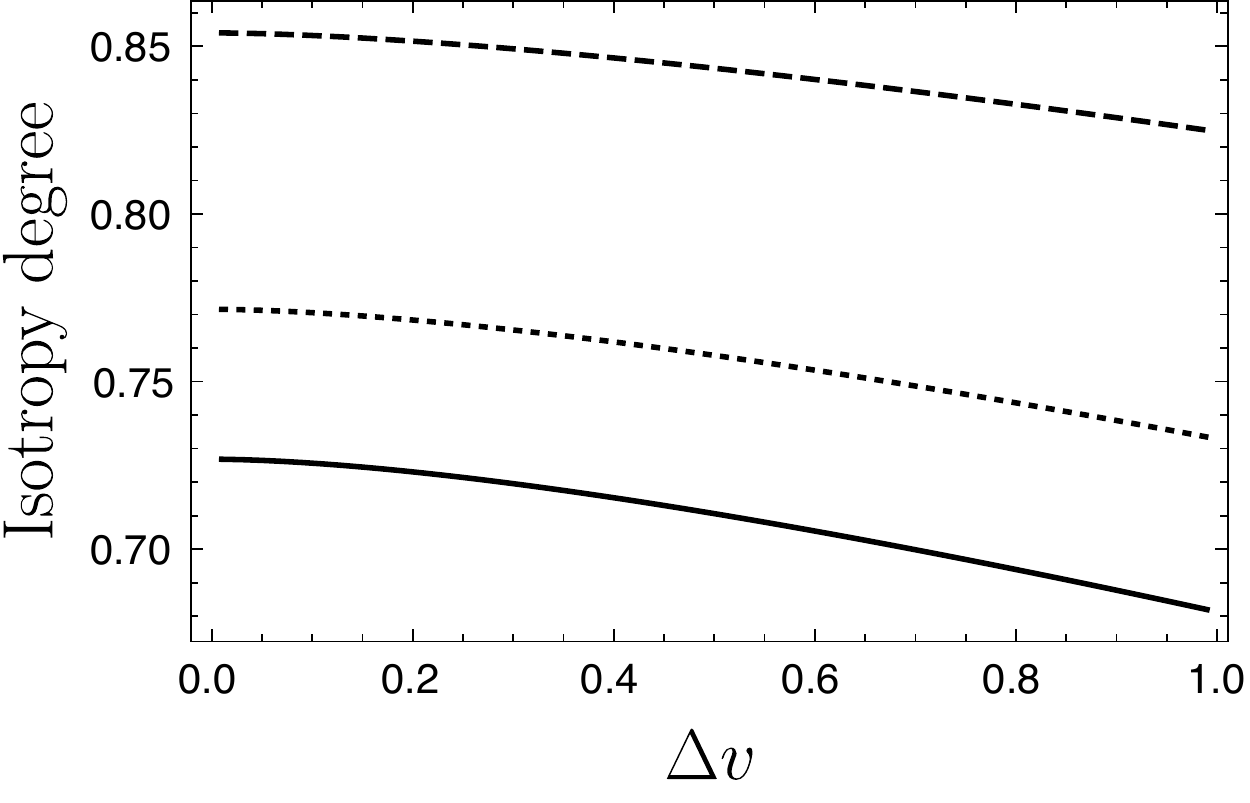}
\caption{Alfv\'en $+$ fast modes at low-$\beta$ at $M_\text{A}=0.7$. Left and centre: monopole and quadrupole for various percentage of mixture. Right: isotropy degree at $\gamma=\upi/2$. The solid curve is for $95\%$ Alfv\'en and $5\%$ fast, dotted curve is for $90\%$ Alv\'en and $10\%$ fast and the dashed curve is for $80\%$ Alv\'en and $20\%$ fast.}
\label{fig:fastmixture}
\end{center}
\end{figure*}

We consider the mixture of Alfv\'en modes and fast modes , as well as  mixture of Alfv\'en and slow modes. Fig. \ref{fig:fastmixture} shows that the mixture of Alfv\'en and mixture of fast mode with Alfv\'en mode in low-$\beta$ increases the isotropy (cf. Fig. \ref{amachp}) when compared with pure Alfv\'en isotropy. This effect can is caused by two factors: first, fast modes are less anisotropic than Alfv\'en and therefore, we expect their combination to be more isotropic than Alfv\'en alone. More important is the second factor: the quadrupole anisotropies of fast (in low-$\beta$) and Alfv\'en modes are opposite in sign. This means the anisotropy effects of the two modes act against each other. Therefore, even a small percentage of fast modes in the mixture can cause a significant difference in the anisotropy level. This has been confirmed in the left-hand and central panels of Fig. \ref{fig:fastmixture}, which shows that while the monopole is relatively unaffected by the composition of mixture, the quadrupole is significantly affected with larger composition of fast modes. Note that we usually expect fast mode to be marginal in the mixture. Fast modes in high $\beta$, however, are {\it isotropic}. Therefore, we again expect the mixture of high-$\beta$ fast and Alfv\'en mode to be more isotropic than Alfv\'en alone. However, unlike low-$\beta$ fast mode , this mode at high-$\beta$ does not have any quadrupole anisotropy to act against the Alfv\'en anisotropy. Therefore, this mixture should be more anisotropic than the mixture of high-$\beta$ fast mode. 
\begin{figure*}
\begin{center}
\includegraphics[scale=0.4]{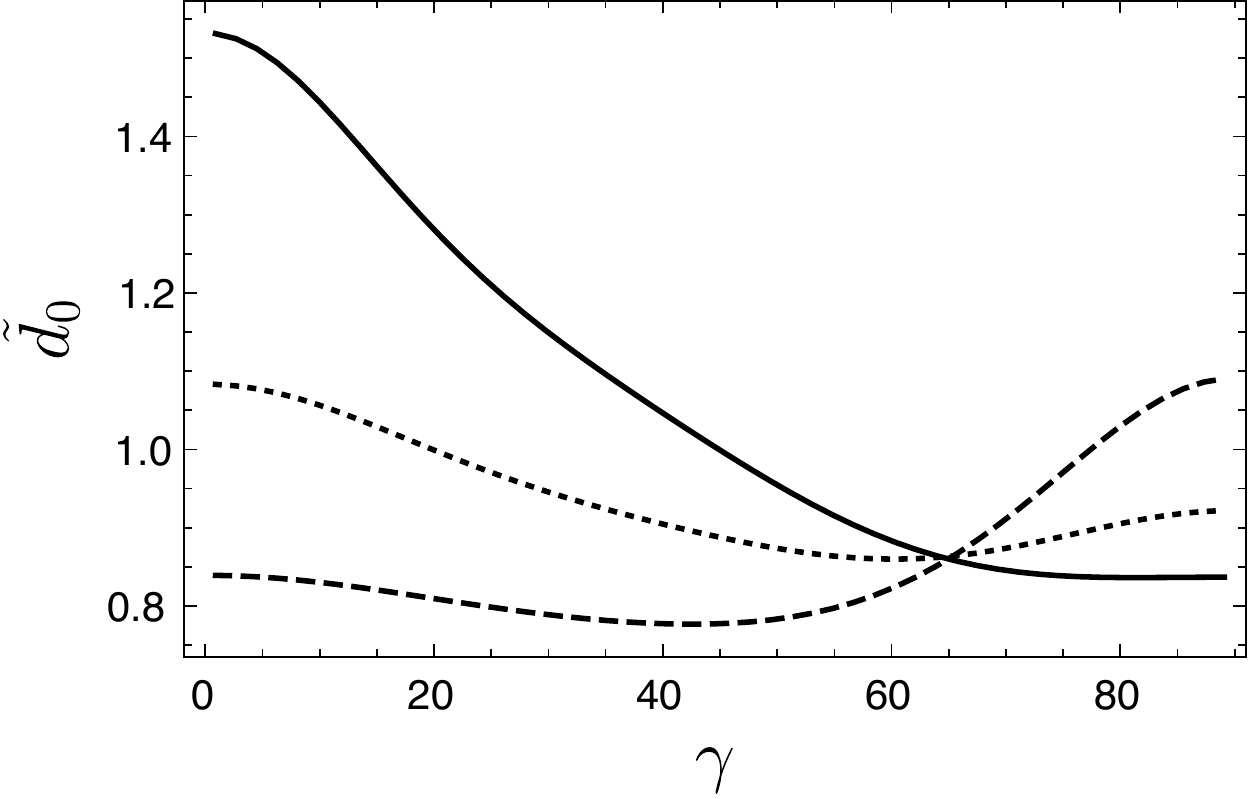}
\hspace*{0.2cm}
\includegraphics[scale=0.4]{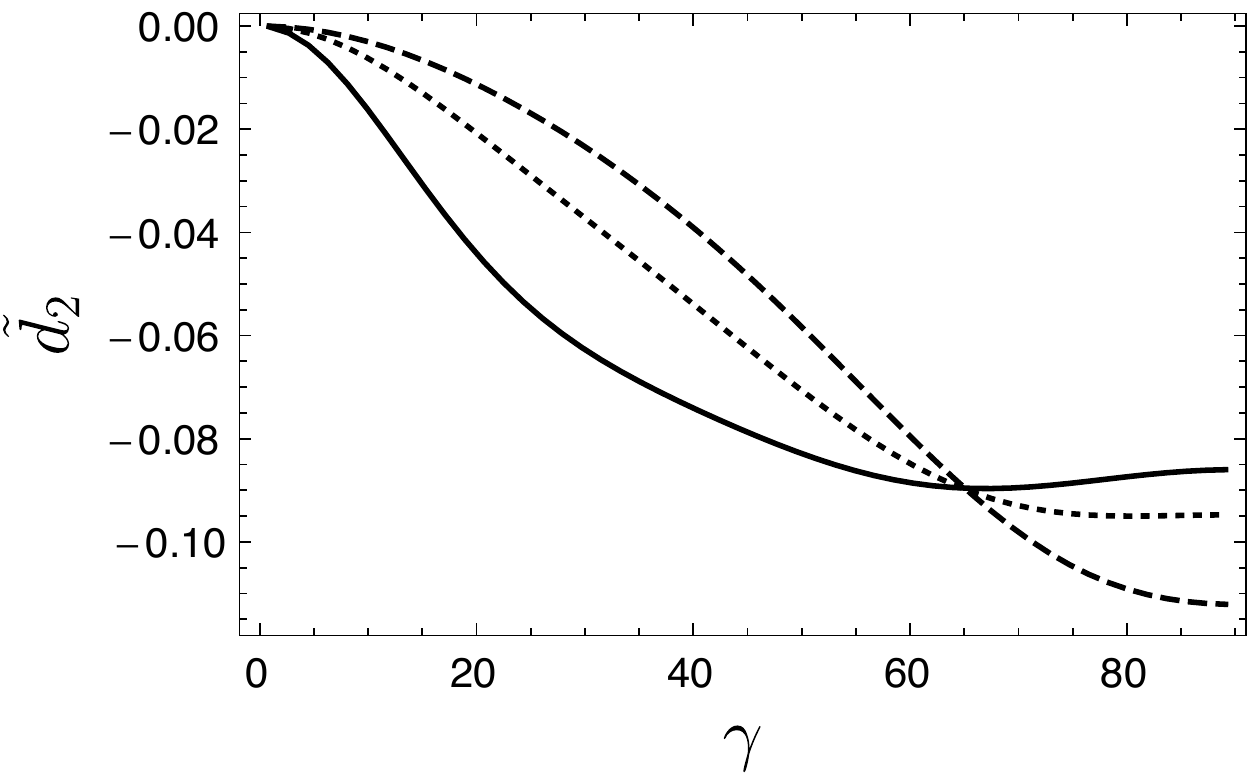}
\hspace*{0.2cm}
\includegraphics[scale=0.4]{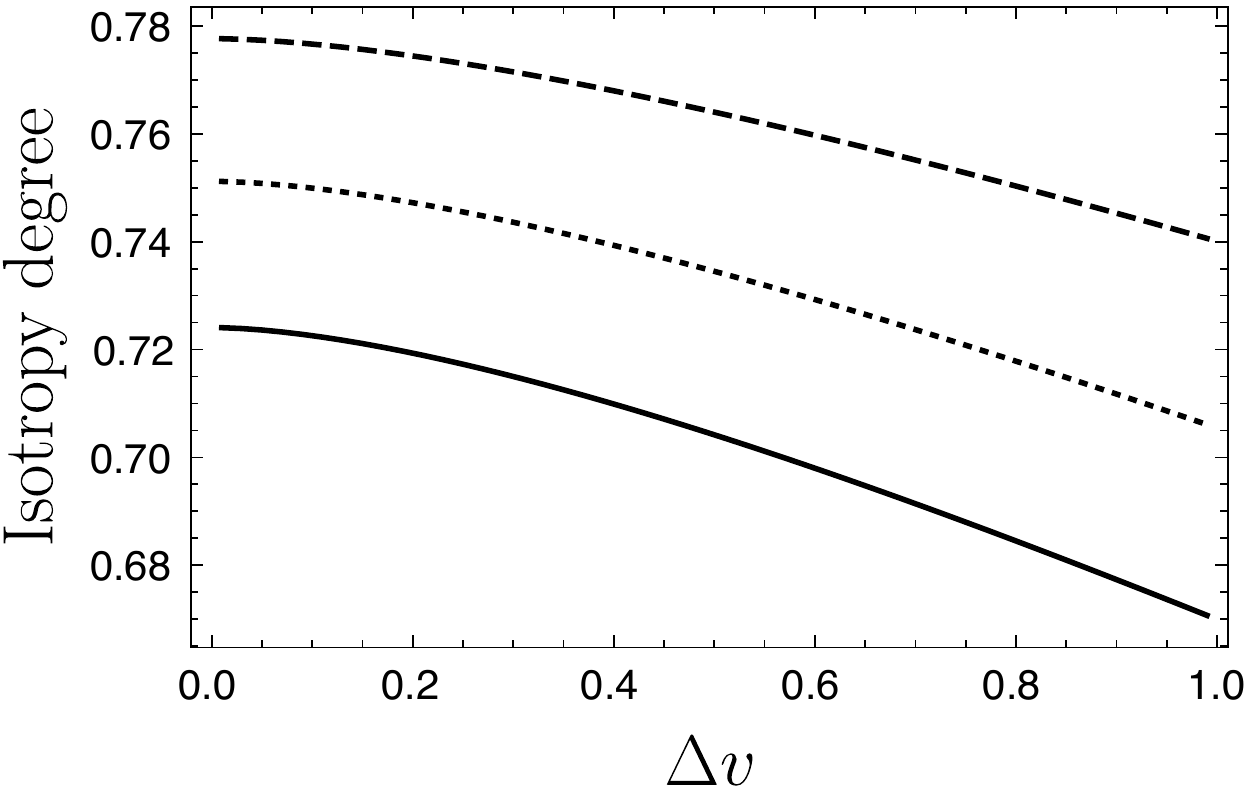}
\caption{Alv\'en $+$ low-$\beta$ slow modes at $M_\text{A}=0.7$. Left and center: monopole and quadrupole for various percentage of mixture. Right: isotropy degree at $\gamma=\upi/4$. The solid curve is for $85\%$ Alfv\'en and $15\%$ slow, dotted curve is for $70\%$ Alv\'en and $30\%$ slow and the dashed curve is for $50\%$ Alv\'en and $50\%$ slow.}
\label{fig:slowmixture}
\end{center}
\end{figure*}

Another interesting mix is Alv\'en and slow modes in low-$\beta$ plasma. We have shown that both of these modes have negative quadrupole moment. Moreover, these modes are different domains of dominance. At $\gamma\sim 0$, slow modes dominate while at $\gamma\sim \upi/2$, Alfv\'en modes dominate. Therefore, we expect the anisotropy level of their mixture to be not too different from the anisotropy level of each individual mode in the region of their dominance. This is shown in Fig. [\ref{fig:slowmixture}]. Note that in that figure, changing percentage of composition has relatively unaffected the level of anisotropy. This result shows that the anisotropy effects come primarily from the power spectra rather than the exact local structure of the spectral tensor (LP12).

It is important to note that for the case of mix between Alfv\'en and slow mode in low $\beta$, the anisotropy level is unaffected at $\gamma\sim \upi/2$ when compared with Alfv\'en mode. This is because of the fact that for low-$\beta$ slow mode, the motions are along the direction of magnetic field, and therefore these motions should not affect the statistics in the direction perpendicular to them. Similarly, at smaller $\gamma$, the mix of Alfv\'en and low-$\beta$ slow mode should produce anisotropy level similar to that of the slow mode alone. This effect is again primarily because of the anisotropy from power spectra.

\subsection{Comparison with Esquivel et al. (2015)}
One of the most interesting and important findings of our study is the decrease of isotropy degree with increasing slice thickness. This matches exactly with the findings of \cite{esquivel2015studying}. We compare our result with their results.  In their study, for $M_\text{A}=0.7$ and $M_S=2.2$, most of the contribution comes from Alfv\'en mode and density effects. Comparing our results for pure Alfv\'en effects and their result at constant density should be reasonable. In our case, at $M_\text{A}=0.7$, isotropy degree at thin slice regime is $\sim 0.65$, while their result shows an isotropy degree of $\sim 0.6$, which is close to our result. At $M_\text{A}=0.4$, however, our result shows an isotropy of $0.59$, while they predicted much less isotropy degree of $\sim 0.3.$ However, the overall trend of decreasing isotropy with increasing slice thickness matches well with our results. 

\subsection{Study on Density Effects}\label{densthickslice}
Besides velocity, density statistics also provide important contribution to intensity statistics. In LP00, the issue of separating density contribution from velocity contribution to the intensity statistics was addressed. For steep spectra (see Sec. \ref{densitystatistics}), it was mentioned in LP04 that density effects are important at large lag $R$ and velocity effects are important at small lags, but this was invalidated in LP06, where it was clarified that velocity statistics are dominant in thin slice regime no matter what the scale $R$ is. In the case of {\it shallow} spectra, however, density effects are important even in the thin velocity slice regime. With this, it is natural to expect that for steep spectra, anisotropy in intensity statistics should be primarily dominated by velocity effects in the thin slice regime, while for shallow spectra, anisotropy is affected by density effects as well in this regime. In the thick slice regime, only density effects are important. 

We tested the effects of density anisotropy at different scales for both steep and shallow spectra. 
\begin{figure*}
\begin{center}
\includegraphics[scale=0.4]{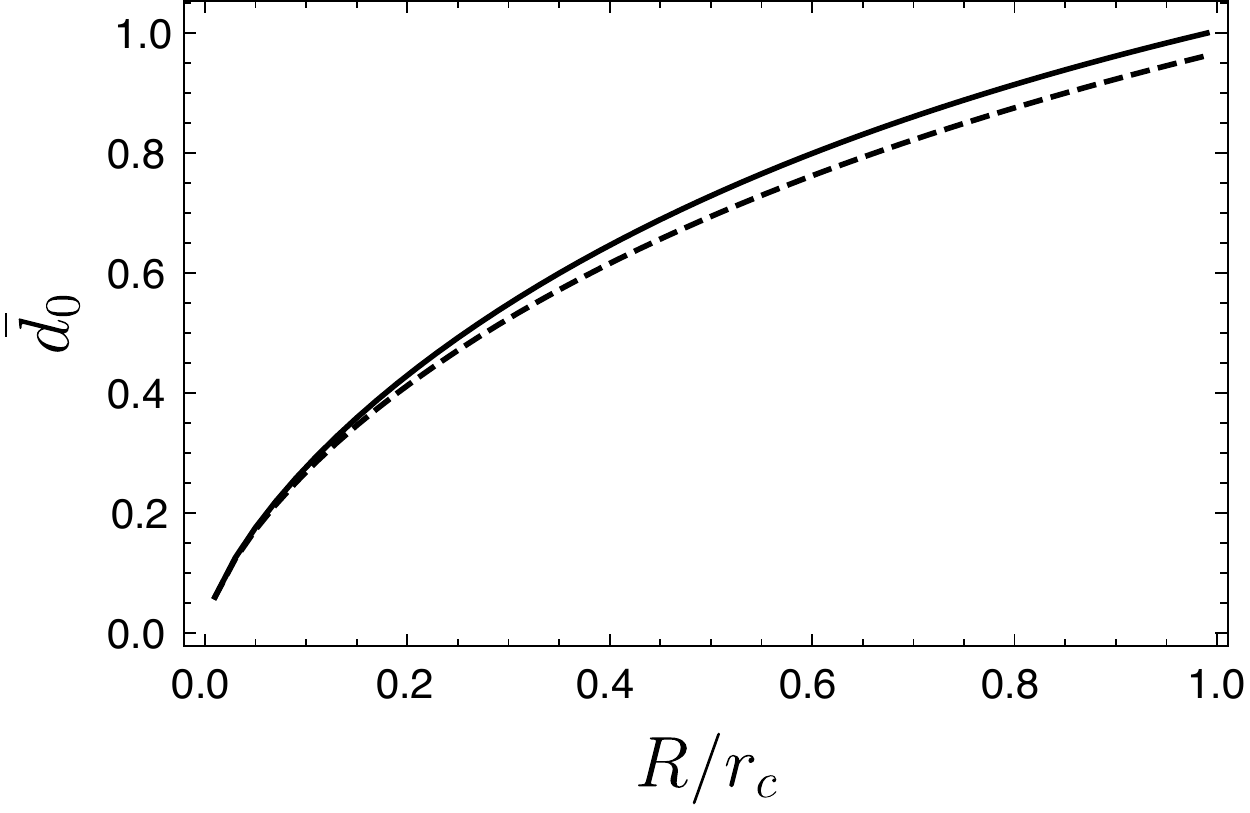}
\hspace*{0.2cm}
\includegraphics[scale=0.4]{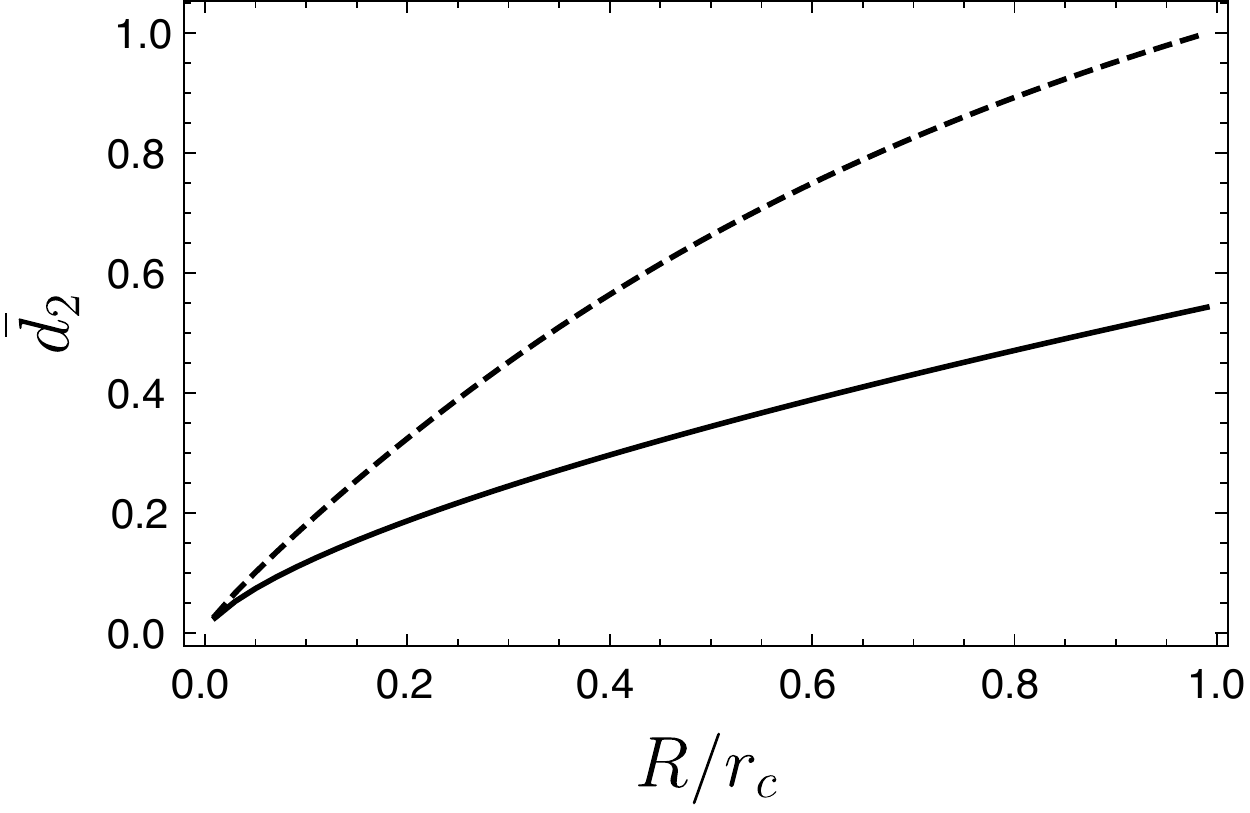}
\hspace*{0.2cm}
\includegraphics[scale=0.4]{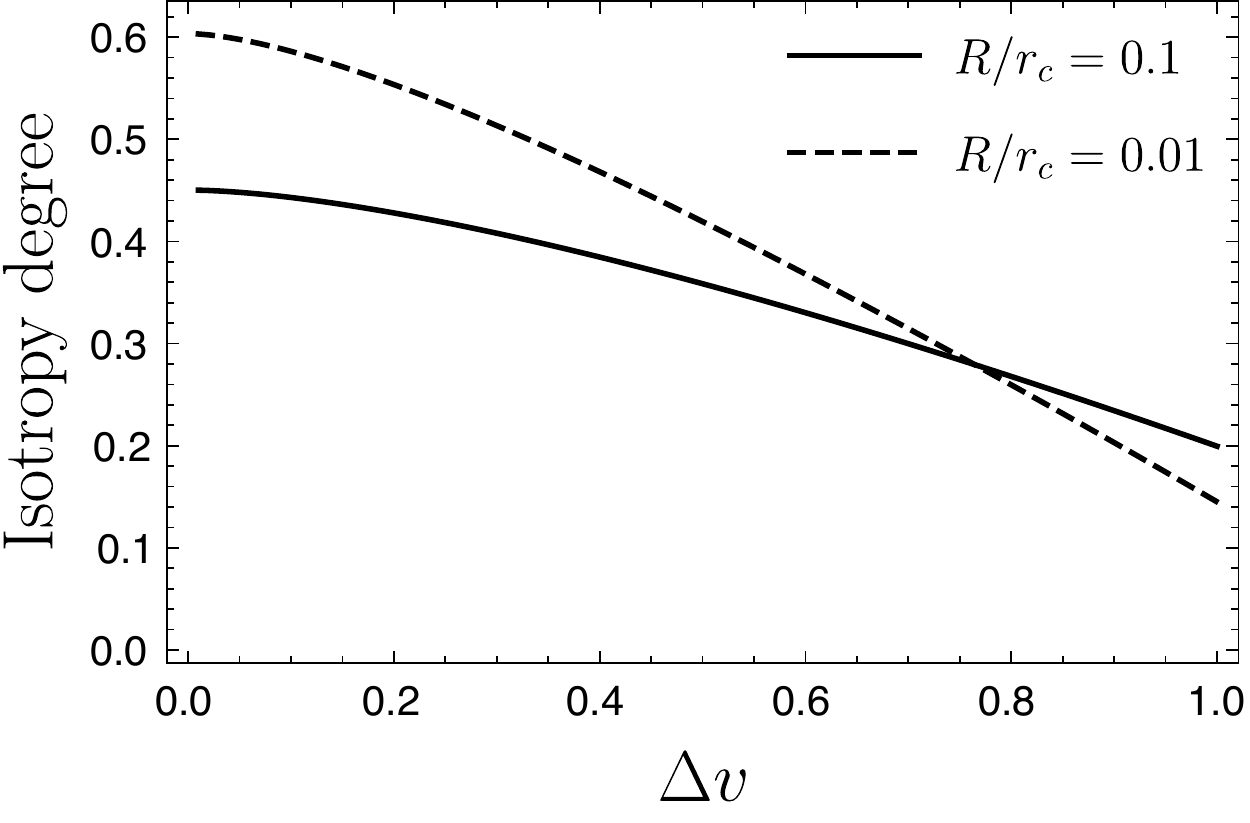}
\caption{Steep density: normalized monopole (left) and normalized quadrupole (centre) for Alfv\'en mode at $\gamma=\upi/2$ and $c_\rho=-0.6$. The solid curve is for pure velocity contribution while the dotted curve is with steep density of Kolmogorov index.  Right: Isotropy degree as a function of velocity slice thickness $\Delta v$ for various $R/r_c$, all the parameters are the same as in the left-hand and central panels.}
\label{fig:density}
\end{center}
\end{figure*}

\begin{figure*}
\begin{center}
\includegraphics[scale=0.4]{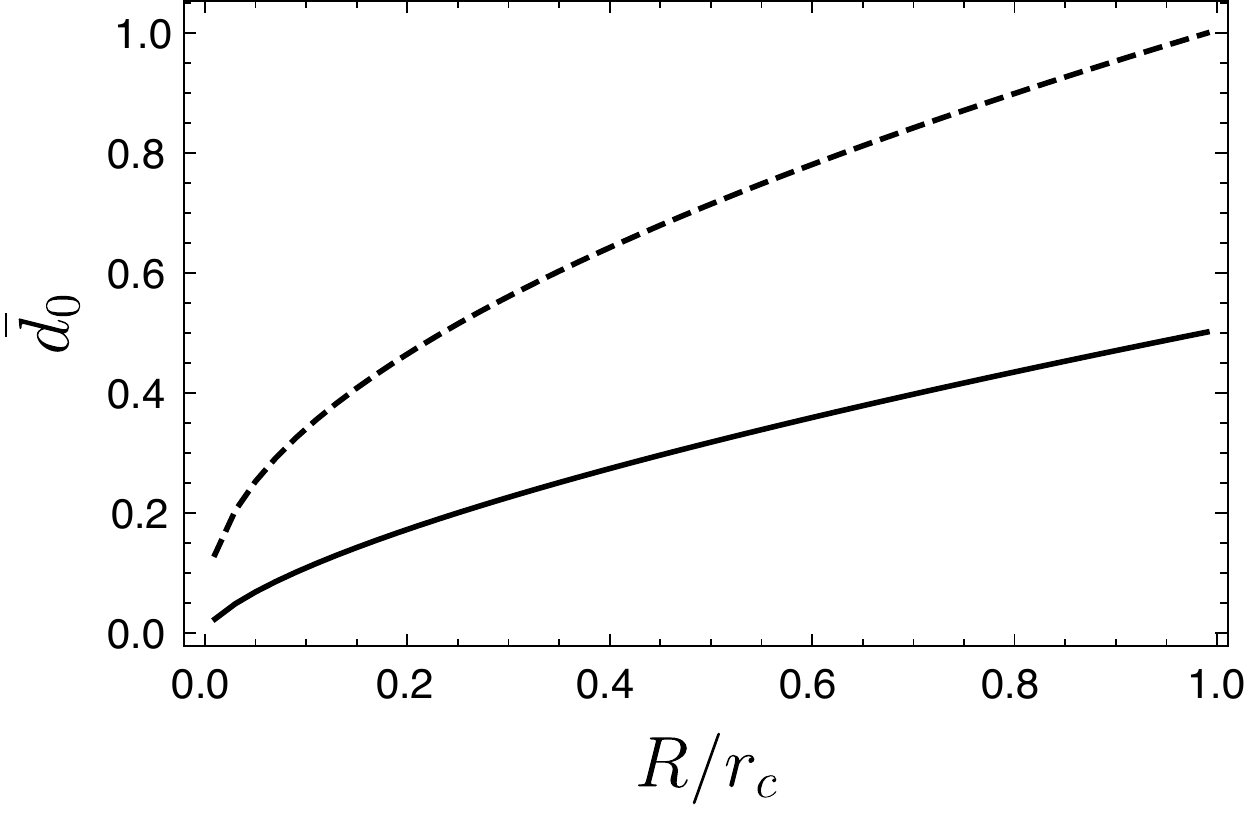}
\hspace*{0.2cm}
\includegraphics[scale=0.4]{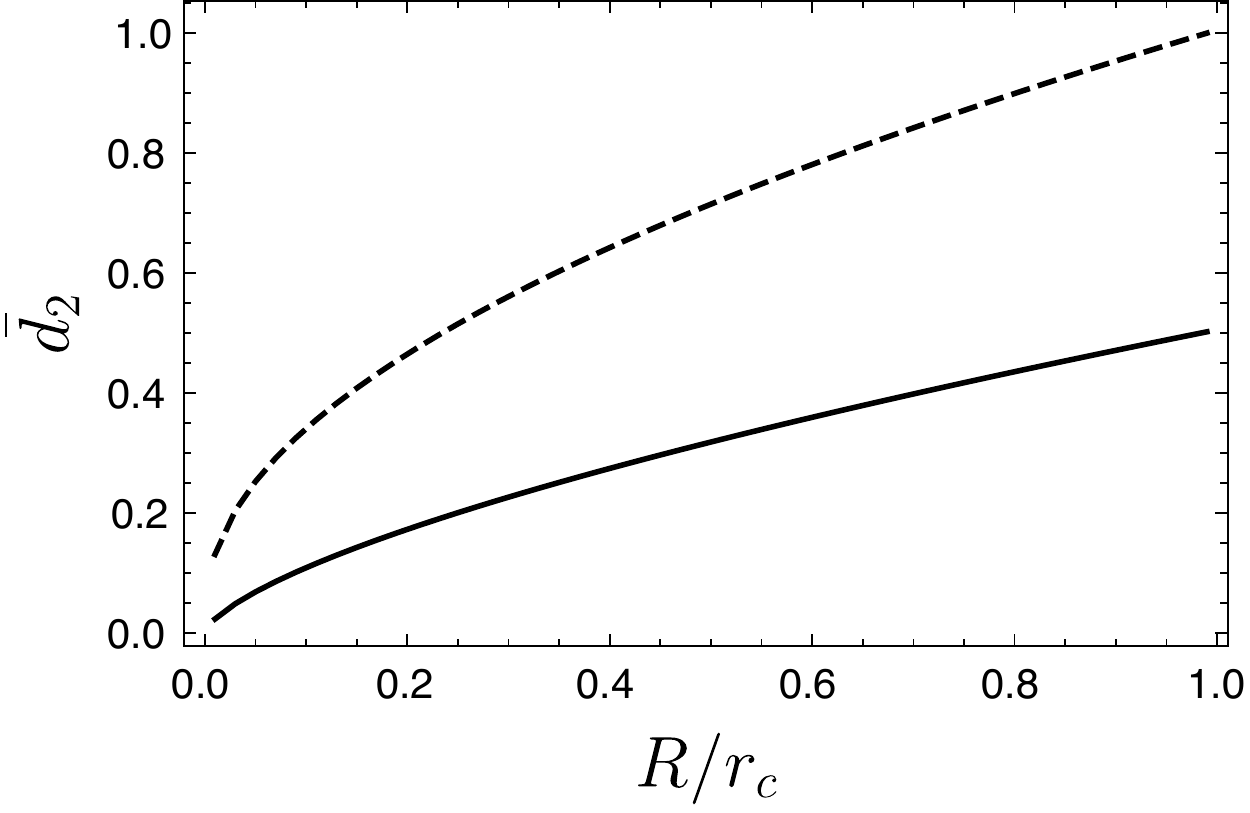}
\hspace*{0.2cm}
\includegraphics[scale=0.4]{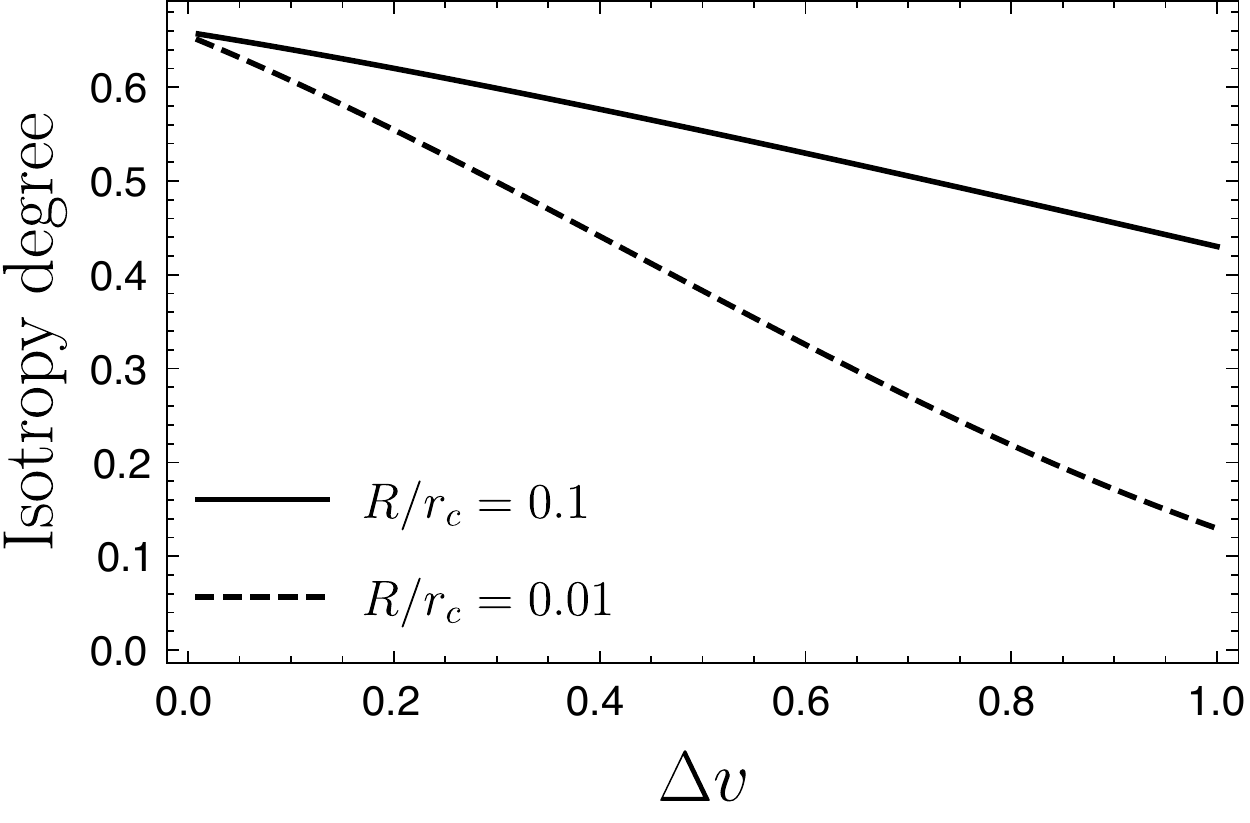}
\caption{Shallow density: normalized monopole (left) and normalized quadrupole (right) for Alfv\'en mode at $\gamma=\upi/2$ and $c_\rho=0.3$. The solid curve is for pure velocity contribution while the dotted curve is for shallow density of index $\nu_\rho=1/3$.  Right: Isotropy degree as a function of velocity slice thickness $\Delta v$ for various $R/r_c$, all the parameters are the same as in the left and central panels.}
\label{fig:shallowdensity}
\end{center}
\end{figure*}
Fig. \ref{fig:density} shows some of the key features shown by density effects. Both quadrupole and monopole for the combination of velocity and density effects are similar to velocity effects alone at $R< r_c$ for a steep spectrum.  This is consistent with the fact that for steep density spectra, the intensity correlation function is dominated by velocity effects at $R< r_c$. Interestingly, the quadrupole moment is affected by density effects, while monopole remains relatively unaffected. The relative importance of density effects in quadrupole moment depends on the degree of density anisotropy, $c_\rho$. For, sufficiently small $c_\rho$, we can not see any significant deviation from the pure velocity contributions. Therefore, studying monopole moment at small $R$ should give us information on velocity statistics, while quadrupole moment will give information about the presence of density effects. Note that at thick velocity slice, the intensity statistics is dominated by density effects alone. 

If the density spectrum is shallow, the density effects become important at small scales. Therefore, we expect significant deviation from pure velocity effects in the case of shallow density spectrum. This is confirmed from Fig. [\ref{fig:shallowdensity}], as the degree of isotropy changes significantly from the pure velocity effects. 
 
\subsection{Absorption line studies}\label{absorptionlinestudy}

The present study is focused on emission lines. Absorption lines present another way to study the turbulence. The theory of PPV description of absorption lines was presented in \cite{lazarian2008studying} (henceforth LP08). There it was suggested to correlate the logarithms of absorbed intensities. This trivially extends our earlier study to the unsaturated absorptions lines. 

The absorption lines are frequently saturated, however. 
For saturated absorption lines, LP08 showed that only wings of the line are available for the analysis. In terms of the analysis this is equivalent to introducing an additional window, whose size decreases with the increase of the optical depth. As a result, only narrow velocity channels carry information on turbulence and only the correlations over small-scale separations within the channel carry meaningful information. In other words, only the information about the small-scale turbulence is available in the case of heavily saturated absorption lines. This conclusion coincides with the one in LP08 obtained for the VCS technique.

The absorption lines may be created by extended sources or a set of discrete sources, e.g. stars in a star cluster.  A big advantage of studying turbulence using absorption lines is that
multiple lines with different optical depths can be used simultaneously. Naturally, noise of a constant level, e.g.  instrumental noise, will affect weaker absorption lines. The strong absorption lines will sample turbulence only for sufficiently small scales.  However, the contrast that is obtained with the strong absorption lines is higher, which provides an opportunity for increasing the signal-to-noise ratio for the small turbulent scales. 
By combining different absorption lines, one can accurately sample turbulence for both large and small scales.  Using absorption sources at different distances from the observer, it is possible to study turbulence in a tomographic manner. 

\begin{figure}
\begin{center}
\includegraphics[scale=0.5]{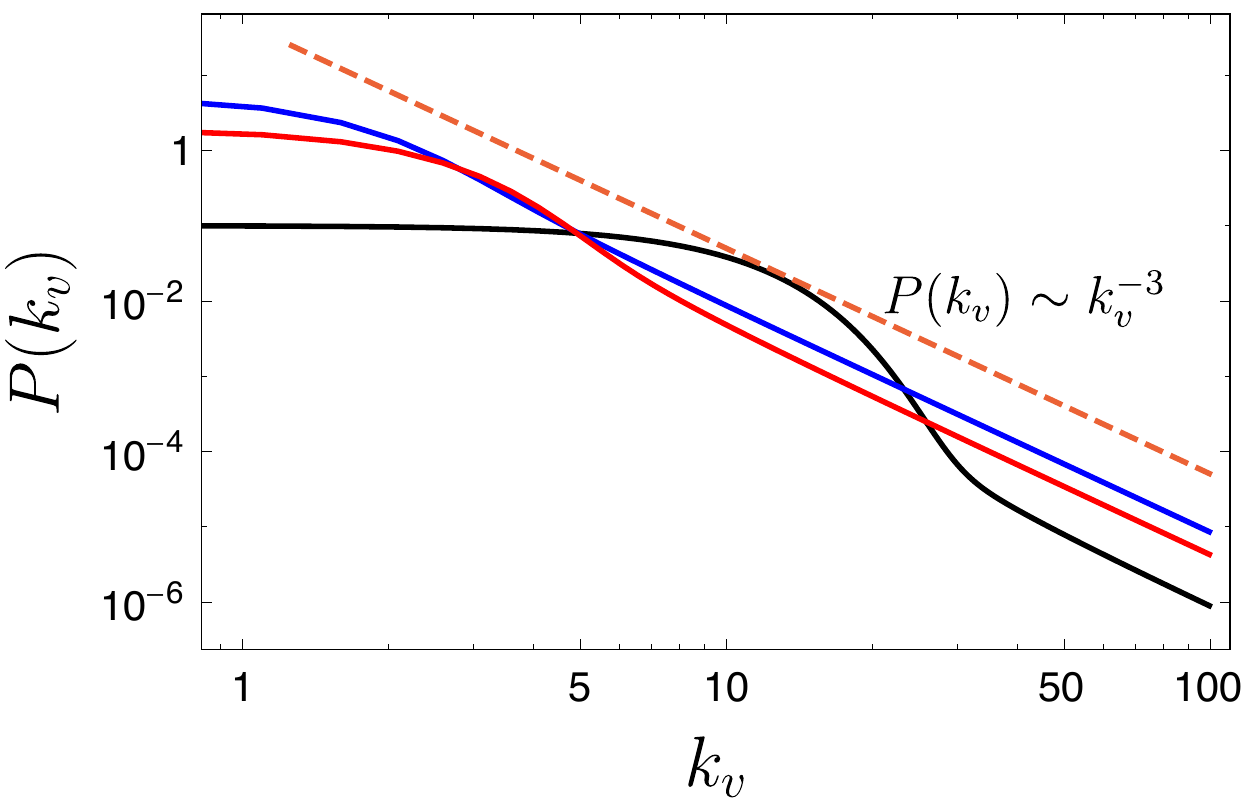}
\caption{Power Spectrum as a function of $k_v$ for pure velocity effect with Kolmogorov index $\nu=2/3$. Curves from top to bottom represent $\Delta =1, 0.5, 0.1$ respectively.}
\label{fig:absorptionline}
\end{center}
\end{figure}

In LP08, it was discussed that the atomic effects introduce an additional mask, which is responsible for the corruption of turbulence spectra at small wavenumber. If the mask is taken to be Gaussian centred at the middle of wing, with the width $\Delta$, it was shown that for wavenumbers $k_v<\Delta^{-1}$, the lines are saturated and the information on turbulence spectra is lost. On the other hand, for $k_v\gg 3\Delta^{-1}$, we can recover the turbulence spectra, as shown in Fig. \ref{fig:absorptionline}. This result would mean that due to atomic effects, we can study anisotropy of eddies only for a velocity slice $\Delta v<1/(3\Delta^{-1})$, meaning that for sufficiently thin slices, one need not worry about atomic effects. 

\subsection{VCA and interferometric studies}

Results obtained in LP00 in terms of the 2D spectra of fluctuations of intensities in velocity slice are important as these 2D spectra can be measured by interferometers. Therefore, using interferometers one does not need to first create intensity maps, but can use the raw interferometric data. This gives a significant advantage for studying turbulence in extragalactic objects as well as for poorly resolved clouds in Milky Way. For obtaining the spectrum, just a few measurements corresponding to different baselines, i.e. for different $|\bm{K}|$, of an interferometer are sufficient\footnote{The procedures are also discussed in LP16 for synchrotron polarization data.}. 

For the anisotropy studies, one can also use raw interferometric data with missing frequencies, but it is important to sample the fluctuations for different direction of the two-dimensional vector $\bm{K}$. This provides more stringent requirements to the interferometric data compared to just studying of velocity and density spectra with the VCA, but still it is much easier than restoring the full spatial distribution of intensity fluctuations. 

A simple estimate of the degree of anisotropy of the interferometric signal can be obtained by taking the Fourier transform of the monopole and quadrupole part of the expansion in Eq. \eqref{eq:multipolemoment}. With this, we have the quadrupole power spectrum 
\begin{align}\label{eq:interferometerpower}
P(\bm{K})=\sum_m&P_m(K)\cos(m\phi_K)\nonumber\\
&=\sum_m\int\mathop{\mathrm{d}^2\bm{R}} \mathrm{e}^{i\bm{K}.\bm{R}}\tilde{d}_m(R)\cos(m\phi),
\end{align}
where $\cos\phi_k=\bm{\hat{K}}\cdot\hat{\Lambda}$ and $P_m(K)$ is the quadrupole moment in Fourier space. After expanding the two-dimensional plane wave as
\begin{equation}
\mathrm{e}^{i K R \cos\theta}=\sum_{n=-\infty}^\infty i^nJ_n(K R)\mathrm{e}^{in\theta},
\end{equation}
where $J_n(z)$ is the Bessel function of first kind, the angular part of the integral in Eq. \eqref{eq:interferometerpower} gives
\begin{equation}
P_m(K)=\int\mathop{\mathrm{d}R}R\tilde{d}_m(R)J_m(KR).
\end{equation}
The above equation provides important information that the anisotropy in real space manifests as anisotropy in Fourier space, and each multipole in real space has one-to-one correspondence with the multipoles in Fourier space. 

The asymptotic form of $P_m(K)$ for large $K$ can be obtained analytically and the result in the case of pure velocity contribution is 
\begin{equation}\label{eq:interferometermoment}
P_m(K)=\frac{2^{2-\frac{\nu }{2}} \Gamma \left(\frac{1}{4} (2 m-\nu +6)\right)}{\Gamma \left(\frac{1}{4} (2 m+\nu -2)\right)}\tilde{d}_m(KS)^{3-\nu/2}
\end{equation}
where $\tilde{d}_m$ is the real space intensity moment after $R$ dependence being explicitly factored out. With this the ratio of quadrupole to monopole moment is 
\begin{equation}
\frac{\tilde{d}_{2K}}{\tilde{d}_{0K}}=-\frac{6-\nu}{2-\nu}\frac{\tilde{d}_2}{\tilde{d}_0}.
\end{equation}
Note that the sign of quadrupole moment changes in Fourier space when compared to real space. Moreover, the ratio of quadrupole to monopole is enhanced by a factor of $(6-\nu)/(2-\nu)$ which for $\nu=2/3$ is 4. Therefore, the anisotropy is much more apparent in Fourier space. This provides an unique way to study turbulence with interferometric signal as we can utilize both the isotropic part and the anisotropic parts (like quadrupole moment) to study turbulence spectra.

\subsection{Effects of spatial and spectroscopic resolution}

The effects of telescope resolution for the VCA ability to get the spectra were considered in LP04. Naturally, the finite resolution of telescopes introduced the uncertainty of the order of $\delta K$ which is inversely proportional to $\delta \theta$ that characterize the resolution of telescopes. For the analysis of anisotropies in the present paper, the requirement is that we study anisotropies at the separation $\gg \delta \theta$. Anisotropies can be studied at large separations, even in the absence of good spectroscopic resolution, as the slices are effectively thin in this scale. 

While the studies of velocity spectra critically depend on the thickness of velocity slices, the velocity resolution is not so critical for studies of the media magnetization. Indeed, even with the limited velocity resolution, it is possible to observe the anisotropy of fluctuations within the velocity slice. This opens ways of using instruments with limited velocity resolution to study magnetic fields. 

On the other hand, in the presence of various velocity slice thicknesses, we have more statistical information that can be studied. Thin velocity slices can be used to study turbulence spectra at small separation, intermediate slices can be used for intermediate scale and thick velocity slices can be used to study spectra at large separation. 

To study effects of finite resolution on intensity anisotropy, we start with some of the equations presented in LP06. The intensity measured by a telescope is $\int \mathop{\mathrm{d}\bm{X}}B(\bm{X}-\bm{X}_0)I(\bm{X},v)$, where $B(\bm{X})$ is the beam of the instrument centred at $\bm{X}=\bm{X}_0$. With some analysis, the intensity structure function is given by (LP06)
\begin{equation}\label{eq:interferometerbeam}
\mathcal{D}(\bm{R}_0,v)\approx \int \mathop{\mathrm{d}^2\bm{R}}B^2(\bm{R}-\bm{R}_0)W_{abs}(R)[d_s(\bm{R},v)-d_s(0,v)],
\end{equation} 
where $W_{abs}(R)$ is the absorption window. We take Gaussian beam
\begin{equation}\label{eq:diagram}
B^2(\bm{R}-\bm{R}_0)=\frac{1}{\upi\theta_0}\mathrm{e}^{-\frac{|\bm{R}-\bm{R}_0|^2}{\theta_0^2}},
\end{equation}
where $\theta_0$ is the diagram of the instrument, relating to the resolution.
$\theta_0$ should be compared with the separation $\bm{R}_0$ 
between LOS at which the correlation is measured. 
If $\theta_0 \gg R_0$, the resolution is poor, and the correlation scale is not
resolved.
If $\theta_0 \ll R_0$, $B^2(\bm{R}) \to \delta(\bm{R}-\bm{R}_0)$, the and resolution is increasingly good, and we return to the VCA regime.

\begin{figure*}
\begin{center}
\includegraphics[scale=0.4]{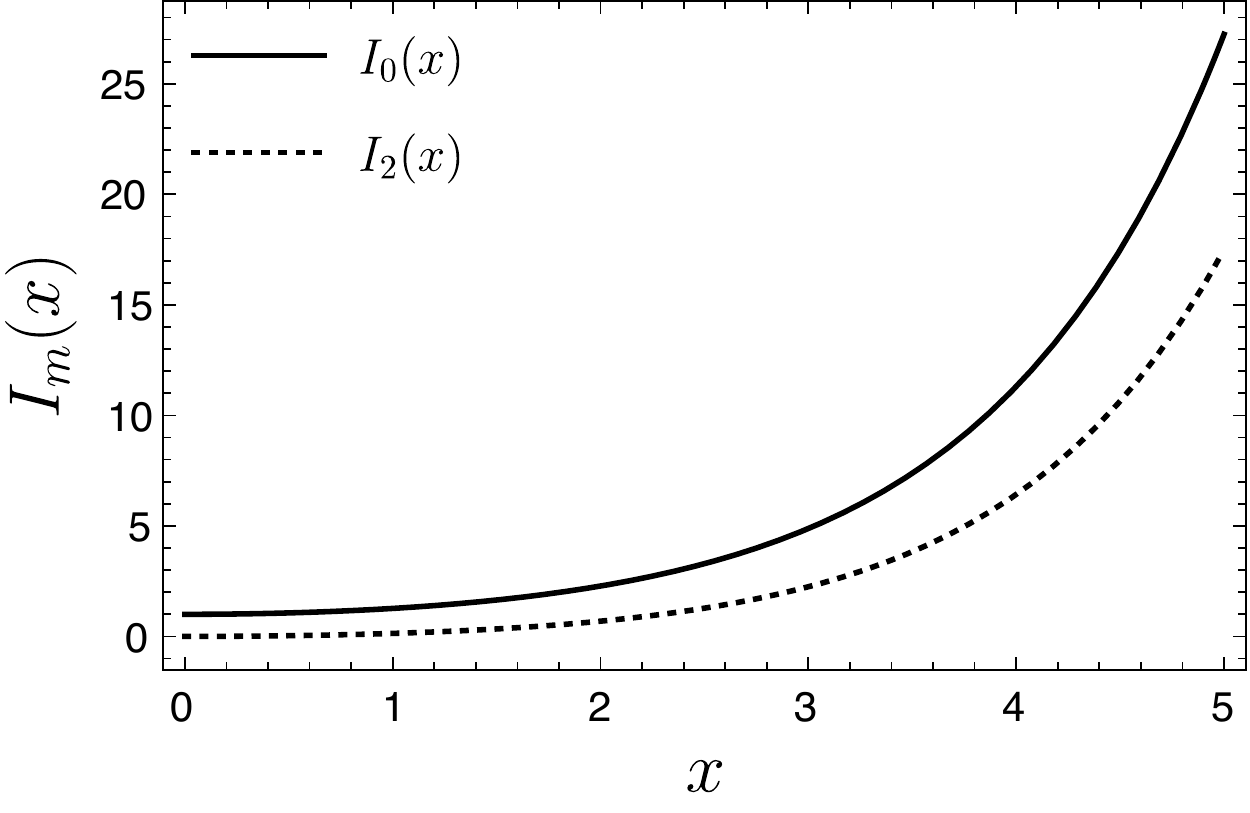}
\hspace*{0.5cm}
\includegraphics[scale=0.4]{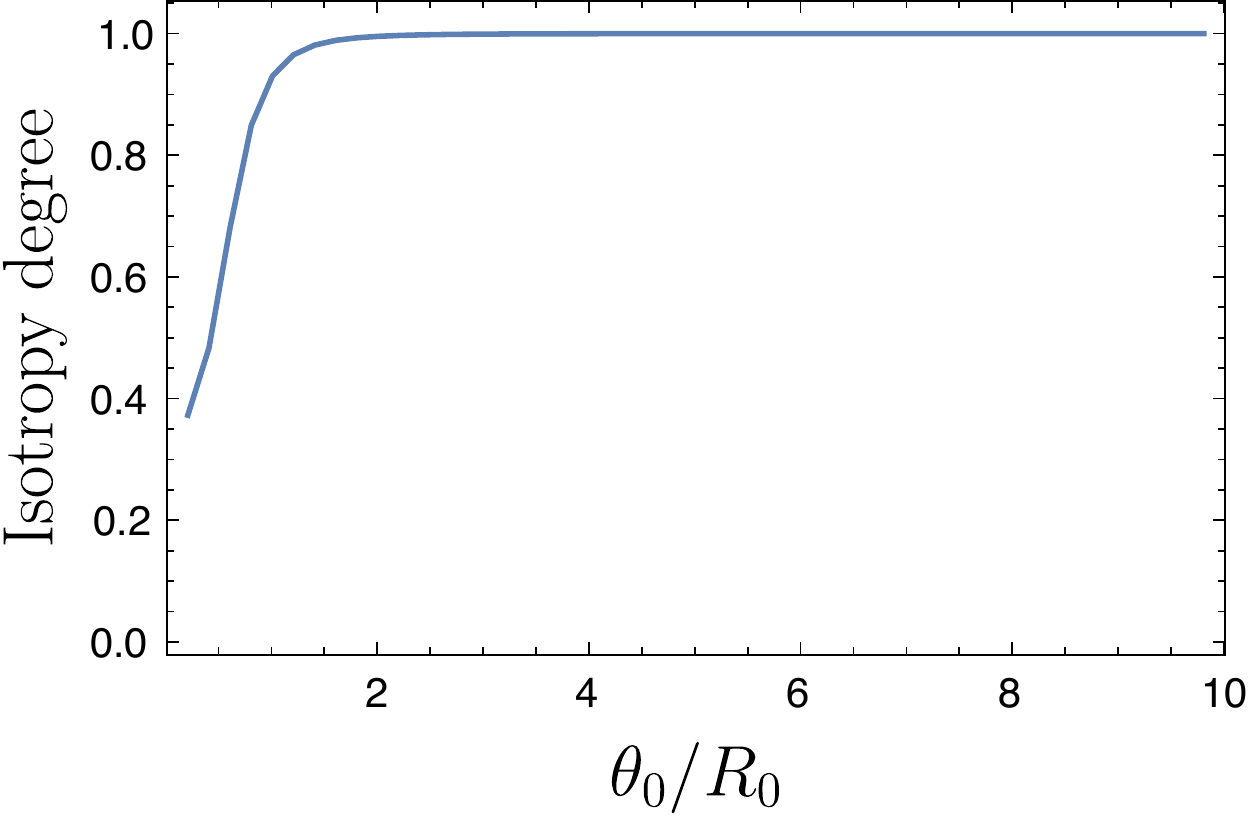}
\hspace*{0.5cm}
\includegraphics[scale=0.4]{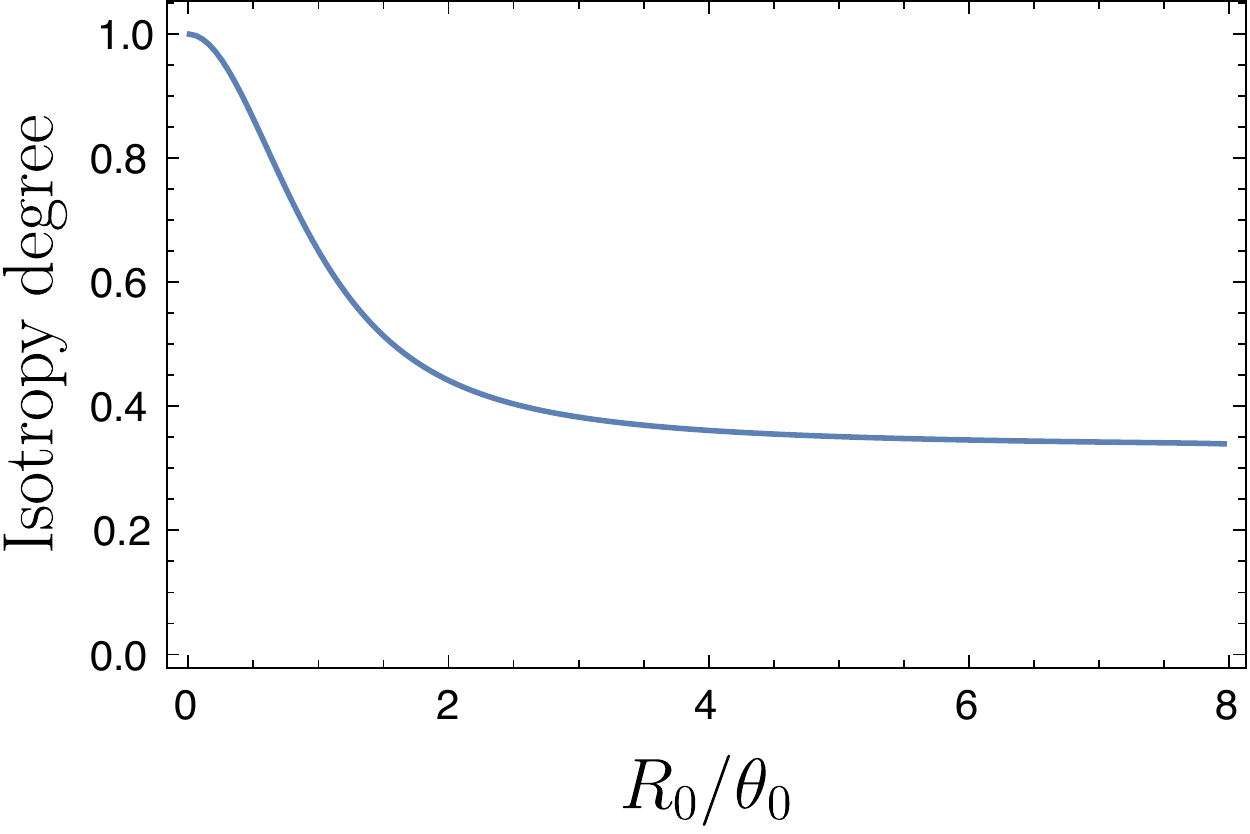}
\caption{Left: hyperbolic Bessel function of the first kind. Center: isotropy degree for varying diagram $\theta_0$ and fixed $R_0$. Right: same for varying lag $R_0$ and fixed $\theta_0$.}
\label{fig:finiteresolution}
\end{center}
\end{figure*}

With decreasing resolution, it is expected that the anisotropy decreases. 
To understand this effect, we consider the multipole expansion of the
intensity structure function.  Contribution to its $m$th multipole moment with
account for a finite resolution is
\begin{align}
&\mathcal{D}_m(\bm{R}_0,v)=\frac{1}{\upi\theta_0}\int \mathop{\mathrm{d}^2\bm{R}}\mathrm{e}^{-\frac{|\bm{R}-\bm{R}_0|^2}{\theta_0^2}}\tilde{d}_m(R)\cos(m\phi)\nonumber\\
&=\frac{2\mathrm{e}^{-R_0^2/\theta_0^2}}{\theta_0}\cos(m\phi_0)\int\mathop{\mathrm{d}R}R\mathrm{e}^{-R^2/\theta_0^2}I_m\left(\frac{2RR_0}{\theta_0^2}\right)\tilde{d}_m(R),
\end{align}
where $I_m(x)$ is the hyperbolic Bessel function of the first kind.
This factor $I_m(2RR_0/\theta_0^2)$ acts as a suppressing factor for increasing
$m$. This has been shown in the left-hand panel of Fig. \ref{fig:finiteresolution},
where $I_2(x)< I_0(x)$ for all $x$. Therefore, we should expect quadrupole to
vanish for $\theta_0\gg R_0$. The change of isotropy with changing diagram has
been illustrated in the central panel of Fig. \ref{fig:finiteresolution}. 
At $\theta_0/R_0\sim 0$, we have a finite anisotropy which corresponds to 
the previous VCA results. With the increasing diagram $\theta_0$,
the statistics become more isotropic and for $\theta_0 > R_0$, information on anisotropy is completely lost. As a function of $R_0$ (right-hand panel), we see that
practically as soon as we start measuring correlations at resolved scales
$R_0 > \theta_0$, the anisotropy can be recovered.

\section{Study on Effects of Self-Absorption}\label{sec:selfabsorption}

In the previous sections, we studied anisotropy of channel maps in optically thin medium. However, knowledge of absorption effects can be important to understand the intensity statistics in various interstellar environments, for instance in molecular clouds. The effects of absorption in the intensity statistics were studied in LP04. Their study suggests that power-law behaviour of intensity statistics  is distorted in the presence of absorption, and the velocity effects are more prominent in this case. 

In this section, we make use of the results of LP04 to study the effect of absorption in the degree of isotropy (cf. equation \ref{eq:isotropydeg}). In the presence of absorption, the intensity structure function is given by (LP04)
\begin{equation}\label{eq: absorption}
\mathcal{D}(\bm{R},\phi,\Delta v)\propto \int_{-\Delta v}^{\Delta v}\mathop{\mathrm{d}v}\tilde{W}(v)\mathrm{e}^{-\frac{\alpha^2}{2}\tilde{d}_s(0,v)}\left[d_s(\bm{R},v)-d_s(0,v)\right],
\end{equation}
where $\tilde{W}(v)$ is the window which defines how integration over velocity is carried out, $\alpha$ is the absorption coefficient, and is zero in the case when absorption effect is absent. The most important feature shown by the above equation is the presence of an exponential factor. Due to the presence of this factor, velocity effects do not get washed out even if we enter thick slice regime, unlike the optically thin case when this factor was absent. Analysis presented in LP04 shows that in the case of Alfv\'en mode (which has a power law index 2/3), 
\begin{equation}\label{eq: absorptionconstant}
d_s(0,v)\propto -v^2\log v,
\end{equation}
which is valid for small argument $v$. With this, for Alfv\'en modes Eq. \eqref{eq: absorption} can be written as
\begin{equation}
\mathcal{D}(\bm{R},\phi)\propto \int\mathop{\mathrm{d}v}\tilde{W}(v)\mathrm{e}^{\frac{\alpha_{\text{eff}}^2}{2}v^2\log v}\left[d_s(\bm{R},v)-d_s(0,v)\right],
\end{equation}
where $\alpha_{\text{eff}}$ is the effective absorption constant, which takes into account the proportionality constant of equation \eqref{eq: absorptionconstant}. 

\begin{figure*}
\begin{center}
\includegraphics[scale=0.4]{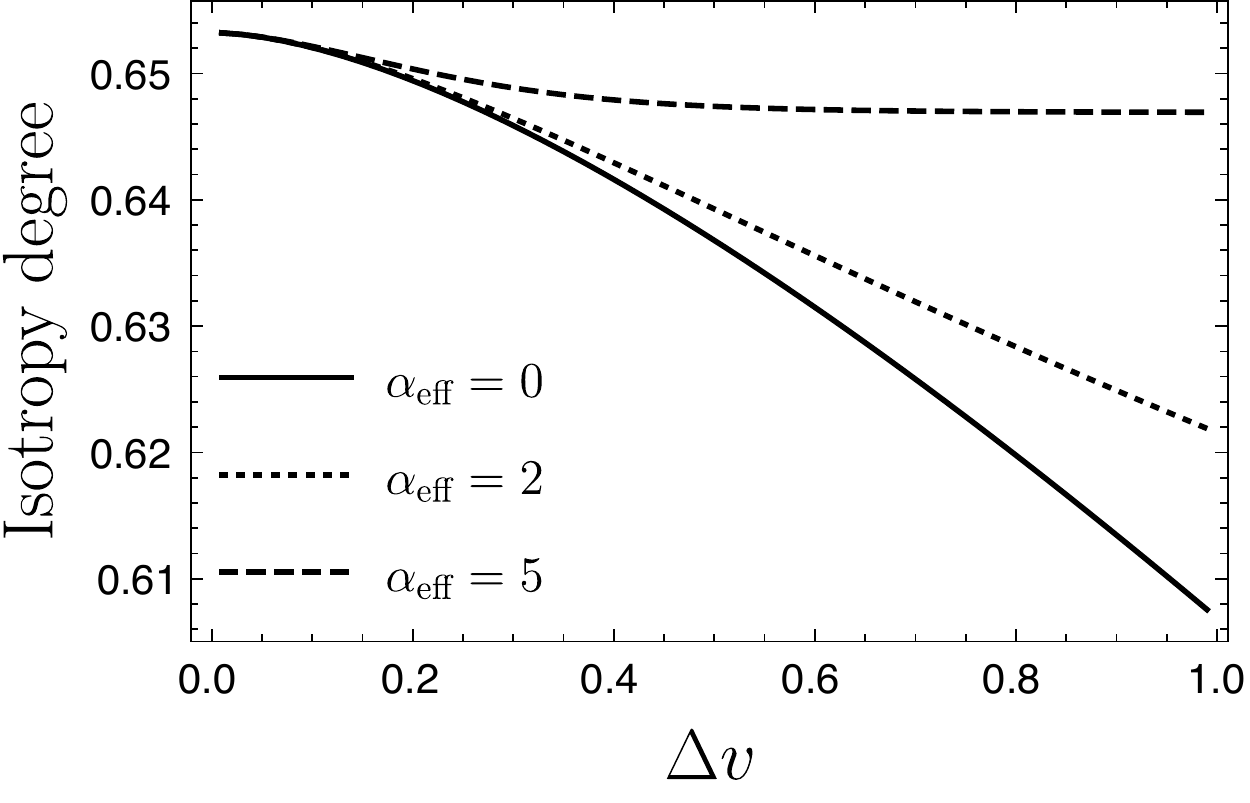}\hspace*{0.2cm}
\includegraphics[scale=0.4]{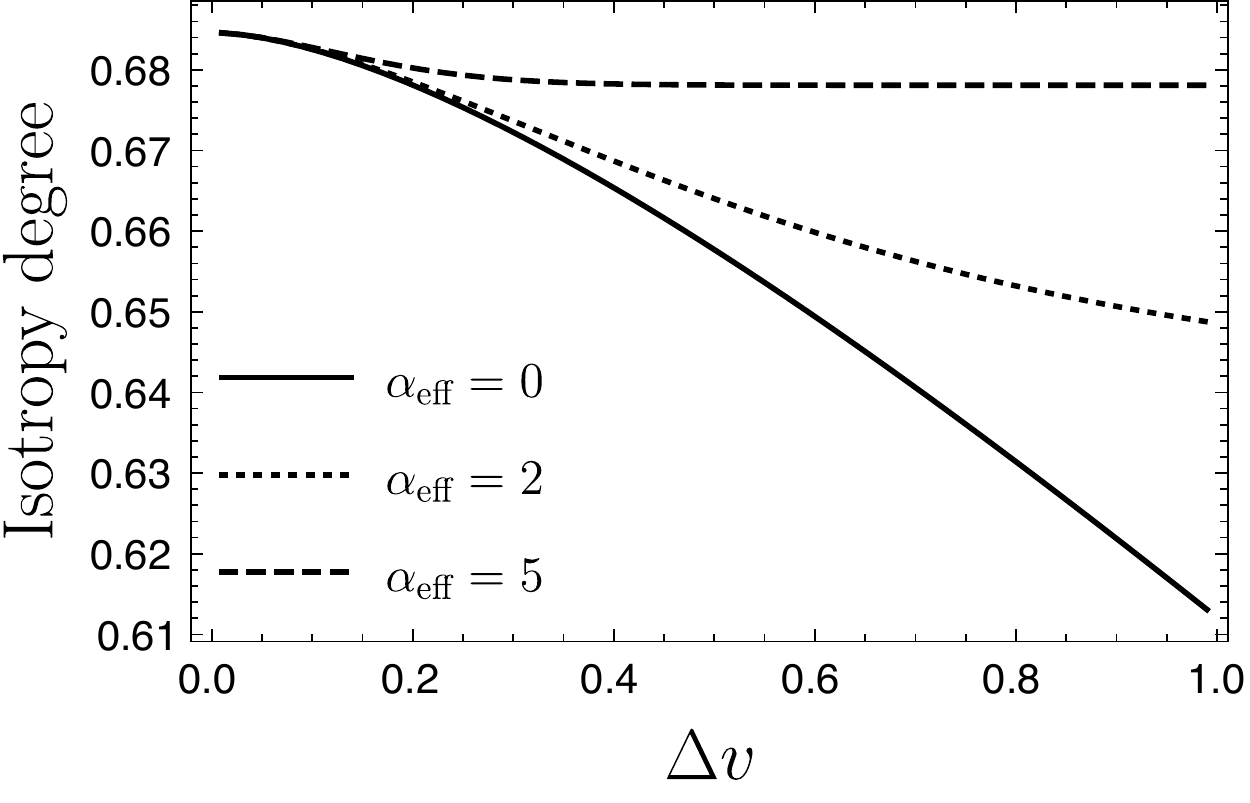}
\hspace*{0.2cm}
\includegraphics[scale=0.4]{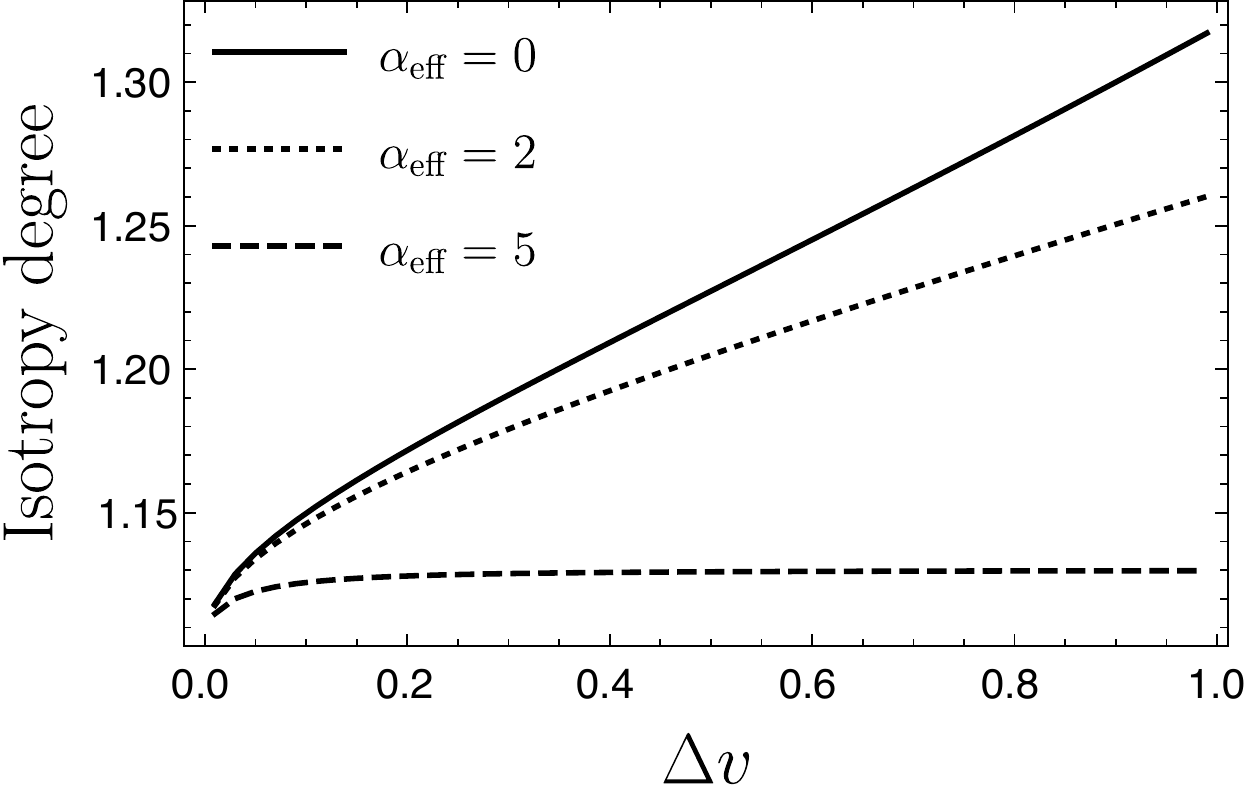}
\caption{From left to right: isotropy degree for Alfv\'en (at $M_\text{A}=0.7$, $\gamma=\upi/2$), slow (low $\beta$ at $M_\text{A}=0.7$, $\gamma=\upi/3$) and fast (low $\beta$ at $\gamma=\upi/2$) in the presence of different degrees of absorptions.} 
\label{fig:absorption}
\end{center}
\end{figure*}

To study the effects of absorption on the anisotropy of channel maps, we performed numerical evaluation for the degree of anisotropy as a function of velocity width which results are shown in Fig. \ref{fig:absorption}. These plots show that with absorption effect included, the intensity statistics become more isotropic. Fig.~[\ref{fig:absorption}] shows  that the deviation of isotropy degree of optically thick case from optically thin case occurs at a critical velocity thickness $\Delta v_c$ roughly given by the relation $
-\alpha^2_{\text{eff}}(\Delta v_c)^2\log (\Delta v_c)=\alpha^2 d_s(0,v)\sim 1$, 
which in the case of $\alpha_{\text{eff}}=5$ gives $\Delta v_c\sim 0.1$, consistent with Fig. [\ref{fig:absorption}]. This is the cut-off beyond which non-linear effects become important while studying the effects of absorption (LP04). Therefore, this implies that although absorption affects the intensity statistics, the degree of isotropy however remains unaffected as long as we are in a regime where absorption is moderate. 

In the regime where absorption is strong, the degree of isotropy decreases less rapidly  in comparison to the case where absorption is absent. This can be understood in the following way: with stronger absorption effects, the thin slice statistics hold for larger range of velocity width and therefore, the degree of isotropy tends to flatten. This is shown by Fig. [\ref{fig:absorption}], where the flattening of this curve is shown in a gradual manner as we increase the absorption coefficient for $\alpha_{\text{eff}}=0$ to $5$.

\begin{figure*}
\begin{center}
\includegraphics[scale=0.4]{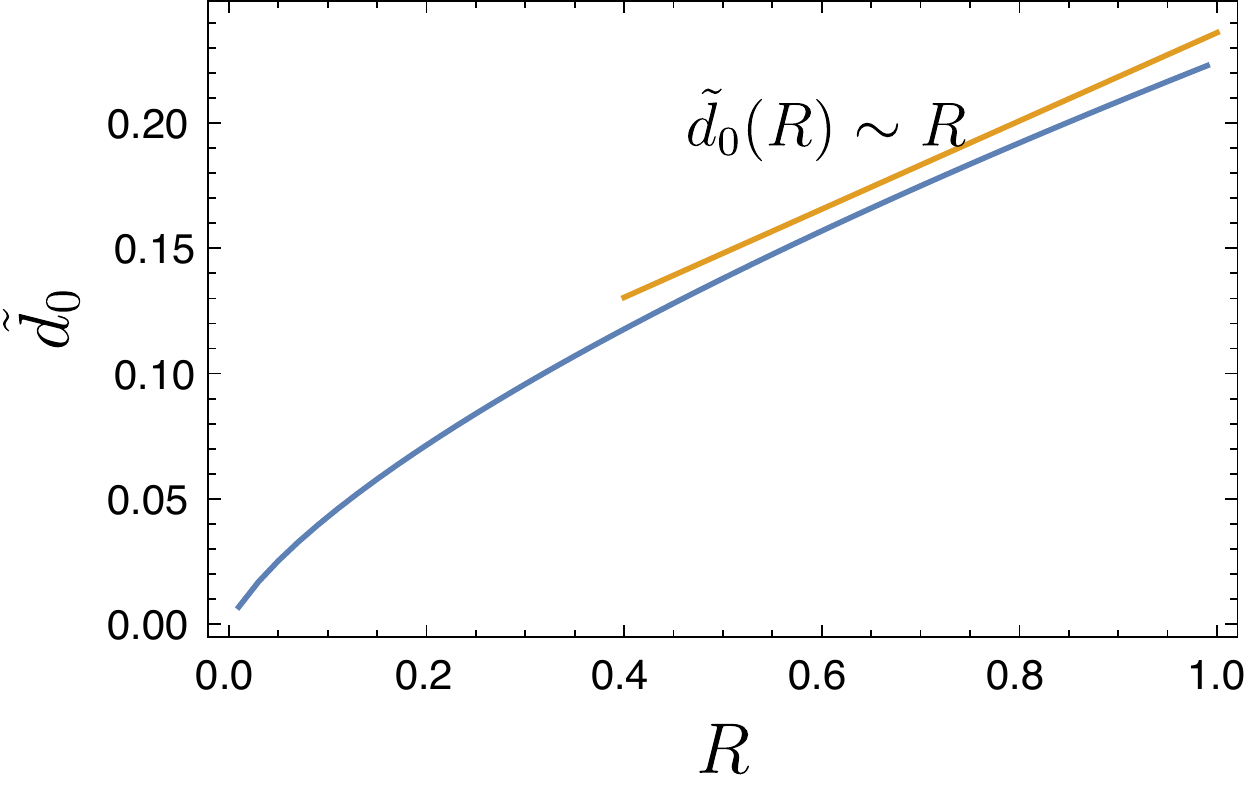}\hspace*{0.2cm}
\includegraphics[scale=0.4]{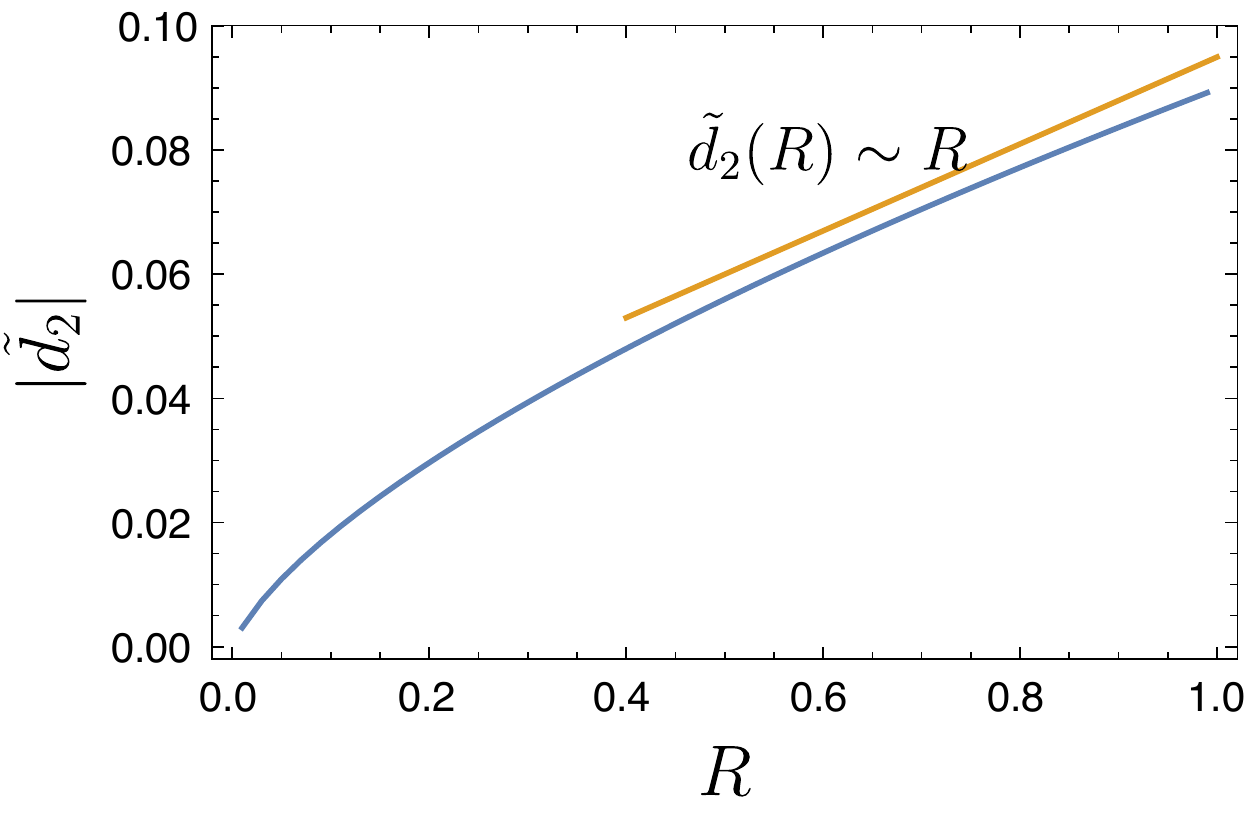}
\caption{From left to right: monopole and quadrupole as a function of $R$ for slow modes at low $\beta$ at $\alpha_{\text{eff}}=5$ at $M_\text{A}=0.4$ and $\gamma=\upi/3$ and $\Delta v=1$. Only pure velocity contribution is considered.} 
\label{fig:universal}
\end{center}
\end{figure*}

The fact that degree of isotropy for optically thick medium  is similar to the degree of isotropy for the optically thin medium in the case when absorption is strong has important consequences that need to be addressed. LP04 showed that for optically thick case, at some intermediate scale $R$, a new asymptotic regime is seen. In this regime, the intensity statistics get independent of the spectrum of the underlying velocity and density field by exhibiting a scaling $\sim R$. This can also be seen in Fig. \ref{fig:universal}, where at large $R$, the scalings for both monopole and quadrupole terms of the intensity structure function vary like $\sim R$. However, what is important is that even though the new intermediate asymptote is established, the imprint of anisotropy is left, which implies that some valuable information about the underlying turbulent field is still left in this regime. In fact, as we discussed earlier, the isotropy degree at this intermediate regime is still around the same as the isotropy degree in the case of thin slice. Therefore, isotropy degree can be an important tool to analyse turbulence in optically thick medium.

\begin{figure*}
\begin{center}
\includegraphics[scale=0.5]{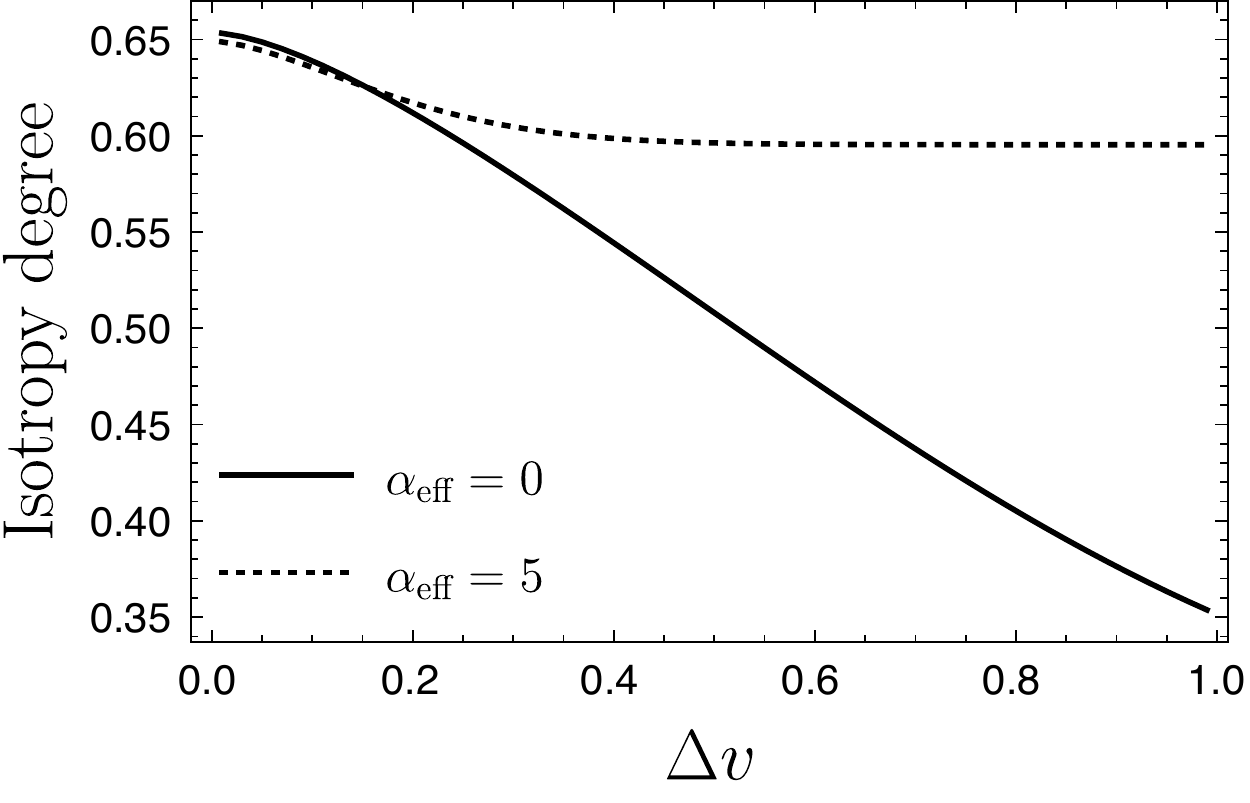}\hspace*{0.2cm}
\includegraphics[scale=0.5]{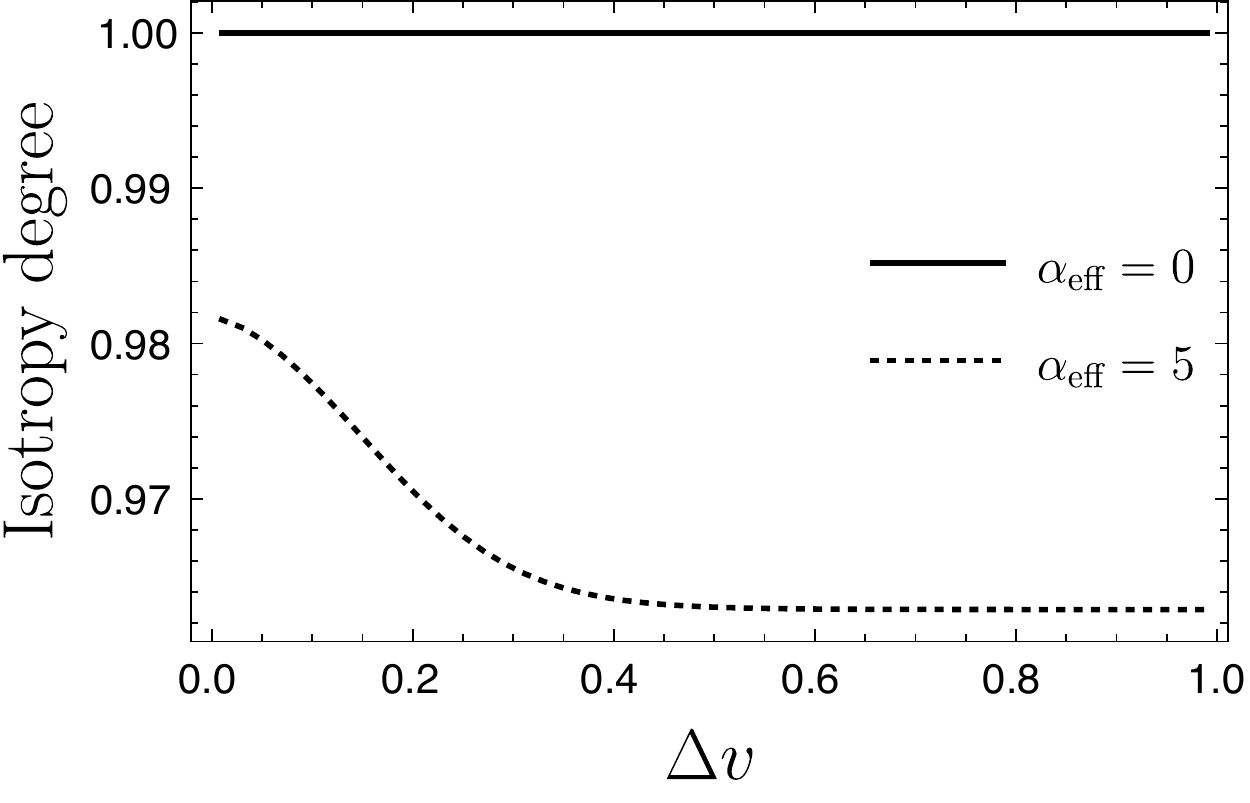}
\caption{Left: isotropy degree for the combination of Alfv\'en modes with steep density of Kolmogorov index $\nu_\rho=-2/3$ and $c_\rho=-0.6$ at $R=0.1$, $r_c=10$, $M_\text{A}=0.7$ and $\gamma=\upi/2$. Right: isotropy degree for the anisotropic density field but isotropic velocity field for the same spectral index as the left figure. LOS angle $\gamma=\upi/2$, but all other parameters are also the same as in left-hand panel. In both panels, the solid curve is in the absence of the absorption while the dotted one is in the presence of absorption at $\alpha_{\text{eff}}=0.5$.} 
\label{fig:vrhodominates}
\end{center}
\end{figure*}

We analyse the relative importance of velocity and density anisotropy when both velocity and density effects are important.  In the absence of absorption, LP00 showed that at small scales, velocity effects are important for thin channel thickness, while density effects become important for thick channels. Naturally, we expect anisotropy to be dominated by velocity effects for thin channels and by density effects by thick channel. Interestingly, in the presence of absorption, this is not true any more. Looking at Fig. \ref{fig:vrhodominates}, we see that in the presence of absorption, the degree of anisotropy is almost the same for both thin and thick channels even when the density effects are present. Note that we have considered a strong anisotropy $c_\rho=-0.6$ for density, and even this anisotropy does not affect much of the isotropy degree in the case of thick slice $\Delta v=1$. Therefore, what we can say is that in the presence of absorption, the anisotropy due to velocity effects is important at small scales even for the thick slice thickness where spectral resolution is absent. 

In the case of isotropic velocity field and anisotropic density field, our previous discussion applies again for both shallow and steep density spectra. Due to the presence of strong absorption, the window $\mathrm{e}^{-\alpha^2\tilde{d}_s(0,v)/2}$ in Eq. \eqref{eq: absorption} suppresses any non-zero $v$ and therefore in the case of strong absorption, it effectively acts like a delta function $\delta(v)$. This explains why even in the optically thick regime $\Delta v=1$, we still have the anisotropy similar to optically thin regime.

\section{Practical guide to the results in the paper}
The main purpose of this paper is to develop a new quantitative way of the spectroscopic data analysis with the goal to study mean magnetic field direction and media magnetization, as well as to pave a way to evaluate the contributions of fundamental MHD modes to the observed turbulence.  
Here we summarize usefulness of our results in a broader picture.

One of the important issues that need to be addressed is the issue of mode separation, i.e. how to separate Alfv\'en, fast and slow modes. Our results partially address this issue. It has been shown that fast modes are more isotropic than Alfv\'en and slow modes. In fact, fast modes in high $\beta$ are acoustic waves, and do not show any anisotropy. On the other hand, fast modes in low $\beta$ do show some anisotropy. However, we have shown that the isocorrelation contours of fast modes in low $\beta$ are elongated towards the direction orthogonal to the sky-projected magnetic field, which is distinctly different from Alfv\'en and slow modes whose iso-correlation contours are elongated towards the direction parallel to the sky projected magnetic field. This provides an important way to separate fast modes from Alfv\'en and slow. However, the issue of separating slow and Alfv\'en modes still persists. This is challenging for two obvious reasons: both of them share same spectral index $\nu$ and same energy spectrum. Therefore, the mixture of slow modes and Alfv\'en modes show about the same level of anisotropy as each of them alone. In the case of strong turbulence, the slow modes and Alfv\'en modes are two `linear polarizations' of transverse displacement waves (LP12), which is again a testament of the similarity of these two modes.

Another issue is deducing LOS angle $\gamma$ and Alfv\'en Mach number $M_\text{A}$ based on observations. One would think that these observables are degenerate. This is because  of the following reason: the anisotropy is maximum at $\gamma=\upi/2$ and decreases consistently with decreasing $\gamma$. On the other hand, the anisotropy decreases with increasing $M_\text{A}$. This has several consequences. As an example, if we see small anisotropy, there are two possibilities: first due to $\gamma\sim 0$ and secondly due to large $M_\text{A}$. However, as shown Fig. [\ref{amachp}], for $\gamma\gtrsim \upi/4$, the monopole and quadrupole are not very sensitive to $\gamma$, meaning that one can deduce $M_\text{A}$ by studying anisotropy in this regime of $\gamma$. This implies that if one has rough information about the range of $\gamma$, one may or may not be able to deduce $M_\text{A}$ based on whether or not $\gamma\gtrsim\upi/4$.

Slice thickness is an important parameter for the study of turbulence. In this paper, we have explained what can be obtained from thin and thick slice, which will be briefly summarized here. In the case of thick velocity slices, one can obtain  information about density spectra and the level of anisotropy of the density field. Since density fields are expected to be isotropic at high sonic Mach numbers, this estimation might help us to deduce the sonic Mach number of a turbulent cloud. On the other hand, velocity effects are important in thin slice regime, whereas density effects become important  if the density spectra are shallow. In the case of steep density spectrum, it was shown in Fig. \ref{fig:density} that although monopole term is unaffected by density in the thin slice regime, the quadrupole is affected at larger lags. Therefore, for a steep density spectra, one can obtain velocity spectra by considering small-scale asymptotes of the monopole term of the intensity structure function. To properly estimate the velocity contribution in quadrupole moment, one needs to account for the density anisotropies. If density is relatively isotropic, one need not worry about density contamination in the quadrupole moment. On the other hand, if density is highly anisotropic, then the quadrupole moment can be substantially contaminated by the density effects even at thin velocity slices. To separate this density contamination, one needs to first estimate the level of density anisotropy (using thick slices) before carrying out thin slice analysis.

Another important study that was carried out in this paper was on effects of absorption on intensity statistics. In the presence of  absorption, the observed anisotropy is less than that without absorption. In fact, if the absorption is strong, the anisotropy remains unchanged with the changing velocity slice. For an optically thick medium, it was shown in LP04 that at some intermediate scale $R$, the intensity statistics get independent of the spectrum of the underlying velocity and density field. However, we have shown that at this intermediate scale, the imprint of anisotropy still exists, and this anisotropy is the same as the anisotropy in thin slice regime provided that the absorption is strong. This can be valuable to study turbulence even in the presence of strong absorption.

The interferometric studies are also a valuable way to study turbulence, because it provides a good resolution even in the presence of sparse data. Another advantage of interferometric studies is that the anisotropies are more enhanced, and therefore easy to observe. We have shown how this anisotropy obtained from interferometric study can be mapped to real space.  

We also carried out absorption line studies, and explained that for sufficiently thin velocity slices, we can study anisotropies even in the presence of atomic effects. However, in the main text we assumed extended source of absorption. In many cases, we only have discrete sources at different lags $R$. In this case, to study anisotropy, one needs to first  obtain turbulent spectra using techniques like VCS. With the information on spectra, one should then construct correlation function between $n$ points (which would imply $n!$ correlation pairs), after which one needs to rescale the lag to $R=1$. However, this rescaling changes the angle $\mu$ between separation between two turbulent points and the direction of magnetic field. With this consideration, one can then apply our analysis to study turbulence.

\subsection{Overview of the results of the paper}

The original formulation of the VCA technique was done in LP00 and LP04 in order to study the spectra of velocity and density. For this purpose, the turbulence anisotropies arising from the presence of magnetic field in the astrophysical media were disregarded. However, more recently, the anisotropies of turbulence have attracted more attention. Therefore, the anisotropies of the channel maps first reported and discussed as a means of studying diffuse media magnetization in \cite{lazarian2001emissivity} became important to quantify. 

In this paper, we used the description of MHD turbulence that is based on the decomposition of turbulent velocity motions into Alfv\'en, slow and fast modes following the prescriptions in \cite{cho2003compressible}. This step is very similar to our decomposition of turbulent magnetic field in the same component that we used in LP12 in order to provide the quantitative description of synchrotron intensity fluctuations. The differences between the two representations arose from the fact that while magnetic fields are subject to the solenoidality constraint, the fast and slow velocity modes have also a potential component. 
\begin{table*}
\centering
\caption{Some useful equations of the paper.}
\begin{tabular}{l c }  
\hline
\hline  
\emph{Parameter} & \emph{Equation}\\
\hline
&\\
LOS projected Velocity Structure Function $D_z(\bm{r},\phi)$ & \eqref{structure}\\
Density correlation function (shallow) $\xi_\rho(\bm{r},\phi)$ & \eqref{shallow}\\
Density correlation function (steep) $\xi_\rho(\bm{r},\phi) $ &  \eqref{steep}\\
&\\
\emph{General case (no absorption)} & \\
&\\
Intensity correlation function $\xi_\text{I}(\bm{R},\phi)$ &  \eqref{eq:intensitycorrelation}\\
Intensity structure function $\mathcal{D}(\bm{R},\phi,\Delta v) $&   \eqref{eq:intesityst} \\
Degree of Isotropy & \eqref{eq:isotropydeg}\\
&\\
\emph{Thin slice (no absorption, constant density)} & \\
&\\
Intensity structure function $\mathcal{D}(\bm{R},\phi) $&   \eqref{eq:narrowintensityst}\\
Intensity multipole moments $\tilde{d}_m(R)$&  \eqref{eq:multipolemoment} \\
Quadrupole-to-monopole ratio $\tilde{d}_2/\tilde{d}_0$ &  \eqref{eq:quadtomono}\\
&\\
\emph{Very thick slice (no absorption)}& \\
&\\
Intensity structure function $\mathcal{D}(\bm{R},\phi)$ &  \eqref{eq:verythickintensity}\\
&\\
\emph{With absorption}& \\
&\\
Intensity structure function $\mathcal{D}(\bm{R},\phi, \Delta v)$ &  \eqref{eq: absorption}\\
\hline
\hline
\end{tabular} 
\label{tab:useful equations}
\end{table*}

The following are some of the important points of our paper with references to the corresponding parts of the main text. Equations useful for the major results are summarised in Table \ref{tab:useful equations}. 

\begin{itemize}
\item The anisotropy in intensity statistics comes from the anisotropy built in the tensor structure and the spectrum of different MHD modes (see sections \ref{mhdmodes} and \ref{anisotropicppvintensity}, and equations [\ref{eq:multipolemoment}-\ref{quadrupole}]).

\item Alfv\'en modes are highly anisotropic at small Alfv\'en Mach number $M_\text{A}$ and the anisotropy decreases with increasing $M_\text{A}$ (see equations \ref{eq:alfvencorr} and \ref{eq:alfvenpower} for the expression of the tensor).

\item Alfv\'en modes (and slow and fast modes as well) become more anisotropic with increasing velocity slice $\Delta v$ (see right-hand panels of Figs. \ref{amachp}, \ref{fig:fast} and \ref{fig:slowhighbeta}]). 

\item  The iso-correlation contours of Alfv\'en and slow modes are elongated towards direction parallel to sky-projected mean magnetic field, while for fast modes the contour is perpendicular to sky-projected mean magnetic field (see Figs. \ref{amachp}, \ref{fig:fast} and equation \ref{eq:isotropydeg}). This effect arises from the anisotropy of the power spectrum of underlying mode (see Section \ref{sec:alvenresult}).

\item In the case of mixture of fast and Alfv\'en modes, the overall intensity structure function becomes more isotropic in comparison to pure Alfv\'en effects (see Fig. \ref{fig:fastmixture}).  The degree of isotropy is more or less the same for the case when there is a mixture of Alfv\'en and slow modes (see Figure [\ref{fig:slowmixture}]).  This result is consistent with LP12, where magnetic field fluctuations were probed.

\item For steep density spectra at thin velocity slice, the contribution of density effects to the monopole is always subdominant (see the left-hand panel of Fig. \ref{fig:density}) However, density contribution to the quadrupole is important if density is anisotropic, and level of density contribution depends on the level of density anisotropy (see the central panel of Figure [\ref{fig:density}]).

\item Absorption effects tend to increase isotropy degree (see Figure [\ref{fig:absorption}] ). At strong absorption, the thin slice approximation holds even for thick velocity slice, and the isotropy degree does not change much with slice thickness (see Figs. \ref{fig:absorption} and \ref{fig:vrhodominates}).

\item  Interferometers allow studies of anisotropy in turbulent media for which the resolution of ordinary telescopes is not sufficient (see equations \ref{eq:interferometerpower}-\ref{eq:interferometermoment}). 
\end{itemize}

\section{Example of possible data handling}
The integral expressions obtained in this paper allow us to develop the procedures of decoupling the contributions from Alfv\'en, slow and fast modes. We believe that by varying the slice thickness, the fitting of the modes to the observational data will be done in future (see the procedures of fitting in \citealt{chepurnov2010velocity, chepurnov2015turbulence}). Below we just sketch the steps of the corresponding procedure. 

Consider a mixture of Alfv\'en ($60\%$), fast ($20\%$) and slow ($20\%$) modes with steep density spectra. What kind of observations can be done to obtain information about the underlying turbulence. First, if one makes observations in thick slice regime, one can obtain information about the spectra and anisotropy level of density field as explained in Sections (\ref{thick}, \ref{densthickslice}) and Eq. \eqref{eq:intensitycorrel}. By decreasing the slice thickness, one starts observing distortions in the intensity statistics due to velocity effects. At thin velocity slice, velocity effects become dominant. As was shown in Fig [\ref{fig:fastmixture}], the mixture of low-$\beta$ fast modes does not distort the monopole moment, as long as fast mode is marginal in the mixture. Since slow modes have same spectral index as that of Alfv\'en modes, we can still obtain spectral index of Alfv\'en modes even in the presence of fast mode. However, the mixture of low-$\beta$ fast modes  can distort the quadrupole anisotropy as shown in the central panel of Fig. \ref{fig:fastmixture}. Note that the degree of anisotropy for the combination of Alfv\'en and slow is not so much affected by the composition of the mixture as shown in Fig. \ref{fig:slowmixture}. Therefore, observing significant distortion of anisotropy from purely Alfv\'en contribution might be a strong signal of the presence of fast modes.

\section{Discussion}

\subsection{Foundations of the technique}

This paper continues the work of quantitative study of the PPV space that was initiated in LP00 for the case of optically thin turbulent medium and later extended in LP04 for the absorbing media. These advances produced the machinery for describing the PPV space that our present study is based upon. 

The next significant advancement is related to the present-day understanding of MHD turbulence theory (see \citealt{2015ASSL..407..163B} for a review). Theoretical and numerical research (GS95,  \citealt{lithwick2001compressible}; \citealt{cho2002compressible, cho2003compressible}; \citealt{kowal2010velocity}) have shown that the MHD turbulence can be viewed as a superposition of the cascades of Alfv\'en, slow and fast modes. The representation of the statistical properties of these cascades in the global frame of reference was obtained in LP12, where the anisotropy analysis of synchrotron fluctuations was quantified. We particularly stress the importance of the observational frame, as this frame is related to the mean magnetic field, and the statistics of fluctuations in this frame is different from the statistics  in the local magnetic field frame in which the Alfv\'enic turbulence is formulated (\citealt{cho2000anisotropy}; LV00;  \citealt{maron2001simulations}; \citealt{cho2002simulations}).  

Additionally, in order to enhance the potency of the proposed anisotropy analysis of spectroscopic data we also employed the description of the absorption spectral lines that was developed in LV08. This allows using a wide variety of the absorption lines.

\subsection{Range of Applicability of VCA}

To apply the VCA, one preferably should have velocity broadening due to turbulence to exceed the thermal line broadening. This is not the strict requirement, as the information on turbulence fluctuations is still present in spite of the thermal broadening, but the extraction of these fluctuations is difficult for realistic noisy observational data. This does not mean that one has to study only supersonic turbulence using the VCA. Indeed, while hydrogen is the most abundant atom in the present universe, one can use both heavier atoms and molecules to trace turbulent motions. 

Originally, the VCA technique was developed for emission lines. However, our study in \S\ref{absorptionlinestudy} makes use of the LP08 description of absorption lines and extends the technique for the absorption line studies. This can be the absorption from a collection of point sources or the absorption from a spatially extended source. 

Molecular clouds and diffuse ISM of the Milky Way are the natural objects for the application of the elaborated version of the VCA that we considered in this paper. However, with the use of interferometers that we also considered in this paper, it seems possible to study turbulence anisotropies and, thus, media magnetization for external galaxies. It is important to note that for the interferometric studies that we have in mind, a few measurements are enough, rather than restoring the entire PPV cube.

\subsection{New Power of VCA}
The present study significantly extends the ability of the VCA technique by augmenting the ability of the technique to measure spectrum of turbulence by providing it with a way to study turbulence magnetization and determining the magnetic field direction. It also outlines the ways for possible separating contribution of Alfv\'en, slow and fast modes. The latter is important due to the fact that different modes have different impact for astrophysical processes. For instance, Alfv\'enic modes are essential for magnetic field reconnection (LV99;
see also \citealt{lazarian2015turbulent} and references therein), superdiffusion of cosmic rays perpendicular to the mean magnetic field direction (\citealt{lazarian2014superdiffusion}), while fast modes dominate resonance scattering of cosmic rays (\citealt{yan2002scattering}). The potential ability of VCA to determine the relative contribution of these different modes for spectroscopic data complements this ability for the technique in LP12 and LP16 for synchrotron data. This has the potential of bringing observational quantitative studies of turbulence to a new level.

\subsection{Model assumptions}

Our analytical studies require adopting different assumptions to perform the analysis. The usual for this sort of studies assumption is that the fluctuations are Gaussian. This assumption is satisfied for an appreciable degree for the turbulent velocity field (see \citealt{MoninYaglomLumley197504}), but it is not good for density fluctuations in high Mach number turbulence. Fortunately, the VCA is mostly focused on studying velocity statistics and for some regimes, e.g. steep density, the density fluctuations do not affect statistics of thin slices. It was also shown in LP00 that the VCA formulae stay valid for the lognormal distribution of density. Thus we do not believe that our Gaussianity assumptions are a serious shortcoming.

The independence of velocity and density fluctuations is another assumption employed in the derivation of the basic equations of the VCA. The effect of this assumption was analysed in LP00 where it was shown that even in the case of the maximal possible velocity-density  cross-correlation that follows from general Cauchy-Schwarz inequality (see \citealt{mathews1970mathematical}), the expressions for the thin slices stay the same for the steep velocity (see more discussion in appendix D of LP00). The measures of anisotropy are expected to be more robust compared to the spectrum. Thus, we expect the VCA not to be affected by this assumption.

The decomposition of MHD turbulence into Alfv\'en, slow and fast modes is also an approximation based on the assumption of small coupling between these different modes. The degree of coupling of the modes was quantified in \cite{cho2002compressible}, and it was shown to be very moderate unless the sonic Mach number of the media is very high. The exact spectral slope of fast modes may change for high sonic Mach number, but this should not change significantly the anisotropy analysis in this paper.

In terms of the turbulent media to be studied by the technique, it is assumed that the media are isothermal. This is an excellent assumption for molecular clouds. The effects of the variations of temperature within atomic hydrogen were discussed in LP00, where it was shown that the effects of the temperature variations there make the contributions of hotter gas subdominant compared to the colder gas. The effects of temperature variations on the turbulence studies using absorption lines are discussed in LP08, where it is shown that those can result in the `renormalization' of density fluctuations, while the velocity remains undistorted.

\subsection{Relation to earlier studies}

Studies of fluctuations of intensity within velocity slices of PPV cubes can be traced back to works of \cite{crovisier1983spatial}, \cite{green1993power} and \cite{stanimirovic1999large}. The fluctuations in the slices were attributed to turbulence, but their relation to the underlying spectra of density and velocity was not clear at all; the relation of the thickness of slice and to the spectrum of the measured fluctuations was not realized. For instance, \cite{green1993power} erroneously attributed the fluctuations of intensity of radiation within the velocity slices to the density fluctuations in ISM. In fact, these fluctuations arise mostly from velocity fluctuations (LP00). Similarly, the thickness of slices in the above analyses was chosen arbitrarily and therefore the differences in the spectral indices obtained in different studies were  attributed to the differences in the underlying interstellar turbulence. In fact, the differences in the measured spectra in \cite{green1993power} and \cite{stanimirovic1999large} were arising mostly due to the difference of the slice thickness adopted in the two studies. This was shown on the basis of LP00 theory in a subsequent study by \cite{stanimirovic2001velocity}. 

The situation changed with the development of theory of VCA technique in LP00. This theory allowed them to solve many puzzles existing in the field and opened ways for studying velocity and density turbulence by observing the changes of spectral slope of intensity fluctuations within velocity slices of PPV cubes. A number of studies on turbulence in interstellar galactic and extragalactic  HI and CO have been successfully performed using the VCA technique (see examples in \citealt{2009SSRv..143..357L}). 

The original LP00 study did not account for the effects of self-absorption of radiation. For self-absorbing CO media, the measurements were usually performed not for velocity slices, but for total intensity arising from the turbulent volume (see \citealt{falgarone1995intermittency}). For this way of studying, it was shown in LP04 that a universal spectrum $P(K)\sim K^{-3}$ is expected, which corresponds to the spectrum measured in a number of studies (see examples in \citealt{2009SSRv..143..357L}). The earlier papers were erroneously attributing the measured spectra of fluctuations to the density fluctuations of the ISM. The washing out of the information of underlying turbulence fluctuations with the increase of absorption and the emergence of the universal predicted spectrum $\sim K^{-3}$ was demonstrated numerically in \cite{burkhart2013turbulence}. LP04 suggested a way of studying the actual turbulent spectrum by using sufficiently thin velocity slices of PPV\footnote{The slice thickness is limited by thermal velocities of the emitting species, however.}.

The studies of anisotropies in velocity channels earlier on were done only empirically. The first study in \cite{lazarian2001emissivity} demonstrated the potential value of velocity anisotropies of the intensity fluctuations within velocity PPV slices as a way to study magnetic field direction and probing media magnetization. The subsequent study in \cite{esquivel2015studying} (henceforth ELP15) allowed us to get an empirical relation between the degree of anisotropy and the Alfv\'en Mach numbers, but some particular issues, e.g. the decrease of the degree of anisotropy with the decrease of the slice thickness, were puzzling. These puzzles are resolved via this study. 

\subsection{Significance of the analytical description}
 
The development of the techniques to study turbulence using the observed Doppler shifted lines which started more than 60 years ago (\citealt{von1951methode}; \citealt{munch1958space}; \citealt{wilson1959internal}) has been significantly impeded by the inability of the researchers to analytically describe turbulent fluctuations in an adequate manner. For instance, a traditional technique to get the information on turbulence spectra is the use of measure of Doppler shift, termed velocity centroids $\sim \int v_z \rho_s \mathrm{d}v_z$, where the integration is carried over the range of the velocities relevant to the object under study\footnote{ While usually the velocity centroids are normalized over the integrated intensity over the LOS, \cite{stenholm1990molecular}; \cite{esquivel2005velocity} showed that this normalization does not change the statistical properties of the measure.}. In the case of the optically thin media, it can be shown that the velocity centroids are also proportional to $\int v_z \rho \mathrm{d}z$, where $\rho$ is an actual three-dimensional density and the integration is performed along the LOS (see \citealt{lazarian2003statistics}). However, this trick does not work as soon as the media have absorption, the fact that was noticed in \cite{munch1958internal}. To get the proper description of a realistic case of turbulent media with velocity centroids, one has to use the description of
$\rho_s$ provided in LP00 and the radiation transfer in PPV provided in LP04. 

Velocity centroids have been used for studying turbulence anisotropy in \cite{esquivel2005velocity}, \cite{esquivel2007statistics} and in \cite{burkhart2014measuring}. At the moment, they present the best way of obtaining the Alfv\'en Mach number $M_\text{A}$ and statistically determining the direction of the magnetic field. The value of the technique is likely to be significantly enhanced using the description $\rho_s$ arising from anisotropic turbulence. In particular, it should be possible to study the effects of absorption on the anisotropy analysis. Indeed, as we show in Section \ref{sec:selfabsorption}, the anisotropy is expected to be decreased for high optical depths, which is the effect not considered for velocity centroids. 

Similarly, as we mentioned earlier, an empirical study by ELP15, did not find an explanation for the observed change of the degree of anisotropy with this thickness of the channel maps, while this change naturally follows from the analytical description above. 

The velocity centroids and analysis of fluctuations in channels are the presently used techniques. However, we expect that with the available analytical description of $\rho_s$ for
a realistic model of anisotropic MHD turbulence, we should expect the emergence of new tools that will benefit from the developed formalism. 
This all shows that the significance of the developed description goes beyond improving the ability of VCA to study magnetization. 

\subsection{Towards quantitative VCA study}

The VCA in its original form suggested in LP00 and LP04 is concerned with the asymptotic regimes for the thin and thick velocity slices. This however does not use the whole information. \cite{chepurnov2015turbulence} have developed a numerical approach that fits the actual gradual change of the spectral index as the thickness of the channel increases. The approach is similar to the one already used with the VCS technique. Indeed, the original LP06 study on the VCS provided the asymptotic behaviour of the spectra. To increase the accuracy of getting the underlying spectra of velocity and density in the analysis of data in \cite{chepurnov2010velocity, chepurnov2015turbulence}, the fitting of full expressions for the correlations of velocity along the LOS (see \citealt{chepurnov2009turbulence}) was used. As a result, not only the corresponding velocity and density spectra were determined, but also the injection scale and thermal velocity of the turbulent gas. This way of using the VCS made it much more powerful in terms of the practical analysis of observational data. 

The extension of the VCA quantitative treatment for describing Alfv\'en, slow and fast modes that we performed in this paper opens avenues for developing the fitting procedures that allow separation of the contributions of the corresponding modes by changing the thickness of the PPV velocity slice. This should also significantly help with separating the anisotropy of velocity and density. We know that the latter quantity is changing its anisotropy with the sonic Mach number $M_s$ (see \citealt{kowal2007density}), and for high $M_s$ the statistics of density gets isotropic. The density, as we know, dominates at thick PPV velocity slices. This shows that comparing the anisotropy of thin and thick velocity slices, one can single out the anisotropy of velocity, which has a clear dependence on magnetization.

\subsection{VCA and other technique}

If one glances at the literature on  the techniques to study turbulence from spectroscopic data, one may come to a conclusion that there are a lot of different techniques available. However, this is mostly due to the fact that the use of different wavelets for the analysis of data is frequently treated as different statistical techniques of turbulence studies (\citealt{gill1990first}; \citealt{stutzki1998fractal}; \citealt{bernard1999pronaos}; \citealt{khalil2006morphological}). In reality, while Fourier transforms use harmonics of $\mathrm{e}^{i\bm{k}\cdot\bm{r}}$, wavelets use more sophisticated basis functions, which may be more appropriate for problems at hand. In previous studies, wavelets have also been used to analyse the results of computations (see \citealt{kowal2006separation}) along with or instead of Fourier transforms or correlation functions. Wavelets may reduce the noise arising from inhomogeneity of data, but it was found that in the situations when correlation functions of centroids were failing, a popular wavelet ($\Delta$ variance) was also failing (cf. \citealt{esquivel2005velocity}; \citealt{ossenkopf2006interstellar}; \citealt{esquivel2007statistics}). While in wavelets the basis functions are fixed, a more sophisticated technique, PCA, chooses basis functions that are, in some sense, the most descriptive. Nevertheless, the empirical relations obtained with PCA for extracting velocity statistics provide, according to \cite{padoan2006power}, an uncertainty of the velocity spectral index of the order of 0.5 (see also \citealt{brunt2003intrinsic}), which is too large for testing most of the turbulence theories.

We have discussed  two techniques of studying  magnetization and the direction of magnetic field from the analysis of spectral line information, namely, the one employing VCA and the other employing velocity centroids. The latter can be improved and modified on the basis of the present study. There is, however, another technique for similar studies that does not look easy to reformulate in view of the available description of the PPV. It is an empirical technique of anisotropy analysis based on the PCA (see \citealt{brunt2002interstellar}). 

Unlike the VCA and the velocity centroids that provide a simple description of turbulence through PPV space, the actions of PCA over PPV data are not easy to quantify. Some current work (\citealt{0004-637X-818-2-118}) show advantages of the analysis of the anisotropy based on the velocity centroids over the PCA technique and these advantages will only increase as the description of the PPV space gets available. However, an interesting property of PCA technique is that it seems to be sensitive to the phase information (\citealt{0004-637X-818-2-118}). The utility of this phase information should be revealed by the further research.

The spectral correlation function (SCF;  \citealt{rosolowsky1999spectral}) is another way to study turbulence using velocity slices of PPV space.  \cite{padoan2001effects} removed the adjustable parameters from the original expression for the SCF which made the technique  similar to VCA in terms of the observational data analysis. Indeed, both SCF and VCA measure correlations of intensity in velocity slices of PPV, but if SCF treats the outcome empirically, the analytical relations in LP00 relate the VCA measures to the underlying velocity and density statistics\footnote{Mathematically, SCF contains additional square roots and normalizations compared to the VCA expressions. Those make the analytical treatment, which is possible for simpler VCA expressions, prohibitive. One might speculate that, similar to the case of conventional centroids and unnormalized centroids introduced in \cite{lazarian2003statistics}, the actual difference between the statistics measured by the VCA and SCF is not significant.}.

 VCA relates the spectral index of intensity fluctuations within channel maps to the thickness of the velocity slice and to the underlying velocity and density in the emitting turbulent volume.
 We believe that the {\it thick} and {\it thin} slice regimes should also be present in the SCF analysis of data, although they have not been reported. We believe that the VCA can be used for all the purposes the SCF is used for (e.g. for an empirical comparisons of simulations and observations), although the opposite is not true. In fact, \cite{padoan2004structure} stressed that VCA eliminates errors inevitable for empirical attempts to calibrate PPV fluctuations in terms of the underlying 3D velocity spectrum. The present study extends the advantages of the VCA technique providing the analytical description of the anisotropy fluctuations and their variations with the thickness of the velocity slice. 
 
 There also exist numerous techniques identifying and analysing clumps and shells in PPV (see  \citealt{stutzki1990high}; \citealt{houlahan1992recognition}; \citealt{williams1994determining}; \citealt{pineda2006complete}; \citealt{ikeda2007survey}). They, however, identify an extended hierarchy of cores/shells arising from velocity crowding even in the synthetic observations obtained using incompressible simulations with no density clumps whatsoever. A more advanced technique to study a hierarchical structure of the PPV, namely 
 dendrogram technique (\citealt{goodman2009role}), can provide a complementary insight to the values of sonic and Alfv\'en Mach numbers $M_s$ and $M_\text{A}$ (\citealt{burkhart2013hierarchical}). 
 
\subsection{Spectroscopic and synchrotron studies of magnetic turbulence}

For synchrotron polarization studies, the analogue of PPV cube is the position-position-frequency (PPF) cube. In LP16, a number of techniques were suggested aimed at obtaining the information about magnetic field and the density of cosmic electrons using these cubes. In terms of anisotropy studies, it was suggested there to make use of the analysis of synchrotron intensity fluctuations in order to determine the same parameters that we are focused in this study, namely the mean magnetic field direction, the degree of magnetization of the media 
and the contribution of Alfv\'en, slow and fast modes. This suggests that studies of turbulence using  synchrotron and spectroscopic data can be very much complementary. Indeed, for the understanding of  dynamics of the ISM as well as for processes of the transport of heat and cosmic rays, it is essential to understand the properties of turbulent cascade in different interstellar phases. Synchrotron emission samples turbulence mostly in the diffuse hot and warm media (see \citealt{draine2010physics} for the table of the interstellar phases), while the turbulence in HI and molecular gas is well sampled via spectroscopic measurements\footnote{Future X-ray spectroscopy should provide a way to study hot plasma turbulence using spectroscopic techniques as well.}.
The correspondence of the properties of magnetic turbulence in different interstellar phases would testify about the single turbulent cascade on the galactic scale, which can be a discovery with important consequences for different branches of astrophysical research, e.g. for cosmic ray physics (see \citealt{2002cra..book.....S}). 

While the properties of turbulent fluctuations of magnetic field and velocity are closely related, there are differences. In particular, magnetic field is solenoidal, while velocity in MHD turbulence can have a potential component. Therefore, the treatments of anisotropy of magnetic turbulence and velocity turbulence in this work and in LP12 are similar, but not completely identical. Potentially, the VCA technique provides way to study compressible motions in a more adequate way. 

The generalization of the anisotropy study from pure synchrotron intensity in LP12 to synchrotron polarization in LP15 opened up ways to study anisotropies of MHD statistics in the PPF space. 
It is also interesting to compare the statistics of the PPV and the PPF. The PPV statistics is homogeneous along the $v$-axis, while the one of PPF is inhomogeneous. As a result, due to the effect of Faraday depolarization, for different frequencies one can sample turbulence at different distances from the observer, which allows study of the spatial distribution of turbulence. Such an effect is not present for the PPV studies. However, the homogeneity of the PPV in the $v$-direction allows one to better separate the contribution of the Alfv\'en, fast and slow modes by varying the slice thickness. Therefore, the statistical information in PPV  is complimentary to the statistics that can be obtained from PPF.   

\subsection{Synergy with other techniques}

Alfv\'en Mach number $M_\text{A}$ can be estimated on the basis of other techniques. For instance, the measures of Tsallis statistics (see \citealt{esquivel2010tsallis}; \citealt{tofflemire2011interstellar}), kurtosis and skewness (\citealt{burkhart2009density, burkhart2009characterizing}) also show sensitivity on $M_\text{A}$ as well as on the sonic Mach number $M_s$. Therefore, combining the VCA technique with other techniques can provide a more reliable determination of both $M_\text{A}$ and $M_s$. Note, that different techniques have their own limitations and uncertainties. Therefore, a combination of different techniques can significantly help. 

One of the parameters that influence the VCA is the density statistics which includes both its spectrum and anisotropy. This statistics can be obtained through the analysis of different data sets, e.g. dust emission or absorption. This should increase the accuracy of the VCA in determining the statistics of velocity. 

Combining the present technique with the synchrotron studies for independently obtaining the magnetization and decomposition of turbulence into the fundamental MHD modes is very advantageous. The intriguing opportunity of obtaining the angle between the mean magnetic field and the LOS on the basis of synchrotron polarization data that were discussed in LP16 allows one to remove the degeneracy between this angle and the Mach number that exists otherwise.  

\section{Summary}

In the paper above we have extended our VCA technique based on the analysis of fluctuation statistics in the velocity slices of the PPV space (i.e. velocity channel maps). Unlike our earlier study in LP00 and LP04, we accounted for the anisotropy of turbulence and provided expressions for the anisotropies in the velocity channel maps that arise from Alfv\'en, slow and fast modes of MHD turbulence. We calculated how these anisotropies change with the thickness of the velocity channel maps and compared our results with the numerical study in ELP15. In addition, we have studied the effects of absorption on the measured anisotropies and discussed the use of absorption spectral lines for studying turbulence anisotropy. 

Our study main results are as follows.
\begin{itemize}
\item  Analytical expressions for the degree of anisotropy of intensity fluctuations in slices of PPV space were obtained, and the variations of the degree of anisotropy as a function of the slice thickness were explored and successfully compared with the available numerical data. 
\item The procedures of separating contributions to anisotropy arising from density fluctuations and velocity fluctuations were studied, and the technique of establishing the anisotropies of density and velocity underlying turbulent field was formulated.
\item The separation of the contributions from Alfv\'en, slow and fast modes was investigated for the thin slice regime, and the ratio of the anisotropic to isotropic part of the slice intensity fluctuations was identified as a measure for fluid magnetization and compressibility.
\end{itemize}

\section*{Acknowledgements}
A.~L. acknowledges the NSF grant AST 1212096 and Center for Magnetic Self Organization (CMSO). 
He also acknowledges a distinguished visitor PVE/CAPES appointment at the Physics Graduate Program of the Federal University of Rio Grande do Norte and thanks the INCT INEspao and Physics Graduate Program/UFRN. D.~P. thanks 
the Department of Physics, 
Universitade Federal do Rio Grande do Norte (Natal, Brazil) for hospitality and stimulating atmosphere which initiated the start of this work. D.~K. and D.~P.~ are grateful to Institut Lagrange de Paris and Institut Astrophysique de Paris, France for hospitality.

\bibliographystyle{mnras}
\bibliography{AVCA_Apr14MNRAS}

\appendix

\section{Turbulence Statistics in PPV Space}\label{ppvspace}
\begin{table}
\centering
\caption{List of special mathematical functions used in this paper}
\begin{tabular}{l c}  
\hline
\hline  
\emph{Function} & \emph{Symbol} \\
\hline
Spherical Bessel function  & $j_n(x)$\\
Bessel function of the first kind & $J_n(x)$\\
Hyperbolic Bessel function of the first kind & $I_n(x)$\\
Spherical harmonics & $Y_{\ell}^{m}(x)$\\
Associated Legendre polynomial & $P_\ell^m(x)$\\
Legendre polynomial & $P_\ell(x)$ \\
Wigner 3-j symbols & $\begin{pmatrix}
\ell&\ell_1&\ell_2\\
m&m_1&m_2
\end{pmatrix}$ \\
Gamma function & $\Gamma[n]$\\
Gauss hypergeometric function & $\text{ }_2F_1[a,b;c;z]$\\
Step function & $\Theta[x]$\\
\hline
\hline
\end{tabular} 
\label{tab:mathsymbols}
\end{table}

Below we present the main expressions of the theory that we are going to use within our study. 

The transformation between real space and PPV space is of the form
$(\textbf{X},z)\rightarrow (\textbf{X},v)$ where $v$ is the LOS
velocity of the gas element.  The PPV density 
$\rho_s(\textbf{X},v)$ is dependent on both density of the gas in the real
space and its velocity, and is written as  (LP04)
\begin{equation}\label{densityexp}
\rho_s(\bm{X},v)=
\int_0^{S}\mathrm{d}z \rho(\bm{x})\Phi_v(v,\bm{x}),
\end{equation}
where $S$ is spatial extent of the turbulent cloud and $\Phi_v$ is the Maxwell
distribution of the thermal component of the turbulent particles
defined by
\begin{equation}
\Phi_v(v,\bm{x})=\frac{1}{\sqrt{2\upi\beta_\text{T}}}
\exp\left[-\frac{(v-u(\bm{x}))^2}{2\beta_\text{T}}\right],
\end{equation}
where $u(\textbf{x})$
is the non-thermal velocity of a particle at position $\textbf{x}$
which consists 
of the contribution of turbulent velocity as coherent velocity with the gas
cloud. If the gas is isolated, and coherent motions are negligible, as we 
adopt in this paper, $u$ is the LOS component of the turbulent
motion, and $\beta_\text{T}$ is the temperature parameter.

Intensity of radiation in an optically thin line is  proportional 
to the `density of emitters' of PPV space. This density is the result of the velocity mapping of emitters from XYZ to XYV space and is, in general, significantly different from the real space density. To describe statistical properties
of PPV density, we use the correlation 
\begin{equation}
\xi_s(R,\phi,v_1,v_2)\equiv \langle\rho_s(\textbf{X}_1,v_1)\rho_s(\textbf{X}_2,v_2)\rangle,
\end{equation}
or structure 
\begin{equation}
d_s(R,\phi,v_1,v_2)=\left\langle\left(\rho_s(\textbf{X}_1,v_1)-\rho_s(\textbf{X}_2,v_2)\right)^2\right\rangle,
\end{equation}
functions, 
where, in contrast to LP00 we take into account the dependence of 
the correlations on
the angle $\phi$ of the separation vector 
$\bm{R}=\bm{X}_1-\bm{X}_2$.
between the two LOS,

The averaging is performed over realizations of two random fields turbulent
velocity $\bm u$ and real space density $\rho$ of the emitters. Statistical
properties of these quantities reflect the properties of the magnetized turbulent
processes.

The turbulent velocity field is assumed to be described by the Gaussian two point probability distribution function (LP00)
\begin{align}\label{gaussian}
P(u_1, u_2)=\frac{1}{\upi\sqrt{2D_z(\infty)-D_z(\textbf{r})}\sqrt{D_z(\textbf{r})}}\exp\left[-\frac{u^2}{2D_z(\textbf{r})}\right]\nonumber\\
\exp\left[-\frac{u_+^2}{D_z(\infty)-D_z(\textbf{r})/2}\right],
\end{align}
where $u_1=u_z(\textbf{r}_1), \ldots$ and $u=u_1-u_2$, $u_+=(u_1+u_2)/2$.

No assumptions about Gaussianity of the density inhomogeneities of the sources
are made. We introduce density correlation function 
$\xi(\bm{r}) \equiv \langle\rho(\bm{x}_1)\rho(\bm{x}_2)\rangle $
whose properties are to be determined or modelled.
Splitting the density into the mean value and fluctuations, 
$\rho = \langle \rho \rangle + \delta \rho$, we have
$\xi(\bm{r}) 
= \langle\rho\rangle^2 +
\langle\delta\rho(\bm{x}_1)\delta\rho(\bm{x}_2)\rangle 
=\langle\rho\rangle^2 + \tilde \xi(r)$.
For specific calculations, we consider two distinct cases. 
If density perturbations have
a \textit{shallow} power spectrum, 
$\langle \delta\rho_k^2 \rangle \propto k^{-3+\nu_\rho},~\nu_\rho > 0$,
the correlation function is modelled as 
\begin{align}
\xi(\bm{r}) = \langle\rho\rangle^2 + 
\langle\delta\rho^2\rangle \frac{r_c^{\nu_\rho}}{r_c^{\nu_\rho}+r^{\nu_\rho}}
{\approx} \langle\rho\rangle^2 + 
\langle\delta\rho^2\rangle \left(r_c/r\right)^{\nu_\rho}
~,~ r > r_c
\label{eq:xi_shallow}
\end{align}
while if the power spectrum is steep, $\nu_\rho < 0$,
the density correlation function is
\begin{equation}
\xi(\bm{r})= 
\langle\rho\rangle^2 + 
\langle\delta\rho^2\rangle \frac{r_c^{-\nu_\rho}}{r_c^{-\nu_\rho}+r^{-\nu_\rho}}
\approx \langle\rho\rangle^2 + \langle\delta\rho^2\rangle 
-\langle\delta\rho^2\rangle \left(r/r_c\right)^{-\nu_\rho}
~,
\label{eq:xi_steep}
\end{equation}
for $ r < r_c$. The difference between the two cases is that for \textit{shallow} density, the
scaling range lies at separations exceeding the correlation length, $r > r_c$,
with $r_c$ associated with short scale damping, 
while for \textit{steep} density it lies at separations shorter than the
correlation length, $r < r_c$, which is now associated with 
the largest energy injection scale. Eq.~[\ref{eq:xi_steep}] shows that for
the \textit{steep} spectrum, scale
dependent part of the correlation function is always subdominant to the 
constant $\langle\rho\rangle^2 + \langle \delta\rho^2\rangle = 
\langle \rho^2 \rangle$.

Using Eq. \eqref{densityexp} and Eq. (\ref{gaussian}), it can be shown that (see LP04)
\begin{align}
\xi_s(\bm{r},& v_1, v_2) \approx \frac{S}{\sqrt{D_z(\infty)+\beta_\text{T}}}
\exp\left[-\frac{v_{+}^2}{D_z(\infty)+\beta_\text{T}}\right] \nonumber \\
&\int_{-S}^{S}\mathrm{d}z\left(1-\frac{|z|}{S}\right)
\frac{\langle\rho\rangle^2+\tilde\xi(\bm{r})}{\sqrt{D_z(\bm{r})+2\beta_\text{T}}}
\exp\left[-\frac{v^2}{2(D_z(\bm{r})+2\beta_\text{T})}\right].
\end{align}
under the assumption that
density fluctuations are uncorrelated with the turbulent velocities.
The arguments can be found in LP00, but importantly this assumption has
been checked in numerical MHD simulations \citep{esquivel2003velocity} and has been found to hold
with sufficient accuracy. 

The first exponential term reflects the amplitude
of correlation depending on the value of the central velocity $v_+$ relative
to the variance of the turbulent velocities $D(\infty)/2$. The LOS
integral term reflects the statistics of the turbulence at different separation
scales $\bm{R}$ and velocity differences $v$.

Since density correlation function has a constant term, there are non-trivial 
correlations in PPV cube even for uniform density of emitters. They arise from
different velocities of the emitters. Thus in our discussion we split the PPV
correlations into \textit{velocity} and \textit{density} contributions
\begin{equation}\label{eq:velden}
\tilde{\xi}_s(\textbf{R},\phi,v)=\tilde{\xi}_v(\textbf{R},\phi,v)+\tilde{\xi}_\rho(\textbf{R},\phi,v),
\end{equation}
where
\begin{equation}
\tilde{\xi}_v(\textbf{R},\phi,v)\propto \int_{-S}^{S}\mathrm{d}z\frac{\bar{\rho}^2(\textbf{r})}{\sqrt{D_z(\textbf{r})}}\exp\left[-\frac{v^2}{2D_z(\textbf{r})}\right],
\end{equation}
and
\begin{equation}\label{dense}
\tilde{\xi}_\rho(\textbf{R},\phi,v)\propto \int_{-S}^{S}\mathrm{d}z\frac{\tilde{\xi}(\textbf{r})}{\sqrt{D_z(\textbf{r})}}\exp\left[-\frac{v^2}{2D_z(\textbf{r})}\right].
\end{equation}
In these expressions we have omitted for brevity the thermal effects and the
finite cloud size effects. We should stress that although \textit{density}
correlation contribution is zero when the gas density is uniform, 
it depends on both density and velocity fluctuations when gas distribution
is inhomogeneous. 

This theory for PPV correlations allows for angular dependence of 
the correlation functions $\xi$ and $D_z$. Consequently,
after the integration over $z$  the anisotropic dependence on polar angle
$\phi$ is still allowed. This allows us, in what follows, to use whole machinery
developed in our earlier works to deal with the anisotropic turbulence. 

Let us turn to the quantity that can be measured in the observations.
The measured intensity of radiation in a velocity channel of width $\Delta v$,
centred at velocty $v_i$ is given by the integral
\begin{equation}
I(\bm{R},v_i)=\epsilon 
\int_{v_i-\Delta v/2}^{v_i+\Delta v/2} dv_1 \rho_s(\bm{R},v_1).
\end{equation}
With this, it can be shown that the intensity correlation function is (LP04)
\begin{align}
\xi_\text{I}(\bm{R},\Delta v)\propto \frac{\epsilon^2\bar{\rho}^2}{2\upi}&\int_{-S}^S\mathop{\mathrm{d}z}
\frac{1+\tilde{\xi}_s(\bm{r})}{D_z^{1/2}(\bm{r})}\nonumber\\
&\int_{-\infty}^{\infty}\mathop{\mathrm{d}v}
W_i(v,\Delta v, \beta_\text{T})\exp\left[-\frac{v^2}{2 D_z(\bm{r})}\right],
\end{align}
where $W_i(v,\Delta v, \beta)$ is a composite window of channel $i$. Its 
properties are such that for zero temperature $\beta_\text{T}=0$ its width is bounded
by $\Delta v$, but for high temperature $\beta_\text{T} > \Delta v^2$ it is given by 
the thermal width $\beta_\text{T}$.  Thus, thermal broadening sets the minimal effective
channel width.  In LP00 we have shown the importance of distinction between
\textit{thin} and \textit{thick} channels. The criterion is set by comparison
of characteristic velocity difference at the scale of separation between the
LOS, $\sqrt{D(R)}$ and the channel width $\Delta v$. 
In \textit{thin} channels such difference is
resolved $\Delta v < \sqrt{D_z(R)}$,  while in \textit{thick} it is not, 
$\Delta v > \sqrt{D_z(R)}$.  Thus in a \textit{thin} channel velocity 
differences along the LOS within the channel can be neglected,
$v=0$ leaving intensity correlations to be sensitive both to density and
velocity differences between the LOS, 
\begin{equation}
\xi_\text{I}(\bm{R},\Delta v)\propto \frac{\epsilon^2\bar{\rho}^2}{2\upi}\int_{-S}^S\mathop{\mathrm{d}z}
\frac{1+\tilde{\xi}_s(\bm{r})}{D_z^{1/2}(\bm{r})} ~.
\end{equation}
In contrast, in \textit{thick} channels, velocities are integrated over,
leaving only density inhomogeneities as the source of intensity fluctuations
\begin{equation}
\xi_\text{I}(\bm{R},\Delta v)\propto \frac{\epsilon^2\bar{\rho}^2}{2\upi}\int_{-S}^S\mathop{\mathrm{d}z}
\left[1+\tilde{\xi}_s(\bm{r}) \right].
\end{equation}
We note that \textit{thick} slicing can be obtained synthetically, by adding 
intensities (before computing the correlations) for adjacent thinner channels.
Our ability to have \textit{thin} velocity channels is limited by the 
instrument spectral resolution and the thermal broadening.

\section{General Approach To Find Velocity Correlation In Real Space }\label{general}
The velocity correlation tensor in the axisymmetric case in Fourier space is
\begin{equation}
\langle v_i(\bm{k}) v_j^*(\bm{k}') \rangle
= \mathcal{A}(k,\hat{\bm{k}}\cdot\hat{\bm{\lambda}})\left(\hat{\xi}_{\bm{k}} \otimes \hat{\xi}^*_{\bm{k}}\right)_{ij},
\end{equation}
where $\mathcal{A}(k,\hat{\bm{k}}\cdot\hat{\bm{\lambda}})$
is the power spectrum and
$\left(\hat{\xi}_{\bm{k}} \otimes \hat{\xi}^*_{\bm{k}}\right)_{ij}$
is a $\hat\lambda$-dependent tensor build from the displacement direction
characteristic for the given mode.
Correspondingly, in real space the velocity correlation function 
can be written as
\begin{equation}
\left\langle v_i(\bm{x}_1) v_j(\bm{x}_1+\bm{r})\right\rangle =\int \mathop{\mathrm{d}k}k^2 \mathop{\text{d$\Omega $}_k}  \mathrm{e}^{i\bm{k}\cdot\bm{r}}\mathcal{A}(k,\hat{\bm{k}}\cdot\hat{\bm{\lambda}})\left(\hat{\xi}_{\bm{k}} \otimes \hat{\xi}^*_{\bm{k}}\right)_{ij}.
\end{equation}

The power spectrum can be represented in terms of spherical harmonics as 
\begin{equation}
\mathcal{A}(k,\hat{\bm{k}}\cdot\hat{\bm{\lambda}})=
\sum_{\ell_1 m_1}\frac{4\upi}{2\ell_1+1}\mathcal{A}_{\ell_1}(k)
Y_{\ell_1 m_1}(\hat{\bm{k}})Y_{\ell_1 m_1}^*(\hat{\bm{\lambda}}),
\end{equation}
and similarly
\begin{equation}
\left(\hat{\xi}_{\bm{k}}\otimes\hat{\xi}^*_{\bm{k}}\right)_{ij}=
\sum_{\ell_2 m_2}c^{ij}_{\ell_2m_2}(\hat\lambda)Y_{\ell_2 m_2}(\hat{\bm{k}})~,
\end{equation}
Using well-known representation for the plane wave
\begin{equation}
\mathrm{e}^{i\bm{k}.\bm{r}}=4\upi\sum_{\ell m}i^{\ell}j_{\ell}(kr)
Y_{\ell m}(\hat{\bm{k}})Y_{\ell m}^{*}(\hat{\bm{r}}).
\end{equation}
we obtain
\begin{align}
\left\langle v_i v_j\right\rangle =
\sum_{\ell m}
4\upi i^{\ell}
Y_{\ell m}^{*}(\hat{\bm{r}})
\sum_{\ell_1 m_1}
\frac{4\upi}{2\ell_1+1}
Y_{\ell_1 m_1}^*(\hat{\bm{\lambda}})
\sum_{\ell_2 m_2}
c^{ij}_{\ell_2 m_2}(\hat{\lambda})\nonumber\\
\int \mathop{\mathrm{d}k}k^2 \mathop{\text{d$\Omega $}_k} j_{\ell}(kr)\mathcal{A}_{\ell_1}(k)
Y_{\ell m}(\hat{\bm{k}})Y_{\ell_1 m_1}(\hat{\bm{k}})Y_{\ell_2 m_2}(\hat{\bm{k}})
\end{align}
Defining
\begin{equation}\label{defineT}
\mathcal{T}_{\ell\ell_1}(r)=\int \mathop{\mathrm{d}k}k^2 j_{\ell}(kr)\mathcal{A}_{\ell_1}(k)
\end{equation}
and the symbol $\Psi$
that can be represented in terms of Wigner-3j symbols
\begin{equation}
\Psi_{\ell m,\ell_1 m_1,\ell_2 m_2}=\int \text{d$\Omega $}_kY_{\ell m}(\theta_k,\phi_k)Y_{\ell_1 m_1}(\theta_k,\phi_k)Y_{\ell_2 m_2}(\theta_k,\phi_k)
\end{equation}
\begin{align}
\Psi_{\ell m,\ell_1 m_1,\ell_2 m_2}=\sqrt{\frac{(2\ell+1)(2\ell_1+1)(2\ell_2+1)}{4\upi}}\begin{pmatrix}
\ell&\ell_1&\ell_2\\
0&0&0\\
\end{pmatrix}\nonumber\\
\begin{pmatrix}
\ell&\ell_1&\ell_2\\
m&m_1&m_2
\end{pmatrix}
\end{align}
we arrive at a suitable for further analysis form for the correlation tensor
\begin{align}\label{generaleq}
\left\langle v_i v_j\right\rangle
=\sum_{\ell m}
4\upi i^{\ell}
Y_{\ell m}^{*}(\hat{\bm{r}})
&\sum_{\ell_1m_1}
\frac{4\upi}{2\ell_1+1}
Y_{\ell_1 m_1}^*(\hat{\bm{\lambda}})\nonumber\\
&\sum_{\ell_2 m_2}
c^{ij}_{\ell_2 m_2}(\hat\lambda)\mathcal{T}_{\ell\ell_1}(r)\Psi_{\ell m,\ell_1 m_1,\ell_2 m_2}~.
\end{align}

For this expression, we will derive $A_\xi, B_\xi, C_\xi$ and $D_\xi$ (cf. Eq.\eqref{correlationfunction}).
We develop the following procedure to find these coefficients.
Using the fact that $A_\xi,B_\xi,C_\xi$ and $D_\xi$  are invariant under rotation of coordinate frame, as these coefficients depend only on $\mu\equiv \hat{\bm{r}}\cdot\hat\lambda$, 
we shall work in $\hat\lambda=\hat{\bm{z}}$ frame.
In this frame,
Eq. \eqref{generaleq} is simplified to 
\begin{align}\label{specialcorr}
\left\langle v_i v_j\right\rangle
=\sum_{\ell m}
4\upi i^{\ell}
(-1)^m Y_{\ell m}(\hat{\bm{r}})
\sum_{\ell_1}
\sqrt{\frac{4\upi}{2\ell_1+1}}
\sum_{\ell_2}
c^{ij}_{\ell_2 m}(\hat{\bm{z}})\nonumber\\
\mathcal{T}_{\ell\ell_1}(r)\Psi_{\ell (-m),\ell_10,\ell_2 m}~.
\end{align}
where we have used the fact that Wigner-3j symbols vanish for $m+m_1+m_2 \ne 0$
and the relation $Y_{\ell (-m)}^*(\hat{\bm{r}}) = (-1)^m Y_{\ell m}(\hat{\bm{r}})$.
We also note that $c^{ij}_{\ell_2 m}(\hat{\bm{z}})$ are non-zero only
for $|m| \le 2$ (since they are the multipole expansion of a direct square of a
vector, we shall see this in explicit calculations further on). For an isotropic power spectrum, only $\ell_1=0$ survives and therefore Eq.\eqref{specialcorr} becomes
\begin{equation}\label{isotropiccorr}
\left\langle v_i v_j\right\rangle
=\sum_{\ell m}
4\upi i^{\ell}
Y_{\ell m}(\hat{\bm{r}})
c^{ij}_{\ell m}(\hat{\bm{z}})\mathcal{T}_{\ell 0}(r).
\end{equation}

As the first step of calculations, we take $\langle v_1v_2\rangle$ which in 
$\hat \lambda = \hat{\bm{z}}$ frame has simple form 
$\langle v_1v_2\rangle = A(r,\cos\theta) \hat{\bm{r}}_1 \hat{\bm{r}}_2
= A(r,\cos\theta) \sin^2\theta \cos\phi\sin\phi$, and use this to find
$A(r,\cos\theta)$. Since the expression obtained for  $A$ should be valid 
in all frames, we replace $\cos\theta\rightarrow\mu$ to arrive to the 
frame-independent $A(r,\mu)$. 
Next, we take $\langle v_1v_1\rangle = A\hat{\bm{r}}_1 \hat{\bm{r}}_1 + B$, and repeat the procedure selecting factor $A$ as what is proportional to $\sin^2\theta\cos^2\phi$, with the remainder being $B$.
After that, we take $\langle v_1v_2\rangle = A \hat{\bm{r}}_1 \hat{\bm{r}}_3 + D \hat{\bm{r}}_1$ and repeat the procedure by factoring out $\sin\theta\cos\theta\cos\phi$ component and looking for the remainder, which is 
$D \sin\theta\cos\phi$. Finally, we take $\langle v_3v_3\rangle = A\hat{\bm{r}}_3 \hat{\bm{r}}_3 + B + C + 2D \hat{\bm{r}}_3$ and use
the previously found coefficients $A$, $B$ and $D$ to obtain $C$.
This technique is applied in the subsequent sections. Since the current section and subsequent sections  heavily use some special mathematical symbols, we summarise these symbols in Table \ref{tab:mathsymbols}.

\section{Velocity Correlation Tensor For Different Turbulent Modes}\label{allstf}
With the technique developed in Appendix [\ref{general}], it is straightforward exercise to obtain the coefficients $A, B, C$ and $D$ of the velocity correlation function provided that we have information about the tensor structure of a turbulent mode. In this section, we will apply the technique developed in the previous section to find these coefficients for Alfv\'en mode, fast mode, slow mode and strong turbulence. During our calculation, we use the knowledge about anisotropy of power spectrum of each particular mode. The power spectrum of Alfv\'en mode, slow mode and strong turbulence is anisotropic, while that of fast mode is isotropic \citep{cho2003compressible}, and this fact will be used in our subsequent calculations.

\subsection{Alfv\'en mode}\label{alfven}
The tensor structure for Alfv\'en mode is
\begin{align}
\left(\hat{\xi}_{\bm{k}}\otimes\hat{\xi}^*_{\bm{k}}\right)_{ij} =&
(\delta_{ij}-\hat{k}_i\hat{k}_j)\nonumber\\
&-
\frac{(\hat{k}\cdot\hat{\lambda})^2\hat{k}_i\hat{k}_j-
(\hat{k} \cdot \hat{\lambda})(\hat{\lambda}_i\hat{k}_j+\hat{\lambda}_j\hat{k}_i)+\hat{\lambda}_i\hat{\lambda}_j}{1-(\hat{k} \cdot \hat{\lambda})^2}.
\end{align}
LP12 labelled the first term of the above tensor structure as $E$-type, and the second term as $F$-type. Therefore, the correlation tensor is $E-F$ type\footnote{LP12 obtained correlation of magnetic field, while here we are talking about correlation of velocity field. In the case of Alfv\'en mode, these correlations are the same, but this is in general not the case. This is because magnetic field are solenoidal, while velocity fields can be potential as well.}. 

The coefficients $c_{\ell m}^{ij}$ which we shall use for the derivation of coefficients $A, B, C$ and $D$ are presented in Table [\ref{tab:alfvenparams}] for  $\hat{\lambda}=\hat{\bm{z}}$. It is important to note that $c_{\ell m}^{ij}$ is zero when $\ell$ is odd.
\begin{table}
\centering
\caption{Mode structure of Alfv\'en modes for $\hat{\lambda}=\hat{\bm{z}}$}
\begin{tabular}{@{} *8l @{}}   
\hline
\hline  
\emph{$c^{ij}_{\ell m}$} & \emph{Equation} (for even $\ell$) \\
\hline
&\\
 $c^{11}_{\ell m}$ & $\sqrt{\upi}\delta_{\ell 0}\delta_{m0}-\Theta(\ell-2)\sqrt{\frac{\upi(2\ell+1)(\ell-2)!}{(\ell+2)!}}(\delta_{m2}+\delta_{m,-2})$ \\ 
    &\\
 $c^{22}_{\ell m}$ & $\sqrt{\upi}\delta_{\ell 0}\delta_{m0}+\Theta(\ell-2)\sqrt{\frac{\upi(2\ell+1)(\ell-2)!}{(\ell+2)!}}(\delta_{m2}+\delta_{m,-2})$\\
  
  &\\
 $c^{12}_{\ell m}$ & $\Theta(\ell-2)i\sqrt{\frac{\upi(2\ell+1)(\ell-2)!}{(\ell+2)!}}(\delta_{m2}-\delta_{m,-2})$ \\ 
  &\\
 $c^{i3}_{\ell m}$ &$0,\qquad i\in(1,2,3)$\\
 &\\
 \hline
\end{tabular}
\label{tab:alfvenparams}
\end{table} 

As a first step of our calculations, we compute
\begin{align}
\langle v_1&v_2\rangle=4\upi\sum_{\ell \ell_1\ell_2}i^l\sqrt{\upi(2\ell+1)}(2\ell_2+1)\sqrt{\frac{(\ell_2-2)!}{(\ell_2+2)!}}\mathcal{T}_{\ell\ell_1}(i)\nonumber\\
&\begin{pmatrix}
\ell&\ell_1&\ell_2\\
0&0&0\\
\end{pmatrix}
\begin{pmatrix}
\ell&\ell_1&\ell_2\\
-2&0&2
\end{pmatrix}
\left(Y_{\ell}^{2}(\hat{\bm{r}})-Y_{\ell} ^{-2}(\hat{\bm{r}})\right)
\end{align}
To separate $r_1 r_2 = \sin^2\theta\cos\phi\sin\phi$ factor
we use the following identities of the spherical harmonics
\begin{align}
\label{eq:Yl2_inangles}
Y_{\ell}^{2}(\hat{\bm{r}})-Y_{\ell} ^{-2}(\hat{\bm{r}})=4i\sqrt{\frac{(2\ell+1)(\ell-2)!}{4\upi(\ell+2)!}}\frac{P_{\ell}^2(\cos\theta)}{\sin^2\theta}
\nonumber\\
\sin^2\theta\cos\phi\sin\phi,
\end{align}
and 
\begin{equation}
\frac{P_{\ell}^2(\cos\theta)}{\sin^2\theta}=\frac{\partial^2 P_{\ell}(\cos\theta)}{\partial(\cos\theta)^2},
\label{eq:Pl2_inderivs}
\end{equation}
thus finding
\begin{align}\label{alfvenA}
A=-8\upi\sum_{\ell \ell_1\ell_2} i^l(2\ell+1)(2\ell_2+1)\sqrt{\frac{(\ell-2)!(\ell_2-2)!}{(\ell+2)!(\ell_2+2)!}}\mathcal{T}_{\ell\ell_1}\nonumber\\
\begin{pmatrix}
\ell&\ell_1&\ell_2\\
0&0&0\\
\end{pmatrix}
\begin{pmatrix}
\ell&\ell_1&\ell_2\\
-2&0&2
\end{pmatrix}\frac{\partial^2 P_{\ell}(\mu)}{\partial\mu^2}.
\end{align}
after generalization to an arbitrary frame by replacing $\cos\theta \to \mu$.

Next step of calculation involves finding $B$, and for that we take 
\begin{align}\label{intermediateB}
\langle v_1v_1\rangle=4\upi\sum_{\ell \ell_1} i^l\sqrt{(2\ell+1)}\sqrt{\upi}\begin{pmatrix}
\ell&\ell_1&0\\
0&0&0\\
\end{pmatrix}
\begin{pmatrix}
\ell&\ell_1&0\\
0&0&0
\end{pmatrix}\nonumber\\
\mathcal{T}_{\ell\ell_1}Y_{\ell}^0(\hat{\bm{r}})-4\upi\sum_{\ell \ell_1\ell_2} i^l\sqrt{\upi(2\ell+1)}(2\ell_2+1)\sqrt{\frac{(\ell_2-2)!}{(\ell_2+2)!}}\nonumber\\
\begin{pmatrix}
\ell&\ell_1&\ell_2\\
0&0&0\\
\end{pmatrix}
\begin{pmatrix}
\ell&\ell_1&\ell_2\\
-2&0&2
\end{pmatrix}
\mathcal{T}_{\ell\ell_1}\left(Y_{\ell}^{2}(\hat{\bm{r}})+Y_{\ell} ^{-2}(\hat{\bm{r}})\right).
\end{align}
The above expression can be simplified by considering the following identities of spherical harmonics
\begin{equation}\label{eq:Yl2plus}
Y_{\ell}^{2}(\hat{\textbf{r}})+Y_{\ell} ^{-2}(\hat{\textbf{r}})=2\sqrt{\frac{(2\ell+1)(\ell-2)!}{4\upi(\ell+2)!}}P_{\ell}^2(\cos\theta)(2\cos^2\phi-1),
\end{equation}
and
\begin{equation}\label{eq:Yl0}
Y_\ell^0(\hat{\textbf{r}})=\sqrt{\frac{2l+1}{4\upi}}P_\ell(\cos\theta).
\end{equation}
The second term in Eq.\eqref{intermediateB} contains contribution from $A$ term that is proportional to $\cos^2\phi$.  Taking that into account, we have
\begin{align}\label{alfvenB}
&B=2\upi\sum_{\ell=0,2} i^\ell\mathcal{T}_{\ell\ell}P_\ell(\mu)+4\upi\sum_{\ell \ell_1\ell_2} i^l(2\ell+1)(2\ell_2+1)\nonumber\\
&\sqrt{\frac{(\ell-2)!(\ell_2-2)!}{(\ell+2)!(\ell_2+2)!}}\begin{pmatrix}
\ell&\ell_1&\ell_2\\
0&0&0\\
\end{pmatrix}
\begin{pmatrix}
\ell&\ell_1&\ell_2\\
-2&0&2
\end{pmatrix}\mathcal{T}_{\ell\ell_1}P_\ell^{2}(\mu).
\end{align}
To find $D$, we note that $\langle v_1v_3\rangle=0$, and therefore, in our choice of frame $D=-A\cos\theta$. Therefore, in general 
\begin{align}\label{alfvenD}
D=8\upi\sum_{\ell \ell_1\ell_2} i^l(2\ell+1)(2\ell_2+1)\sqrt{\frac{(\ell-2)!(\ell_2-2)!}{(\ell+2)!(\ell_2+2)!}}\nonumber\\
\begin{pmatrix}
\ell&\ell_1&\ell_2\\
0&0&0\\
\end{pmatrix}
\begin{pmatrix}
\ell&\ell_1&\ell_2\\
-2&0&2
\end{pmatrix}\mathcal{T}_{\ell\ell_1}\mu\frac{\partial^2 P_{\ell}(\mu)}{\partial\mu^2}
\end{align}
Similarly, we note that $\langle v_3v_3\rangle=0$, and therefore, $C=-A\cos^2\theta-B-2D\cos\theta=A\cos^2\theta-B$ which gives
\begin{align}\label{alfvenC}
C=-2\upi\sum_{\ell=0,2} i^\ell\mathcal{T}_{\ell\ell}P_\ell(\mu)-4\upi\sum_{\ell \ell_1\ell_2} i^l(2\ell+1)(2\ell_2+1)\nonumber\\
\sqrt{\frac{(\ell-2)!(\ell_2-2)!}{(\ell+2)!(\ell_2+2)!}}\begin{pmatrix}
\ell&\ell_1&\ell_2\\
0&0&0\\
\end{pmatrix}
\begin{pmatrix}
\ell&\ell_1&\ell_2\\
-2&0&2
\end{pmatrix}\nonumber\\
\mathcal{T}_{\ell\ell_1}\left(2\mu^2\frac{\partial^2 P_{\ell}(\mu)}{\partial\mu^2}+P_\ell^{2}(\mu)\right).
\end{align}
\subsection{Fast modes high-$\beta$}\label{FMHBe}
Fast modes in high $\beta$ regime are purely compressional type of modes, and their tensor structure is 
\begin{equation}
\left(\hat{\xi}_{\bm{k}}\otimes\hat{\xi}^*_{\bm{k}}\right)_{ij} = \hat{k}_i\hat{k_j}.
\end{equation}
The power spectrum of this mode is isotropic, and therefore  we utilise Eq. \eqref{isotropiccorr} for our calculations.
Our first step involves computation of 
\begin{equation}
\langle v_1v_2\rangle=
4\upi i^2(-i)\sqrt{\frac{2\upi}{15}}\mathcal{T}_{20}(r)\left(Y_{2}^{2}(\hat{\textbf{r}})-Y_{2} ^{-2}(\hat{\textbf{r}})\right),
\end{equation}
which yields an isotropic form for $A$
\begin{equation}\label{FMHbA}
A=-4\upi\mathcal{T}_{20}.
\end{equation}
The next step is to compute
\begin{align}
\langle v_1v_1\rangle=4\upi\frac{2\sqrt{\upi}}{3}\mathcal{T}_{00}&Y_{0} ^{0}(\hat{\textbf{r}})+4\upi\mathcal{T}_{20}\frac{2}{3}\sqrt{\frac{\upi}{5}}Y_{2} ^{0}(\hat{\textbf{r}})\nonumber\\
&+4\upi i^2\sqrt{\frac{2\upi}{15}}\mathcal{T}_{20}(r)\left(Y_{2}^{2}(\hat{\textbf{r}})+Y_{2} ^{-2}(\hat{\textbf{r}})\right),
\end{align} 
which after subtracting $A \sin^2\theta\cos^2\phi$ contribution gives
\begin{equation}\label{FMHbB}
B=\frac{4\upi}{3}\mathcal{T}_{00}(r)+\frac{4\upi}{3}\mathcal{T}_{20}(r).
\end{equation}
It is easy to check that $C=D=0$ for this mode. This is expected because both  tensor structure as well as  power spectrum are isotropic.
\subsection{Fast modes low-$\beta$}\label{FMLB}
For fast Modes in low $\beta$ regime, the tensor structure of velocity field in Fourier space is 
\begin{equation}
\left(\hat{\xi}_{\bm{k}}\otimes\hat{\xi}^*_{\bm{k}}\right)_{ij} =
\frac{\hat{k}_i\hat{k_j}-
(\hat{k} \cdot \hat{\lambda})(\hat{k}_i\hat{\lambda}_j+\hat{k}_j\hat{\lambda}_i)
+(\hat{k} \cdot \hat{\lambda})^2\hat{\lambda}_i\hat{\lambda}_j}
{1-(\hat{k} \cdot \hat{\lambda})^2}.
\end{equation}

\begin{table}
\centering
\caption{Mode structure of fast Modes in low-$\beta$ For $\hat{\lambda}=\hat{z}$.}
\begin{tabular}{@{} *8l @{}}  
\hline
\hline  
\emph{$c^{ij}_{\ell m}$} & \emph{Equation}(for even $\ell$) \\
\hline
$c^{11}_{\ell m}$ & $\sqrt{\upi}\delta_{\ell 0}\delta_{m0}+\Theta(\ell-2)\sqrt{\frac{\upi(2\ell+1)(\ell-2)!}{(\ell+2)!}}(\delta_{m2}+\delta_{m,-2})$\\
  
  &\\
 $c^{22}_{\ell m}$ & $\sqrt{\upi}\delta_{\ell 0}\delta_{m0}-\Theta(\ell-2)\sqrt{\frac{\upi(2\ell+1)(\ell-2)!}{(\ell+2)!}}(\delta_{m2}+\delta_{m,-2})$ \\ 
    &\\
 
 $c^{12}_{\ell m}$ & $-i\Theta(\ell-2)\sqrt{\frac{\upi(2\ell+1)(\ell-2)!}{(\ell+2)!}}(\delta_{m2}-\delta_{m,-2})$ \\ 
  &\\
 $c^{i3}_{\ell m}$ &$0,\qquad i\in(1,2,3)$\\
 &\\
 \hline
\end{tabular}
\label{tab:Fastparams}
\end{table}

The power spectrum of this mode is isotropic. To find the coefficients $A, B, C$ and $D$, our starting point is to utilise the table presented above.  Noting the similarity of Table [\ref{tab:Fastparams}] with Table [\ref{tab:alfvenparams}], it is easy to derive these coefficients just by considering the previous results. Similar to the previous section, the velocity has no component along the symmetry axis, so that $D=-A\mu$, and $C=A\mu^2-B$. Due to the fact that the power spectrum is isotropic in this case, the results heavily simplify, and we have the final result for the correlation coefficients as
\begin{equation}
\label{FMLBA}A=8\upi\sum_{\ell} i^l(2\ell+1)\frac{(\ell-2)!}{(\ell+2)!}\mathcal{T}_{\ell 0}\frac{\partial^2 P_{\ell}(\mu)}{\partial\mu^2}
\end{equation}
\begin{align}
\label{FMLBB}B&=2\upi\mathcal{T}_{00}+4\upi\sum_{n=2,2}^\infty i^n(2n+1)\frac{n!}{(n+2)!}\mathcal{T}_{n 0}P_{n}(\mu)\nonumber\\
&-8\upi\sum_{n=0,2}(2n+1)P_n(\mu)\sum_{l=n+2,2}^\infty i^l(2\ell+1)\frac{(\ell-2)!}{(\ell+2)!}\mathcal{T}_{\ell 0}
\end{align}
\begin{align}
\label{FMLBC}C=-2\upi T_{00}+4\upi\sum_{\ell} i^l(2\ell+1)\frac{(\ell-2)!}{(\ell+2)!}&\mathcal{T}_{\ell 0}\Bigg(P_{\ell}^2(\mu)\nonumber\\
&+2\mu^2\frac{\partial^2 P_{\ell}(\mu)}{\partial\mu^2}\Bigg)
\end{align}
\begin{equation}
\label{FMLBD}D=-8\upi\sum_{\ell} i^l(2\ell+1)\frac{(\ell-2)!}{(\ell+2)!}\mathcal{T}_{\ell 0}\mu\frac{\partial^2 P_{\ell}(\mu)}{\partial\mu^2}.
\end{equation}

\subsection{Slow modes high-$\beta$}\label{vcorfsm}
The velocity correlation tensor in Fourier Space for slow Modes in high $\beta$ is
\begin{align}
\left(\hat{\xi}_{\bm{k}}\otimes\hat{\xi}^*_{\bm{k}}\right)_{ij} =
\left(\frac{(\hat{k} \cdot \hat{\lambda})^2\hat{k}_i\hat{k}_j
+\hat{\lambda}_i\hat{\lambda}_j
-(\hat{k} \cdot \hat{\lambda})(\hat{k}_i\hat{\lambda}_j+\hat{k}_j\hat{\lambda}_i)}{1-(\hat{k} \cdot \hat{\lambda})^2}\right).
\end{align}
This was identified to be $F$-type in LP12. Besides the tensor structure, the power spectrum of slow modes is also anisotropic, and is the same as for Alfv\'en mode. 
\begin{table*}
\centering
\caption{Mode structure of slow modes in high-$\beta$ For $\hat{\lambda}=\hat{z}$.}
\begin{tabular}{@{} *8l @{}}  
\hline
\hline  
\emph{$c^{ij}_{\ell m}$} & \emph{Equation}(for even $\ell$) \\
\hline
&\\
$c^{11}_{\ell m}$ & $\frac{\sqrt{\upi}}{3}\delta_{\ell 0}\delta_{m 0}+\frac{2}{3}\sqrt{\frac{\upi}{5}}\delta_{\ell 2}\delta_{m 0}+\left(\frac{1}{2}\sqrt{\frac{\upi}{30}}\delta_{\ell 2}+\Theta(\ell-4)\sqrt{\frac{\upi(2\ell_2+1)(\ell_2-2)!}{(\ell_2+2)!}}\right)(\delta_{m2}+\delta_{m,-2}).$\\
  
  &\\

 $c^{22}_{\ell m}$ & $\frac{\sqrt{\upi}}{3}\delta_{\ell 0}\delta_{m 0}+\frac{2}{3}\sqrt{\frac{\upi}{5}}\delta_{\ell 2}\delta_{m 0}-\left(\frac{1}{2}\sqrt{\frac{\upi}{30}}\delta_{\ell 2}+\Theta(\ell-4)\sqrt{\frac{\upi(2\ell_2+1)(\ell_2-2)!}{(\ell_2+2)!}}\right)(\delta_{m2}+\delta_{m,-2}).$ \\ 
    &\\
 $c^{33}_{\ell m}$ & $\frac{4\sqrt{\upi}}{3}\left(\delta_{\ell 0}-\frac{1}{\sqrt{5}}\delta_{\ell 2}\right)\delta_{m0}$\\
   &\\
 $c^{12}_{\ell m}$ & $-i\left(\frac{1}{2}\sqrt{\frac{\upi}{30}}\delta_{\ell 2}+\Theta(\ell-4)\sqrt{\frac{\upi(2\ell+1)(\ell-2)!}{(\ell+2)!}}\right)(\delta_{m2}-\delta_{m,-2})$ \\ 
  &\\
 $c^{13}_{\ell m}$ &$\sqrt{\frac{2\upi}{15}}\delta_{\ell 2}(\delta_{m1}-\delta_{m,-1})$\\
 &\\
 $c^{23}_{\ell m}$ &$-i\sqrt{\frac{2\upi}{15}}\delta_{\ell 2}(\delta_{m1}+\delta_{m,-1})$\\
 &\\
 \hline
\end{tabular}
\label{tab:slowparams}
\end{table*}

All the coefficients $c^{ij}_{\ell m}$ relevant for our calculations for this mode are summarised in Table \ref{tab:slowparams}. The first step as usual is to compute the following element 
\begin{align}
\langle v_1v_2\rangle=
4\upi\sum_{\ell \ell_1} i^l\sqrt{\upi(2\ell+1)}\Bigg(\sqrt{\frac{1}{24}}
\begin{pmatrix}
\ell&\ell_1&2\\
0&0&0\\
\end{pmatrix}
\begin{pmatrix}
\ell&\ell_1&2\\
-2&0&2
\end{pmatrix}\nonumber\\
+\sum_{\ell_2=4,2}^\infty(2\ell_2+1)\sqrt{\frac{(\ell_2-2)!}{(\ell_2+2)!}}\begin{pmatrix}
\ell&\ell_1&\ell_2\\
0&0&0\\
\end{pmatrix}
\begin{pmatrix}
\ell&\ell_1&\ell_2\\
-2&0&2
\end{pmatrix}
\Bigg)\nonumber\\
\mathcal{T}_{\ell\ell_1}(-i)\left(Y_{\ell}^2(\hat{\bm{r}})-Y_{\ell}^{-2}(\hat{\bm{r}})\right)
\end{align}
Using the relations for spherical harmonics Eq.~(\ref{eq:Yl2_inangles})
and Eq.~(\ref{eq:Pl2_inderivs}), we arrive to the general form 
\begin{align}\label{SMHBA}
A=8\upi\sum_{\ell \ell_1} i^l(2\ell+1)\sqrt{\frac{(\ell-2)!}{(\ell+2)!}}\Bigg(\sqrt{\frac{1}{24}}\begin{pmatrix}
\ell&\ell_1&2\\
0&0&0\\
\end{pmatrix}
\begin{pmatrix}
\ell&\ell_1&2\\
-2&0&2
\end{pmatrix} \nonumber\\
+\sum_{\ell_2=4,2}^\infty(2\ell_2+1)\sqrt{\frac{(\ell_2-2)!}{(\ell_2+2)!}}\begin{pmatrix}
\ell&\ell_1&\ell_2\\
0&0&0\\
\end{pmatrix}
\begin{pmatrix}
\ell&\ell_1&\ell_2\\
-2&0&2
\end{pmatrix}\Bigg)\nonumber\\
\mathcal{T}_{\ell\ell_1}\frac{\partial^2 P_{\ell}(\mu)}{\partial\mu^2},
\end{align}

As a next step, we compute 
\begin{align}
\langle v_1&v_1\rangle=4\upi\sum_{\ell \ell_1} i^\ell\sqrt{(2\ell+1)}\Bigg(\frac{\sqrt{\upi}}{3}\begin{pmatrix}
\ell&\ell_1&0\\
0&0&0\\
\end{pmatrix}^2
\nonumber\\
&+\frac{2}{3}\sqrt{\upi}\begin{pmatrix}
\ell&\ell_1&2\\
0&0&0
\end{pmatrix}^2\Bigg)\mathcal{T}_{\ell\ell_1}Y_{\ell}^0(\hat{\bm{r}})
+4\upi\sum_{\ell \ell_1} i^l\sqrt{\upi(2\ell+1)}\nonumber\\&\Bigg(\sqrt{\frac{1}{24}}\begin{pmatrix}
\ell&\ell_1&2\\
0&0&0\\
\end{pmatrix}
\begin{pmatrix}
\ell&\ell_1&2\\
-2&0&2
\end{pmatrix} +\sum_{\ell_2=4,2}^\infty(2\ell_2+1)\sqrt{\frac{(\ell_2-2)!}{(\ell_2+2)!}}\nonumber\\
&\begin{pmatrix}
\ell&\ell_1&\ell_2\\
0&0&0\\
\end{pmatrix}\begin{pmatrix}
\ell&\ell_1&\ell_2\\
-2&0&2
\end{pmatrix}
\Bigg)\mathcal{T}_{\ell\ell_1}\left(Y_{\ell}^{ 2}(\hat{\bm{r}})+Y_{\ell}^{-2}(\hat{\bm{r}})\right).
\end{align}
Upon using the identities for spherical harmonics Eq.\eqref{eq:Yl2plus}, it is easy to see that the second term in the above equation partially contains $A$ contribution. Therefore, after some manipulations, we obtain an expression for $B$ valid in a general frame:
\begin{align}\label{SMHBB}
B=\frac{2\upi}{3}\sum_{\ell=0,2} i^\ell\mathcal{T}_{\ell\ell}P_\ell(\mu)+\frac{4\upi}{3}\sum_{\ell\ell_1}i^\ell(2\ell+1)
\begin{pmatrix}
\ell&\ell_1&2\\
0&0&0
\end{pmatrix}^2\nonumber\\
\mathcal{T}_{\ell\ell_1}P_\ell(\mu)
-4\upi\sum_{\ell \ell_1} i^l(2\ell+1)\sqrt{\frac{(\ell-2)!}{(\ell+2)!}}
\Bigg(\sqrt{\frac{1}{24}}\begin{pmatrix}
\ell&\ell_1&2\\
0&0&0\\
\end{pmatrix}\nonumber\\
\begin{pmatrix}
\ell&\ell_1&2\\
-2&0&2
\end{pmatrix}
+\sum_{\ell_2=4,2}^\infty(2\ell_2+1)\sqrt{\frac{(\ell_2-2)!}{(\ell_2+2)!}}\begin{pmatrix}
\ell&\ell_1&\ell_2\\
0&0&0\\
\end{pmatrix}\nonumber\\
\begin{pmatrix}
\ell&\ell_1&\ell_2\\
-2&0&2
\end{pmatrix}
\Bigg)\mathcal{T}_{\ell\ell_1}P_\ell^{2}(\mu).
\end{align}
Next, to find out $D$, we compute
\begin{align}
\langle v_1v_3\rangle=4\upi \sum_{\ell\ell_1}i^{\ell}\sqrt{\frac{2\upi}{15}}\sqrt{5(2\ell+1)}\begin{pmatrix}
\ell&\ell_1&\ell_2\\
0&0&0\\
\end{pmatrix}\nonumber\\
\begin{pmatrix}
\ell&\ell_1&\ell_2\\
-1&0&1
\end{pmatrix}
\mathcal{T}_{\ell\ell_1}(-1)\left(Y_\ell^1(\hat{\bm{r}})+Y_\ell^{-1}(\hat{\bm{r}})\right).
\end{align}

With some simplifications, and following the general procedure of subtracting the contribution from $A$, we finally obtain
\begin{align}\label{SMHBD}
D=4\upi \sum_{\ell\ell_1}i^{\ell}(2\ell+1)\sqrt{\frac{2(\ell-1)!}{3(\ell+1)!}}\begin{pmatrix}
\ell&\ell_1&2\\
0&0&0\\
\end{pmatrix}
\begin{pmatrix}
\ell&\ell_1&2\\
-1&0&1
\end{pmatrix}\nonumber\\
\mathcal{T}_{\ell\ell_1}\frac{\partial P_{\ell}(\mu)}{\partial\mu}-8\upi\sum_{\ell \ell_1} i^l(2\ell+1)\sqrt{\frac{(\ell-2)!}{(\ell+2)!}}
\Bigg(\sqrt{\frac{1}{24}}\begin{pmatrix}
\ell&\ell_1&2\\
0&0&0\\
\end{pmatrix}\nonumber\\
\begin{pmatrix}
\ell&\ell_1&2\\
-2&0&2
\end{pmatrix} +\sum_{\ell_2=4,2}^\infty(2\ell_2+1)\sqrt{\frac{(\ell_2-2)!}{(\ell_2+2)!}}\begin{pmatrix}
\ell&\ell_1&\ell_2\\
0&0&0
\end{pmatrix}\nonumber\\
\begin{pmatrix}
\ell&\ell_1&\ell_2\\
-2&0&2
\end{pmatrix}\Bigg)\mathcal{T}_{\ell\ell_1}\mu\frac{\partial^2 P_{\ell}(\mu)}{\partial\mu^2}.
\end{align}
Final set of calculation involves computing 
\begin{align}
\langle v_3v_3\rangle=4\upi \sum_{\ell\ell_1}i^{\ell}\frac{4\sqrt{(2\ell+1)}}{3}\sqrt{\upi}\begin{pmatrix}
\ell&\ell_1&0\\
0&0&0\\
\end{pmatrix}^2\mathcal{T}_{\ell\ell_1}Y_\ell^0(\hat{\bm{r}})\nonumber\\
-4\upi \sum_{\ell\ell_1}i^{\ell}\frac{4\sqrt{(2\ell+1)}}{3}\sqrt{\upi}\begin{pmatrix}
\ell&\ell_1&2\\
0&0&0\\
\end{pmatrix}^2\mathcal{T}_{\ell\ell_1}Y_\ell^0(\hat{\bm{r}}),
\end{align}
which after considering possible contribution from all other coefficients, we finally arrive to an expression for $C$ valid at all frames:
\begin{align}\label{SMHBC}
&C=2\upi \sum_{\ell}i^{\ell}\mathcal{T}_{\ell\ell}P_\ell(\mu)-4\upi\sum_{\ell\ell_1}i^{\ell}(2\ell+1)\begin{pmatrix}
\ell&\ell_1&2\\
0&0&0
\end{pmatrix}^{2}
\mathcal{T}_{\ell\ell_1}P_\ell(\mu)\nonumber\\
&-8\upi \sum_{\ell\ell_1}i^{\ell}(2\ell+1)\sqrt{\frac{2(\ell-1)!}{3(\ell+1)!}}\begin{pmatrix}
\ell&\ell_1&2\\
0&0&0
\end{pmatrix}
\begin{pmatrix}
\ell&\ell_1&2\\
-1&0&1
\end{pmatrix}\mathcal{T}_{\ell\ell_1}\nonumber\\
&\mu\frac{\partial P_{\ell}(\mu)}{\partial\mu}+4\upi\sum_{\ell \ell_1} i^l(2\ell+1)\sqrt{\frac{(\ell-2)!}{(\ell+2)!}}\Bigg(\sqrt{\frac{1}{24}}\begin{pmatrix}
\ell&\ell_1&2\\
0&0&0
\end{pmatrix} \nonumber\\
&\begin{pmatrix}
\ell&\ell_1&2\\
-2&0&2
\end{pmatrix}+\sum_{\ell_2=4,2}^\infty(2\ell_2+1)\sqrt{\frac{(\ell_2-2)!}{(\ell_2+2)!}}\begin{pmatrix}
\ell&\ell_1&\ell_2\\
0&0&0
\end{pmatrix}
\nonumber\\
&\begin{pmatrix}
\ell&\ell_1&\ell_2\\
-2&0&2
\end{pmatrix}\Bigg)\mathcal{T}_{\ell\ell_1}\left(P_\ell^2(\mu)+2\mu^2\frac{\partial^2 P_{\ell}(\mu)}{\partial\mu^2}\right).
\end{align}

\subsection{Slow modes low-$\beta$}\label{SMLB}
Slow modes in low $\beta$ have the tensor structure 
\begin{equation}
\left(\hat{\xi}_{\bm{k}}\otimes\hat{\xi}^*_{\bm{k}}\right)_{ij}\propto\hat\lambda_i\hat\lambda_j.
\end{equation}
It is clear from the above tensor structure that $c^{ij}_{\ell_2m_2}$ is only non-zero for $\ell_2=m_2=0$. This heavily simplifies Eq.\ref{specialcorr}, and subsequent calculation allows us to compute $C$ and arrive to a general form
\begin{equation}\label{SMLBC}
C=\sum_\ell 4\upi i^\ell \mathcal{T}_{\ell\ell}P_\ell(\mu).
\end{equation}
All other coefficients vanish in this mode.
\subsection{Strong turbulence}
In a strong turbulence with the uncorrelated mix of equal power Alfv\'en and slow modes, we expect pure $E$-type correlation, which has a Fourier component 
\begin{equation}\label{eq:strong}
\left(\hat{\xi}_{\bm{k}}\otimes\hat{\xi}^*_{\bm{k}}\right)_{ij}= (\delta_{ij}-\hat{\bm{k}}_i\hat{\bm{k}}_j).
\end{equation}
The real space correlation function has been already derived in LP12 using Chandrashekhar's notations, but here we derive it using the formalism we developed in the previous section.
\begin{table}
\centering
\caption{Mode structure of strong turbulence}
\begin{tabular}{@{} *8l @{}}   
\hline
\hline  
\emph{$c^{ij}_{\ell m}$} & \emph{Equation} \\
\hline
 $c^{11}_{\ell m}$ & $\left(\frac{4\sqrt{\upi}}{3}\delta_{l0}+\frac{2\sqrt{\upi}}{3\sqrt{5}}\delta_{l2}\right)\delta_{m0}-\sqrt{\frac{2\upi}{15}}\delta_{l2}(\delta_{m2}+\delta_{m,-2})$ \\ 
    &\\
 $c^{22}_{\ell m}$ & $\left(\frac{4\sqrt{\upi}}{3}\delta_{l0}+\frac{2\sqrt{\upi}}{3\sqrt{5}}\delta_{l2}\right)\delta_{m0}+\sqrt{\frac{2\upi}{15}}\delta_{l2}(\delta_{m2}+\delta_{m,-2})$\\
  
  &\\
 $c^{33}_{\ell m}$ & $\left(\frac{4\sqrt{\upi}}{3}\delta_{l0} -\frac{4\sqrt{\upi}}{3\sqrt{5}}\delta_{l2}\right)\delta_{m0}$\\
   &\\
 $c^{12}_{\ell m}$ & $i\sqrt{\frac{2 \upi }{15}}\delta_{l2}(\delta_{m2}-\delta_{m,-2})$ \\ 
  &\\
 $c^{13}_{\ell m}$ &$\sqrt{\frac{2 \upi }{15}}\delta_{l2}(\delta_{m1}-\delta_{m,-1})$\\
 &\\
 $c^{23}_{\ell m}$ &$-i\sqrt{\frac{2 \upi }{15}}\delta_{l2}(\delta_{m1}+\delta_{m,-1})$\\
 \hline
\end{tabular}
\label{tab:strongparams}
\end{table}
We will use Table \ref{tab:strongparams} in the subsequent calculations in this section. To find the coefficients $A, B, C$ and $D$, we follow the procedure described in Appendix \ref{general}. Consider
\begin{align}
\langle v_1v_2\rangle=
4\upi\sum_\ell i^{\ell}\sum_{\ell_1}\sqrt{\frac{2\upi}{3}}\sqrt{(2\ell+1)}\mathcal{T}_{\ell\ell_1}\begin{pmatrix}
\ell&\ell_1&2\\
0&0&0\\
\end{pmatrix}\nonumber\\
\begin{pmatrix}
\ell&\ell_1&2\\
-2&0&2
\end{pmatrix}i(Y_\ell^2(\hat{\textbf{r}})-Y_\ell^{-2}(\hat{\textbf{r}})).
\end{align}
In the frame we are dealing with, only $\hat{\textbf{r}}_1\hat{\textbf{r}}_2$ contributes, and therefore upon simplification, we obtain (after considering that $A$ should only depend on $\mu=\hat{r}.\hat\lambda$),
\begin{align}
A=-8\upi\sum_\ell i^{\ell}\sum_{\ell_1}(2\ell+1)\sqrt{\frac{2(\ell-2)!}{3(\ell+2)!}}\mathcal{T}_{\ell\ell_1}\begin{pmatrix}
\ell&\ell_1&2\\
0&0&0\\
\end{pmatrix}\nonumber\\
\begin{pmatrix}
\ell&\ell_1&2\\
-2&0&2
\end{pmatrix}\frac{\partial^2P_\ell(\mu)}{\partial \mu^2}.
\end{align}
Similarly, we compute  
\begin{align}
\langle v_1v_1\rangle=
\frac{8\upi}{3}\sum_\ell i^\ell\mathcal{T}_{\ell\ell}P_\ell(\cos\theta)+\frac{4\upi}{3}\sum_{\ell\ell_1}i^\ell(2\ell+1)\mathcal{T}_{\ell\ell_1}\nonumber\\\begin{pmatrix}
\ell&\ell_1&2\\
0&0&0
\end{pmatrix}^2P_\ell(\cos\theta)
-4\upi\sum_{\ell\ell_1} i^{\ell}\sqrt{\frac{2\upi}{3}}\sqrt{(2\ell+1)}\mathcal{T}_{\ell\ell_1}\nonumber\\
\begin{pmatrix}
\ell&\ell_1&2\\
0&0&0
\end{pmatrix}
\begin{pmatrix}
\ell&\ell_1&2\\
-2&0&2
\end{pmatrix}(Y_\ell^2(\hat{\textbf{r}})+Y_\ell^{-2}(\hat{\textbf{r}})).
\end{align}
In our choice of co-ordinates, $\langle v_1v_1^*\rangle$ is contributed by $A$ and $B$. Therefore to find $B$, we need to subtract the contribution of $A$. This subtraction affects the last factor in the above equation, and we are left with
\begin{align}
B=\frac{8\upi}{3}\sum_\ell i^\ell\mathcal{T}_{\ell\ell}P_\ell(\mu)+\frac{4\upi}{3}\sum_{\ell\ell_1}i^\ell(2\ell+1)\mathcal{T}_{\ell\ell_1}\begin{pmatrix}
\ell&\ell_1&2\\
0&0&0\\
\end{pmatrix}^2\nonumber\\
P_\ell(\mu)+4\upi\sum_{\ell\ell_1} i^{\ell}(2\ell+1)\sqrt{\frac{2(\ell-2)!}{3(\ell+2)!}}\mathcal{T}_{\ell\ell_1}\begin{pmatrix}
\ell&\ell_1&2\\
0&0&0\\
\end{pmatrix}\nonumber\\
\begin{pmatrix}
\ell&\ell_1&2\\
-2&0&2
\end{pmatrix}P_\ell^2(\mu).
\end{align}
Next, we compute 
\begin{align}
\langle v_1v_3\rangle=4\upi\sum_{\ell\ell_1} i^{\ell}\sqrt{\frac{2\upi}{3}}\sqrt{(2\ell+1)}\mathcal{T}_{\ell\ell_1}\begin{pmatrix}
\ell&\ell_1&2\\
0&0&0\\
\end{pmatrix}\nonumber\\
\begin{pmatrix}
\ell&\ell_1&2\\
-1&0&1
\end{pmatrix}(-1)(Y_\ell^1(\hat{\textbf{r}})+Y_\ell^{*1}(\hat{\textbf{r}})).
\end{align}
Using the fact that the above correlation is contributed by $A$ and $D$, and subtracting the contribution of $A$, we finally obtain
\begin{align}
D=4\upi\sum_{\ell\ell_1} i^{\ell}(2\ell+1)\sqrt{\frac{2(\ell-1)!}{3(\ell+1)!}}\mathcal{T}_{\ell\ell_1}\begin{pmatrix}
\ell&\ell_1&2\\
0&0&0\\
\end{pmatrix}\nonumber\\
\begin{pmatrix}
\ell&\ell_1&2\\
-1&0&1
\end{pmatrix}
\frac{\partial P_\ell(\mu)}{\partial\mu}
+8\upi\sum_{\ell\ell_1} i^{\ell}(2\ell+1)\sqrt{\frac{2(\ell-2)!}{3(\ell+2)!}}\mathcal{T}_{\ell\ell_1}\nonumber\\
\begin{pmatrix}
\ell&\ell_1&2\\
0&0&0\\
\end{pmatrix}
\begin{pmatrix}
\ell&\ell_1&2\\
-2&0&2
\end{pmatrix}\mu\frac{\partial^2P_\ell(\mu)}{\partial \mu^2}.
\end{align}
Finally, to obtain $C$ we compute
\begin{align}
\langle v_3v_3\rangle=
\frac{8\upi}{3}\sum_\ell i^\ell\mathcal{T}_{\ell\ell}P_\ell(\cos\theta)-\frac{8\upi}{3}\sum_{\ell\ell_1}i^\ell(2\ell+1)\mathcal{T}_{\ell\ell_1}\nonumber\\\begin{pmatrix}
\ell&\ell_1&2\\
0&0&0\\
\end{pmatrix}^2P_\ell(\cos\theta)
\end{align}
The above correlation comes from the contribution of $A$, $B$, $C$ and $D$. Therefore, to find $C$, we subtract all other contributions to obtain
\begin{align}
C=-4\upi\sum_{\ell\ell_1}i^\ell(2\ell+1)\mathcal{T}_{\ell\ell_1}\begin{pmatrix}
\ell&\ell_1&2\\
0&0&0\\
\end{pmatrix}^2P_\ell(\mu)-8\upi\nonumber\\
\sum_{\ell\ell_1} i^{\ell}(2\ell+1)
\sqrt{\frac{2(\ell-1)!}{3(\ell+1)!}}\mathcal{T}_{\ell\ell_1}\begin{pmatrix}
\ell&\ell_1&2\\
0&0&0\\
\end{pmatrix}
\begin{pmatrix}
\ell&\ell_1&2\\
-1&0&1
\end{pmatrix}
\nonumber\\
\mu\frac{\partial P_\ell(\mu)}{\partial\mu}
-4\upi\sum_{\ell\ell_1} i^{\ell}(2\ell+1)\sqrt{\frac{2(\ell-2)!}{3(\ell+2)!}}\mathcal{T}_{\ell\ell_1}\begin{pmatrix}
\ell&\ell_1&2\\
0&0&0\\
\end{pmatrix}\nonumber\\
\begin{pmatrix}
\ell&\ell_1&2\\
-2&0&2
\end{pmatrix}\left(P_\ell^2(\mu)+2\mu^2\frac{\partial^2P_\ell(\mu)}{\partial \mu^2}\right)
\end{align}
\section{Approximate Expression For The $z-$ Projection Of The Velocity Structure Function}\label{sta}
To study intensity maps analytically, we require knowledge of the $z-$ projection of velocity structure function. An anisotropic velocity structure function manifests in the anisotropy of intensity channel maps. Therefore, for our analytical calculation, we first study how anisotropy is built in the $z-$ projection of the velocity structure function. 
The projection structure function is given
\begin{align}\label{strd1}
D_z(\bm{r})=2[(B(0)-B)+(C(0)-C)\cos^2\gamma-A\cos^2\theta \nonumber\\
-2D\cos\theta\cos\gamma]~,
\end{align}
where $A, B, C$ and $D$ depend on the particular mode of turbulence and has been derived in Appendix [\ref{allstf}] for different modes of turbulence. For the analysis we carry out, it is particularly useful to do the multipole decomposition of these coefficients in Legendre polynomials, so that
\begin{equation}
A=\sum_n A_n(r)P_n(\mu)~,
\end{equation}
and so on, where $A_n(r)$ can be easily obtained with the knowledge of $A$. The expression above is particularly useful to obtain approximate expression for $D_z(r,\mu)$, as the coefficients $A_n(r)$ are usually a decreasing function of $n$. This motivates us to write $D_z(r,\mu)$ by considering the coefficients only upto second order in $n$, i.e. $A=A_0+A_2P_2(\mu)$ and so on.
We define that the power spectrum  $\mathcal{A}_{\ell_1} \propto k^{-m}$ (cf. Eq. \eqref{randomcorr}). Keeping this in mind, it can be shown that each regularised coefficients $A, B(0)-B, C(0)-C, D$ are proportional to $r^{m-3}\equiv r^\nu$. Since $A, B,..$ are functions of $r$ in the same fashion, we explicitly factor out $r^\nu$ from them, so that in the following analysis, it is to be understood that any $r^\nu$ factor comes from these coefficients, and $A_n,..$ are simply some numerical constants.  With these approximations and definitions,  Eq. \eqref{strd1}  can be written as
\begin{align}\label{strapp}
D_z(\bm{r})\approx 2[(B_0(0)-B_0(r)-B_2(r)P_2(\mu))+(C_0(0)-C_0(r)\nonumber\\
-C_2(r)P_2(\mu))\cos^2\gamma
-(A_0(r)+A_2(r)P_2(\mu))\cos^2\theta \nonumber\\-2D_1(r)\mu\cos\theta\cos\gamma]
\end{align}
To obtain the explicit dependence of $D_z$ on $\phi$, we use the relation for $\mu$ in-terms of different angles involved in our setup 
\begin{equation}
\mu=\sin\gamma\sin\theta\cos\phi+\cos\gamma\cos\theta,
\end{equation}
which when used in Eq. \eqref{strapp} shows that $D_z$ is related to $\phi$ only upto $\cos^2\phi$: 
\begin{equation}\label{fappstr}
D_z(\bm{r})\approx c_1-c_2\cos\phi-c_3\cos^2\phi.
\end{equation}
The detailed relations involving parameters $c_1, c_2$ and $c_3$ are presented in the Table [\ref{tab:structurepara}]. For the sake of clarity, these parameters are themselves broken into different pieces. As we will show later, this representation will be useful when carrying out the $z-$ integral to find the intensity structure function. 
\begin{table}
\centering
\caption{Different parameters in the approximate $D_z$}
\begin{tabular}{@{} *8l @{}}  
\hline
\hline  
\emph{Parameters} & \emph{Equation} \\
\hline
&\\
$c_1 $& $(q_1 +q_2\cos^2\theta+q_3\cos^4\theta)r^\nu$\\
  &\\
 $c_2$ & $(s_1+s_2\cos^2\theta)r^\nu\sin\theta\cos\theta\sin \gamma\cos\gamma$ \\ 
    &\\
 
 $c_3$ & $(u_1+u_2\cos^2\theta)r^\nu\sin^2\theta\sin^2\gamma$ \\ 
 &\\
 \hline
 &\\
$q_1 $&$2(B_0(0)-B_0)+2(C_0(0)-C_0)\cos^2\gamma+B_2+C_2\cos^2\gamma. $\\
 &\\
 $q_2$ & $-2A_0+A_2-4D_1\cos\gamma-3(B_2+C_2\cos^2\gamma)\cos^2\gamma$\\
 &\\
 $q_3$ &$-3A_2\cos^2\gamma$\\
 &\\
 \hline
&\\
$s_1$ & $ 6(B_2+C_2\cos^2\gamma)+ 4D_1 $\\
&\\
$s_2$ & $ 6A_2 $\\
&\\
\hline
&\\
$u_1$ & $3(B_2+C_2\cos^2\gamma)$\\
&\\
$u_2$ & $3A_2$\\
&\\
\hline
\end{tabular}
\label{tab:structurepara}
\end{table}

For the sake of convenience for further analysis, we write Eq. \eqref{fappstr} as
\begin{equation}
D_z(\bm{r})\approx f_1(1-f_2\cos\phi-f_3\cos^2\phi),
\end{equation}
where,
\begin{equation}
f_1=c_1,\qquad f_2=\frac{c_2}{c_1},\qquad f_3=\frac{c_3}{c_1}.
\end{equation}

\section{Evaluating $\phi$ Integral for Pure Velocity Term}\label{phia}
In order to fully obtain multipole moments of intensity structure function (cf. \ref{multipolevelocity}), we need to evaluate the integral of the  form  $$\int_0^{2\upi}d\phi\frac{\mathrm{e}^{-im\phi}}{\sqrt{f_1(1-f_2\cos\phi-f_3\cos^2\phi)}}.$$ To evaluate this integral, we will use generalised Gegenbauer polynomial expansion \citep{plunkett1975three}  defined as 
\begin{equation}
\frac{1}{(1-\rho x-\zeta x^2)^\alpha}=\sum_{n=0}^\infty C_n^{(\alpha)}(\rho,\zeta)x^n,
\end{equation}
where, $\rho+\zeta<1$, and 
\begin{align}
C_n^{(\alpha)}(\rho,\zeta)=&\frac{\rho^n\Gamma[\alpha+n-1]}{\Gamma[\alpha]n!}\nonumber\\
&\text{ }_2F_1\left(-\frac{n}{2},\frac{-n+1}{2};-\alpha-n+2;\frac{-4\zeta}{\rho^2}\right)
\end{align} 
or equivalently 
\begin{equation}
C_n^{(\alpha)}(\rho)=\sum_{j=0}^{\lfloor n/2\rfloor}\frac{\Gamma(n-j+\alpha)}{\Gamma(\alpha)j!\Gamma[n-2j+1]}\zeta^j\rho^{n-2j}.
\end{equation}

Using the above equations we can write
\begin{align}\label{jainintegral}
\int_0^{2\upi}\mathrm{d}\phi&\frac{\mathrm{e}^{-im\phi}}{\sqrt{f_1(1-f_2\cos\phi-f_3\cos^2\phi)}}\nonumber\\
&=\sum_{n=0}^\infty\frac{C_n^{(1/2)}(f_2,f_3)}{\sqrt{f_1}}\int_0^{2\upi}d\phi \mathrm{e}^{-im\phi}\cos^n\phi\nonumber\\
&=\sum_{n=m,2}^\infty\frac{2^{-n}C_n^{(1/2)}(f_2,f_3)}{\sqrt{f_1}}\frac{2\upi\Gamma[n+1]}{\Gamma\left[\frac{n-m}{2}+1\right]\Gamma\left[\frac{n+m}{2}+1\right]},
\end{align}
where the sum in $n$ starts at $m$ and proceeds at a step of 2, which implies that $m$ and $n$ should have the same parity. This parity information is particularly useful later to arrive to the conclusion that only even multipoles survive. For any $n<m$, the integral is zero, therefore, these terms have no contribution. 
Upon using definition of $C_n^{(1/2)}$, and considering the fact that $n$ is positive to write $\Gamma[n+1]=n!$, we have
\begin{align}
\int_0^{2\upi}\mathrm{d}\phi\frac{\mathrm{e}^{-im\phi}}{\sqrt{f_1(1-f_2\cos\phi-f_3\cos^2\phi)}}
\nonumber\\
=\sum_{n=m,2}^\infty \frac{2\sqrt{\upi}}{\sqrt{f_1}}\frac{2^{-n}\sin^n\gamma\Gamma\left[n+1\right]}{\Gamma\left[\frac{n-m}{2}+1\right]\Gamma\left[\frac{n+m}{2}+1\right]}\sum_{j=0}^{\lfloor n/2\rfloor}\frac{\Gamma\left[n-j+\frac{1}{2}\right]}{j!\Gamma[n-2j+1]}\nonumber\\(\cos\gamma)^{n-2j}f_3^jf_2^{n-2j}.
\end{align}
\section{Evaluating $z$ Integral for Pure Velocity Term}\label{za}
To obtain multipole moments of the intensity structure function, we now carry out the $z-$ integral (cf. Eq.  \ref{multipolevelocity})
\begin{align}
&\int_{-\infty}^\infty \mathop{dz}\frac{1}{\sqrt{f_1}}f_3^jf_2^{n-2j}=\int_{-\infty}^\infty \mathop{dz}c_1^{-n+j-1/2}c_2^{n-2j}c_3^j\nonumber\\
&=\int_{-\infty}^\infty \mathop{dz}\cos^{n-2j}\theta(q_1+q_2\cos^2\theta+q_3\cos^4\theta)^{-n+j-1/2}\nonumber\\
&(s_1+s_2\cos^2\theta)^{n-2j}(u_1+u_2\cos^2\theta)^j\sin^{n}\theta r^{-\nu/2}.
\end{align}
Using $\sin\theta=R/r$, and $\cos\theta=z/r$, we have
\begin{align}\label{Imprtint}
\int_{-\infty}^\infty \mathop{dz}\frac{1}{\sqrt{f_1}}f_3^jf_2^{n-2j}=\int_{-\infty}^\infty \mathop{\mathrm{d}z}r^{-\nu/2-2(n-j)}R^{n}z^{n-2j}\nonumber\\
(q_1+q_2r^{-2}z^2
+q_3r^{-4}z^4)^{-n+j-1/2}\nonumber\\
(s_1+s_2r^{-2}z^2)^{n-2j}(u_1+u_2r^{-2}z^2)^j.
\end{align}
One of the most important points to note at this stage is that for odd $n$, the above integral vanishes, since for odd $n$, $z^{n-2j}$ is an odd function in $z$, while all other functions involved in this problem are even. This implies that the  multipole contribution, which is the weight of $e^{-im\phi}$, comes only from even $m$, which is consistent with the symmetry of our problem. 

Note that the above is valid only when $n\geq 2$, for $\nu>0$. When $n=0$, we have to consider regularization of the integral (cf. Eq. \eqref{multipolevelocity}). The integral we are interested at, when $n=0$, is 
\begin{equation}\label{Izero}
\mathcal{I}_0=\int_{-\infty}^\infty\mathop{dz}\left[\frac{1}{\sqrt{q_1+q_2+q_3}z^{\nu/2}}-\frac{1}{\sqrt{q_1+q_2\cos^2\theta+q_3\cos^4\theta}r^{\nu/2}}\right]
\end{equation}
which after change of variable $z=R\cot\theta$ can also be written as
\begin{align}\label{zeroI}
\mathcal{I}_0=R^{1-\nu/2}\int_{0}^\upi\mathrm{d}\theta\frac{1}{\sin^2\theta}\Bigg[&\frac{(\tan\theta)^{\nu/2}}{\sqrt{q_1+q_2+q_3}}\nonumber\\
&-\frac{(\sin\theta)^{\nu/2}}{\sqrt{q_1+q_2\cos^2\theta+q_3\cos^4\theta}}\Bigg].
\end{align}

An approximate form of \eqref{Izero} can be obtained by method of series expansion. For that we write the integrand as
\begin{align}
\frac{1}{\sqrt{q_1+q_2+q_3}z^{\nu/2}}-\frac{1}{\sqrt{q_1+q_2\cos^2\theta+q_3\cos^4\theta}r^{\nu/2}}\nonumber\\
\approx \frac{1}{\sqrt{q_1+q_2}z^{\nu/2}}-\frac{1}{\sqrt{q_1+q_2\frac{z^2}{R^2+z^2}}r^{\nu/2}}\nonumber\\
\approx\left[\frac{1}{\sqrt{q_1+q_2}z^{\nu/2}}-\frac{1}{\sqrt{q_1+q_2}r^{\nu/2}}\right]+\frac{q_2R^2}{2(q_1+q_2)^{3/2}r^{2+\nu/2}},
\end{align}
where in the first step, we used the fact that $q_3$ contribution is negligible\footnote{We verified this numerically. Analytically this can be understood by noting that $q_1$ and $q_2$ consists of monopole contribution while $q_3$ consists of only quadrupole contribution (cf. Table. \ref{tab:structurepara}).}. The above approximation is fairly good as long as $q_1+q_2>q_2$. With this approximation, we finally arrive to 
\begin{equation}\label{HI0result}
\mathcal{I}_0\approx -R^{1-\nu/2}\sqrt{\frac{\upi}{q_1+q_2}}\left[\frac{\Gamma \left(\frac{\nu}{4}-\frac{1}{2}\right)}{\Gamma \left(\frac{\nu}{4}\right)}-\frac{q_2}{2(q_1+q_2)}\frac{\Gamma \left(\frac{\nu}{4}+\frac{1}{2}\right)}{\Gamma \left(\frac{\nu}{4}+1\right)}\right].
\end{equation}
To evaluate Eq.\eqref{Imprtint} we first note the the following: due to the presence of a factor $z^{n-2j}$, which is a suppressing factor for small $z$, and for $n\neq 2j$, the integral in Eq. \eqref{Imprtint} gives significantly small value when $n\neq 2j$ in comparison to the case when $n=2j$. Therefore, we will only consider the case when $n=2j$. To make further simplifications, we  approximate the integrand as
\begin{align}
r^{-\nu/2-n}R^{n}(q_1+q_2r^{-2}z^2
+q_3r^{-4}z^4)^{-n/2-1/2}(u_1+u_2r^{-2}z^2)^{n/2}\nonumber\\\approx r^{-\nu/2-n}R^{n}(q_1+q_2r^{-2}z^2)^{-n/2-1/2}u_1^{n/2}\nonumber\\
\approx r^{-\nu/2-n}R^{n}q_1^{-n/2-1/2}u_1^{n/2}\left(1-\frac{n+1}{2}\frac{q_2}{q_1}\frac{z^2}{r^2}\right),
\end{align}
where we have carried out expansion valid for $q_1>q_2$. Therefore, we finally have
\begin{align}\label{HIresult}
\mathcal{I}\approx R^{1-\nu/2}\frac{\sqrt{\upi }}{q_1^{(n+1)/2}} \left(\frac{\Gamma \left(\frac{\nu}{4}+\frac{n-1}{2}\right)}{\Gamma \left(\frac{\nu}{4}+\frac{n}{2}\right)}-\frac{(n+1)}{4}\frac{q_2}{q_1} \frac{\Gamma \left(\frac{\nu}{4}+\frac{n+1}{2}\right)}{\Gamma \left(\frac{\nu}{4}+\frac{n}{2}+1\right)}\right)u_1^{n/2}.
\end{align}
Eqs. (\ref{HI0result}, \ref{HIresult}) allow us to obtain multipole moment of any even order.
\bsp
\label{lastpage}
\end{document}